%% file: attractive_trashcan_draft.tex
\documentclass[prb,twocolumn,notitlepage,superscriptaddress,nofootinbib]{revtex4-2}
\usepackage{bbm}
\usepackage{amsmath,amssymb}
\usepackage[hidelinks,colorlinks,linkcolor=blue,
citecolor=blue,urlcolor=blue]{hyperref}
\usepackage{graphicx,siunitx}
\usepackage{makecell}
\usepackage{etoolbox}
\usepackage[dvipsnames]{xcolor}
\usepackage{flafter}
\usepackage{simpler-wick}
\usepackage{pifont}

\usepackage[normalem]{ulem}

\usepackage{xcolor}
\DeclareUnicodeCharacter{3000}{\textcolor{red}{BAD!!}}

\newcommand{\be}{\begin{equation}}
\newcommand{\ee}{\end{equation}}
\newcommand{\bea}{\begin{equation} \begin{aligned}}
\newcommand{\eea}{\end{aligned} \end{equation} }
\newcommand{\bi}{\begin{itemize}}
\newcommand{\ei}{\end{itemize}}

\renewcommand{\be}{\beta}

\newcommand{\bpm}{\begin{pmatrix}}
\newcommand{\epm}{\end{pmatrix}}

\newcommand{\bk}{\mathbf{k}}
\newcommand{\br}{\mathbf{r}}

\usepackage{amsmath,amsfonts,amssymb,amsthm,epsfig,array}
\usepackage{dsfont}
\usepackage{slashed}
\usepackage{graphics}

\usepackage{float}
\usepackage{verbatim}

\usepackage{color}
\usepackage{tabularx}
\usepackage[mathscr]{euscript}
\usepackage{mathtools}

\newcommand{\mbf}[1]{\boldsymbol{#1}}

\newcommand{\bra}[1]{\langle #1|}
\newcommand{\ket}[1]{|#1 \rangle}

\usepackage{bm}
\usepackage{environ}
\NewEnviron{eqs}{%
\begin{equation}\begin{split}
    \BODY
\end{split}\end{equation}
}

\let\oldAA\AA
\renewcommand{\AA}{\text{\normalfont\oldAA}}

\makeatletter
\newcommand{\appendixlocaltoc}{%
\begingroup
\let\addcontentsline\oldaddcontentsline
\let\clearpage\relax
\section*{Appendix Contents}
\endgroup
\@starttoc{atc}%
}
\newcommand{\startappendixlocaltoc}{%
\addtocontents{atc}{\protect\setcounter{tocdepth}{3}}
\global\let\oldaddcontentsline\addcontentsline
\renewcommand{\addcontentsline}[3]{%
\ifstrequal{##1}{toc}{%
\ifstrequal{##2}{section}{\oldaddcontentsline{atc}{##2}{##3}}{%
\ifstrequal{##2}{subsection}{\oldaddcontentsline{atc}{##2}{##3}}{%
\ifstrequal{##2}{subsubsection}{\oldaddcontentsline{atc}{##2}{##3}}{}%
}%
}%
}{%
\oldaddcontentsline{##1}{##2}{##3}%
}%
}%
}
\makeatother

\begin{document}

\title{Berry Trashcan With Short Range Attraction: \\ 
Exact $p_x+ip_y$ Superconductivity in Rhombohedral Graphene }

\author{Ming-Rui Li}
\thanks{These authors contributed equally in this work}
\affiliation{Institute for Advanced Study, Tsinghua University, Beijing 100084, China}
\affiliation{Department of Physics, Princeton University, Princeton, New Jersey 08544, USA}
\author{Yves H. Kwan}
\thanks{These authors contributed equally in this work}
\affiliation{Princeton Center for Theoretical Science, Princeton University, Princeton NJ 08544, USA}
\affiliation{Department of Physics, University of Texas at Dallas, Richardson, Texas 75080, USA}
\author{Hong Yao}
\affiliation{Institute for Advanced Study, Tsinghua University, Beijing 100084, China}
\author{B. Andrei Bernevig}
\email{bernevig@princeton.edu}
\affiliation{Department of Physics, Princeton University, Princeton, New Jersey 08544, USA}
\affiliation{Donostia International Physics Center, P. Manuel de Lardizabal 4, 20018 Donostia-San Sebastian, Spain}
\affiliation{IKERBASQUE, Basque Foundation for Science, Bilbao, Spain}

\date{\today}

\begin{abstract}
We show the presence of analytic $p_x + i p_y$ superconducting ground states in the Berry Trashcan --- a minimal model of rhombohedral graphene valid for $n \ge 4$ layers --- under short-range attractive interactions. We demonstrate that the model, whose dispersion consists of a flat bottom surrounded by steep walls of prohibitive kinetic energy, serves as a building block to understand superconductivity in the moir\'e-free limit. We find that the ground-state chirality has a ``ferromagnetic'' coupling to that of the uniform Berry curvature of the model, and compare the analytically obtained binding energies, excitation spectra and off-diagonal long-range order (ODLRO) with numerical exact diagonalization results. We show that the analytic structure of this model is that of a restricted spectrum generating algebra (RSGA), initially developed for quantum scars, and build a variety of other exact (but contrived) models with exact chiral superconductivity based on a method developed in Ref.~\cite{herzogarbeitman2022manybodysuperconductivity}. However, under short range attraction, we show that the Berry Trashcan is the optimal and only realistic point in the class of GMP-like algebras to host a chiral superconductor state. A toy model in 1D and its related physics is also investigated. Our results reveal that chiral superconductivity is natural under attractive interactions in the Berry trashcan model of rhombohedral graphene in displacement field, although we make no claim about the origin of the attraction. 
\end{abstract}

\maketitle

\section{Introduction}


\begin{figure}[ht]
    \centering
    \includegraphics[width=0.65\linewidth]{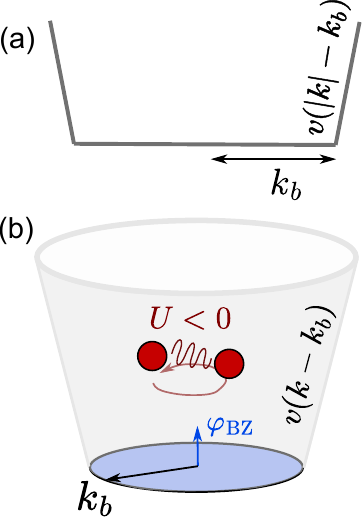}
    \caption{(a) The dispersion of the 1D toy trashcan model consists of a flat bottom with momentum width $2k_b$, surrounded by steep walls with velocity $v$. (b) The 2D Berry Trashcan has a flat bottom with radius $k_b$, which encloses a flux $\varphi_\text{BZ}$ of Berry curvature. Attractive $U<0$ interactions stabilize a superconductor whose chirality is aligned with the underlying Berry curvature.}
    \label{fig:main_splash}
\end{figure}

Recent experiments on $n$-layer rhombohedral graphene (R$n$G) under a displacement field $D$ have uncovered a wide variety of strongly-correlated phenomena~\cite{zhou2021half,zhou2021superconductivity,shi2020electronic,myhro2018large,han2023orbital,han2024correlated,han2024large,sha2024observation,Han2025Signatures,Patterson2025Superconductivity,Yang2025,liu2024spontaneous,zhang2024layer,zhang2024graphite,morissette2025stripedsuperconductorrhombohedralhexalayer,morissette2025coulombdriven,kumar2025superconductivitydualsurfacecarriersrhombohedral,nguyen2025hierarchytopologicalsuperconductingstates,deng2025superconductivityferroelectricorbitalmagnetism,auerbach2025visualizingisospinmagnetictexture,li2025transdimensional,seo2025familyunconventionalsuperconductivitiescrystalline}, such as superconductivity (SC), symmetry-breaking, and correlated insulators. While the mechanisms underlying the multitude of observed phases remain poorly understood, experiments consistently produce contrasting results depending on the alignment with the encapsulating hBN. If the R$n$G is aligned with one of the hBN substrates, thereby forming a moir\'e pattern, integer and fractional Chern insulators~\cite{neupert2011fci,regnault2011fci,sheng2011fractional,sun2011nearly,tang2011high} (CI/FCI) have been found at commensurate filling factors, but only, puzzlingly, when the doped electrons are driven away from the moir\'e interface~\cite{Lu2024fractional,xie2024even,choi2024electricfieldcontrolsuperconductivity,WatersChern2025,75gl-jzl6,lu2025extended,zhou2024layer,chen2020tunable,han2024engineering,ding2024electricalswitchingchiralityrhombohedral,zheng2024switchablecherninsulatorisospin,xiang2025continuouslytunableanomaloushall,wang2025electricalswitchingcherninsulators,li2025tunablechern,xie2025unconventionalorbital}. The nature of the moir\'e potential and its role in stabilizing such topological states have been subject to intense debate~\cite{dong2024AHC1,zhou2024fractional,dong2024theorypentalayer,guo2024fractional,jonahMFCI2,kwan2023MFCI3,dong2024stability,soejima2024AHC2,tan2024parent,zeng2024sublattice,xie2024integerfractional,crepel2024efficientpredictionsuperlatticeanomalous,kudo2024quantumanomalousquantumspin,sarma2024thermal,huang2024selfconsistent,yu2024MFCI4,shavit2024entropy,huang2025displacement,huang2024impurityinducedthermalcrossoverfractional,guo2024beyondmeanfieldstudieswignercrystal,wei2025edge,tan2024wavefunctionapproachfractionalanomalous,zhou2024newclassesquantumanomalous,zeng2024berryphasedynamicssliding,li2025multiband,uzan2025hbnalignmentorientationcontrols}. A small set of theories~\cite{kwan2023MFCI3,yu2024MFCI4} predicted that the moir\'e potential is essential to both the CI and the FCI, and that the moir\'e-less Wigner crystal obtained in Hartree-Fock studies~\cite{kwan2023MFCI3,dong2024AHC1,dong2024theorypentalayer,zhou2024fractional,bernevig2025berrytrashcanmodelinteracting,zhou2024newclassesquantumanomalous} is unstable~\cite{kwan2023MFCI3,yu2024MFCI4,zhou2024newclassesquantumanomalous,dong2025phononselectroncrystalsberry,desrochers2025elasticresponseinstabilitiesanomalous}.  It was then experimentally found that, in the \emph{absence} of hBN alignment, R$n$G exhibits SC, without any signatures of CI/FCI, for a similar regime of displacement field $D$ and electronic density $n_e$~\cite{Han2025Signatures,morissette2025stripedsuperconductorrhombohedralhexalayer,nguyen2025hierarchytopologicalsuperconductingstates,seo2025familyunconventionalsuperconductivitiescrystalline}. The detection of an anomalous Hall effect in the normal state and magnetic hysteresis point to the possibility of chiral, nonzero-momentum (FFLO~\cite{fulde1964superconductivity,larkin1965inhomogeneous}) SC, which has triggered significant theoretical attention~\cite{geier2024chiraltopological,yang2024incommensurate,kim2025beyondpairing,ChouIntravalley2025,wang2024chiralsuperconductivity,qin2024chiralfinite,dong2023sc,PangburnscgrapheneI,AdelinescgrapheneII,PangburnscgrapheneIII,yu2025quantumgeometryquantummaterials,jahin2025enhancedkohnluttinger,sau2024theoryanomalous,parramartinez2025bandrenormalization,dong2025controllabletheory,yoon2025quartermetal,kolář2025singlegatetrackingbehaviorflatband,sedov2025probing,maymann2025pairingmechanism,christos2025finitemomentum,chen2025diode,patri2025family,jiang2025quantumgeometrysurfacestates,raines2025superconductivityinducedspinorbitcoupling,levitan2025trigonalwarping,raines2025superconductivityparamagnonmagnonexchange,han2025exactmodels}.


A key theoretical challenge is to determine whether a reduced effective model of R$n$G can, independent of microscopic band-structure details, simultaneously support chiral finite-momentum SC in the absence of moiré alignment \emph{and} stabilize CI/FCI once moiré effects are included. The Berry Trashcan, first introduced in Ref.~\cite{bernevig2025berrytrashcanmodelinteracting}, is an idealized interacting continuum model that captures the important low-energy features of R$n$G at large $D$. Within a spin-valley sector, the Berry Trashcan has a single conduction band whose dispersion  consists of a flat region with momentum scale $k_b$ (the trashcan bottom),  surrounded by steeply dispersing walls (Fig.~\ref{fig:main_splash}b). 
This naturally leads to the notion of a `filling factor' $\nu$, defined as the electronic density relative to the area of the trashcan bottom.
The single-particle wavefunctions exhibit uniform Berry curvature, whose form factors satisfy the Girvin-MacDonald-Platzman (GMP) algebra obeyed by the lowest Landau level~\cite{girvin1986magnetoroton}. The non-trivial quantum geometry is known to have significant effects on the potential superconductivity~\cite{peotta2015superfluidity,wangweakpairing2018,PhysRevLett.123.237002,10.21468/SciPostPhys.6.5.060,Schindlerpairing2020,xietopology2020,PhysRevLett.128.087002,Ma2021,Törmä2022,peotta2023quantumgeometrysuperfluiditysuperconductivity,PhysRevB.106.184507,Hofmann2023sccdw,Tianevidence2023,jiang2025quantumgeometrysurfacestates,tanaka2025superfluid,banerjee2025superfluid,yu2025quantumgeometrynbse2family}. In the moir\'e-free case, we neglect the valence bands (though there can be inter-band polarization effects~\cite{repulsive_unpub}) owing to the sizable displacement-field-induced gap.
Provided the electronic density remains below $\pi k_b^2$, the flat bottom promotes full spin-valley ferromagnetism~\cite{antebi2024stoner,bernevig2021TBGIII,lian2021TBGIV,bernevig2021TBGV,kang2019strong}, a prerequisite for FCIs and chiral SC.
The critical question that then arises is: what many-body states are realized within the spin- and valley-polarized Berry Trashcan?

Previously, a mean-field analytical study~\cite{bernevig2025berrytrashcanmodelinteracting} of Wigner crystals for repulsive interactions in the Berry Trashcan at $\nu\simeq 1$ produced a Hartree–Fock phase diagram closely resembling those obtained from more detailed models~\cite{kwan2023MFCI3,dong2024AHC1,dong2024theorypentalayer,zhou2024fractional,bernevig2025berrytrashcanmodelinteracting,zhou2024newclassesquantumanomalous}. 
In particular, it explained why the Chern number of Wigner crystal is ferromagnetically coupled to the Berry curvature of the conduction band.
Here, to address the experimentally observed SC in R$n$G, we examine the Berry Trashcan with short-range attractive density-density interactions using analytics and exact diagonalization (ED) for all $\nu\leq 1$. We also introduce and study a related toy 1D model. We uncover exact (ground) states that (approximately) descend from a restricted spectrum generating algebra (RSGA), first introduced in the context of quantum scars~\cite{PhysRevB.102.085140,moudgalya2022review}. For the 2D Berry Trashcan, we demonstrate that these ground states (GS) exhibit $p_x+ip_y$ pairing that is ferromagnetic with the intrinsic Berry curvature of R$n$G, and off-diagonal long-range order (ODLRO)~\cite{yang1962ODLRO}. While we do not yet assert a microscopic mechanism for the attraction, our results show that R$n$G --- in particular its Berry curvature and dispersion-less bottom --- provides a natural host for chiral $p_x+ i p_y$ SC.

\section{1D Toy Model}

\begin{figure*}[t]
    \centering
    \includegraphics[width=1.0\linewidth]{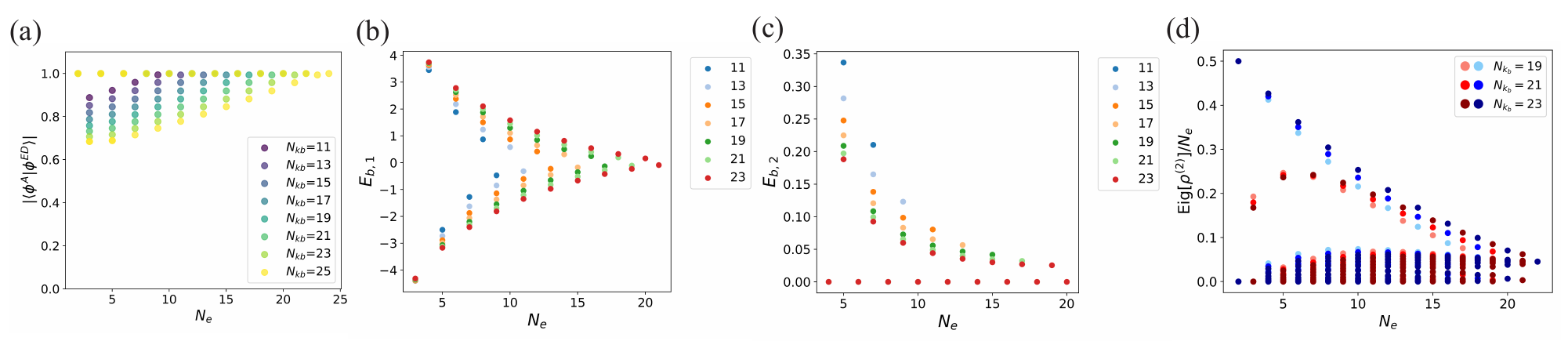}
    \caption{ Superconductivity of the $p=0$ GS in the attractive 1D toy trashcan model with $v=\infty$. (a) Wavefunction overlap between the analytic ansatz $\ket{\phi^A}$ 
    and the GS $\ket{\phi^{ED}}$ obtained numerically with ED. The ansatz is exact for even $N_e$. 
    (b),(c) Binding energies $E_{b,1},E_{b,2}$ (Eq.~\ref{eq:binding_1d_main}) extracted from ED with $U=-1, \,k_b=\pi$, for different $L+1=N_{k_b}$ indicated in the legend. $E_{b,1}$ exhibits even/odd oscillations with $N_e$, indicating electron pairing. $E_{b,2}=0$ for even $N_e$, reflecting the exact tower of states due to the RSGA-1 (Eq.~\ref{eq:RSGA-1}). (d) Eigenvalues of the two-particle density matrix (Eq.~\ref{eq:2RDM}) of the ED GS for $U=-1, \,k_b=\pi$ and $L+1=N_{k_b}=19,21,23$, normalized by $N_e$. The red (blue) dots correspond to odd (even) $N_e$. The presence of a finite eigenvalue illustrates ODLRO.}
    \label{fig:1d_main}
\end{figure*}

As a warmup for the 2D Berry Trashcan, we first consider a toy 1D trashcan model that shares partial similarities. The Hamiltonian is (see App.~\ref{appsec:1d_Hamiltonian} for more details)
\begin{equation}
    \hat{H}=\sum_{k}\epsilon_k\gamma^\dagger_k\gamma_k+\frac{1}{2L}\sum_{q,k,k'}V_q \gamma^\dagger_{k+q}\gamma^\dagger_{k'-q}\gamma_{k'}\gamma_{k},
\end{equation}
where $\gamma^\dagger_k$ is the fermionic plane wave creation operator, and the system length $L$ quantizes the momentum $k$ to integer multiples of $\Delta k=\frac{2\pi}{L}$. 
The first term $\hat{H}^\text{kin}$ describes the kinetic dispersion $\epsilon_k=\theta(|k|-k_b)v(|k|-k_b)$,
with $v>0$ the velocity of the trashcan wall, and $2k_b$ the size of the flat trashcan bottom (Fig.~\ref{fig:main_splash}a). We let $N_{k_b}$ be the number of plane waves inside the trashcan bottom (i.e.~$2k_b=(N_{k_b}-1)\Delta k$), and define a `filling factor' $\nu=\frac{N_e}{N_{k_b}}$ for $N_e$ electrons. We further impose a cutoff $\Lambda$ so that momenta with $|k|>k_b+\Lambda$ are forbidden from being occupied. 
The density-density interaction potential in the second term $\hat{H}^\text{int}$ is chosen as $V_q=-Uq^2$ which corresponds to a short-range interaction $V(x)\sim \frac{d^2}{dx^2}\delta(x)$. Note that this interaction, when repulsive ($U>0$), is the limit of a short screening length $\xi < k_b^{-1}$ Coulomb interaction.  Owing to continuous translation symmetry, we can work within symmetry sectors of fixed total momentum $p$. 

In the following, we restrict to $v=\infty$ where the kinetic term simply restricts the allowed single-particle momenta to $|k|\leq k_b$. The interaction term for $N_e=2$ is separable with rank 1\footnote{In App.~\ref{appsubsec:1d_q2gamma}, we consider more general polynomial $V_q$ which still has finite rank for two electrons.}. As a result, the two-electron spectrum at fixed $p$ consists of a single finite energy eigenstate, with all others being zero modes. For $p=0$, the finite energy solution has energy $E_2=\frac{4U}{L}\left(\sum_{0<k\leq k_b}k^2\right)$, which is the ground state across all momentum sectors for attractive $U<0$. The corresponding (non-normalized) wavefunction can be expressed as a two-particle $p$-wave operator 
\begin{equation}\label{eq:O2dagger}
    \hat{O}^\dagger_2=\sum_{0<k\leq k_b}k\gamma^\dagger_k\gamma^\dagger_{-k}
\end{equation}
acting on the vacuum state $\ket{\text{vac}}$. In App.~\ref{secapp:1d_Hamiltonian_2e_spectrum}, we prove the above statements and generalize them to $p\neq 0$ and finite $v$. We also address the scenario of two holes on top of the fully filled trashcan bottom.

The construction of exact \emph{many-body} eigenstates is enabled by a special algebraic structure of the interaction (see App.~\ref{appsubsec:evenNe_vFinfty})
\begin{align}
\begin{aligned}
[\hat{H}^\text{int}, \hat{O}^\dagger_2]|\text{vac} \rangle &= E_2 \hat{O}^\dagger_2 |\text{vac} \rangle \\
[[\hat{H}^\text{int}, \hat{O}^\dagger_2], \hat{O}^\dagger_2] &= 0,
\end{aligned}\label{eq:RSGA-1}
\end{align}
which corresponds to a restricted spectrum generating algebra of order 1 (RSGA-1)~\cite{PhysRevB.102.085140,moudgalya2022review}. This leads to a tower of eigenstates $\ket{\phi_{2N}}\propto {\hat{O}^{\dagger N}_2}\ket{\text{vac}}$ with even particle number $N_e=2N$ and energy $E_{2N}=N E_2$, all with $p=0$. In App.~\ref{appsubsec:RSGA_generalization}, we discuss the RSGA-1 for more general Hamiltonians and with finite momentum two-body operators. We remark that these models are almost solvable~\cite{repulsive_unpub}.

We can express the interaction as
\begin{align}\label{eq:1d_ham_MM_RR}
    \hat{H}^{\text{int}}&=-\frac{U}{L}\sum_{q}M_q^\dagger M_q+\frac{E_2}{2}\hat{N}_e=\frac{U}{L}\sum_qR_q^\dagger R_q,
\end{align}
where $\hat{N}_e$ is the number operator and we have defined
\begin{equation}
    M_q=\sum_{k}^{\{k,k+q\}}k\gamma^{\dagger}_{k}\gamma_{k+q},\quad R_{q}=\sum_{k}^{\{k,q-k\}}k\gamma_{q-k}\gamma_{k}.
\end{equation}
The summations are restricted such that the momenta in angular brackets lie within the trashcan bottom. For attractive $U<0$, the positive semidefinitness of $M_q^\dagger M_q$ bounds the GS energy from below by $\frac{{N}_eE_2 }{2}$, which implies that $\ket{\phi_{2N}}$ is a GS of the Hamiltonian. We also note that $R_q^\dagger$ creates the two-body GS for total momentum $q$. In App.~\ref{appsubsec:oddNe_vFinfty} we derive the GS ansatz $\ket{\phi^A_{2N+1}}=\gamma^\dagger_0\ket{\phi_{2N}}$ for odd particle numbers $N_e=2N+1$, and we show their high overlap with the numerical GS obtained from ED in Fig.~\ref{fig:1d_main}(a). The overlap is close to 1 at the full-filling side ($\nu\rightarrow1$) and decreases as $N_e$ decreases.

The construction of the many-body GS by repeated application of the pairing operator $\hat{O}^\dagger_2$ (Eq.~\ref{eq:O2dagger}) suggests its interpretation as a condensate of Cooper pairs. Pairing can be quantified through the binding energies 
\begin{align}
E_{b,m}(N_e) = E(N_e - m) + E(N_e + m)-2E(N_e) , \label{eq:binding_1d_main}
\end{align}
where $E(N_e)$ denotes the GS energy for $N_e$ electrons, and $m = 1,2$ correspond to the pair and quartet binding energies, respectively. As shown in Fig.~\ref{fig:1d_main}(b,c), $E_{b,1}$ and $E_{b,2}$ exhibit a pronounced even–odd effect. $E_{b,1}$ is positive (negative) for $N_e$ even (odd), indicating binding of electron pairs. For even $N_e=2N$, $E_{b,2}$ vanishes since $E_{2N}=N E_2$, enabling condensation of Cooper pairs. We find that the binding energy decreases as we approach full filling.

The presence of ODLRO~\cite{PhysRevLett.63.2144,yang1962ODLRO} can be diagnosed by a large eigenvalue (that scales with $N_e$) of the two-particle density matrix~\cite{PhysRevB.37.7359}
\begin{equation}\label{eq:2RDM}
    \rho^{(2)}_{(k_1,k_2),(k_3,k_4)}=\bra{\text{GS}}\gamma^\dagger_{k_1}\gamma^\dagger_{k_2}\gamma_{k_4}\gamma_{k_3}\ket{\text{GS}}.
\end{equation}
As shown in Fig.~\ref{fig:1d_main}(d), the dominant eigenvalue of $\rho^{(2)}/N_e$ remains finite with even/odd oscillations, and decays for larger $\nu$. In App.~\ref{secapp:1d_Hamiltonian_ODLRO}, we demonstrate analytically in real-space the presence of long-range pairing correlations for the exact GS wavefunctions.

\section{2D Berry Trashcan}

We now consider the 2D Berry Trashcan model~\cite{bernevig2025berrytrashcanmodelinteracting} of interacting spin-valley polarized conduction electrons, inspired by the low-energy physics of R$n$G in a displacement field. Analogously to the 1D model, the kinetic dispersion $\epsilon_{\bm{k}}=\theta(|\bm{k}|-k_b)v(|\bm{k}|-k_b)$ captures a flat trashcan bottom with radius $k_b$ surrounded by steeply dispersive walls (Fig.~\ref{fig:main_splash}b). For most of the discussion below, we will use $v=\infty$. The filling factor $\nu$ is again defined as the density relative to full filling of the trashcan bottom. A finite real-space area $\Omega_{tot}$ quantizes the momenta, leading to a finite number of momenta $N_{k_b}$ within the flat bottom. 

The density-density interaction term
\begin{align}
    \hat{H}^\text{int}=\frac{1}{2\Omega_{tot}}\sum_{\bm{k,k',q}}V_{\bm{q}}\mathcal{M}_{\bm{k,q}}\mathcal{M}_{\bm{k',q}}^*\gamma_{\bm{k+q}}^\dagger\gamma_{\bm{k'-q}}^\dagger\gamma_{\bm{k'}}\gamma_{\bm{k}},
\end{align}
inherits form factors $\mathcal{M}_{\bm{k},\bm{q}}$ owing to the non-trivial structure of the underlying R$n$G Bloch wavefunctions. For the Berry Trashcan, the choice $\mathcal{M}_{\bm{k},\bm{q}}=e^{-\frac{|\beta|\bm{q}^2}{2}}e^{-i\beta\bm{q}\times\bm{k}}$ obeys the GMP algebra~\cite{girvin1986magnetoroton} (see App.~\ref{app:subsec:gmp}) and encodes a uniform Berry curvature $2\beta$. 
In the Berry Trashcan parameterization of R5G, the flat bottom encloses a Berry flux $\varphi_\text{BZ}=2\beta A_{b}\approx\pi/2$~\cite{bernevig2025berrytrashcanmodelinteracting}, where $A_{b}$ is the momentum area of the trashcan bottom. Hence, we will mostly use $\varphi_\text{BZ}=\pi/2$ in the numerics.
We consider a Gaussian interaction potential 
\begin{equation}\label{eq:2d_Vq}
    V_{\bm{q}}=Ue^{-(\alpha-|\beta|)\bm{q}^2},
\end{equation}
which is purely attractive for $U<0$ and $\alpha\geq |\beta|$. The limit $\alpha=|\beta|$ corresponds to an on-site attraction in real-space, which due to the non-trivial form factors, gives rise to a non-vanishing $\hat{H}^\text{int}$ for $\beta\neq0$.
Remarkably, we uncover emergent solvable structures in this Hamiltonian.

We first discuss the two-electron problem for $v=\infty$ and total momentum $\bm{p}=0$, which captures the essential pairing physics. Due to the $SO(2)$ symmetry, the solutions are classified by angular momentum $m$, which is odd-integer due to fermionic statistics. As we demonstrate in App.~\ref{sec:2d_berry_2e}, a remarkable simplification occurs for $\alpha=|\beta|$ where the interaction matrix vanishes for $m\beta<0$, and has rank 1 for every angular momentum with $m\beta>0$. The latter implies that for each of these channels with $m\beta>0$, the spectrum consists of a single, gapped eigenstate $\hat{O}^\dagger_{2,m}\ket{\text{vac}}$ with non-zero energy $E_{2,m}$, while all other states are zero modes. For $\beta>0$ and attractive $U<0$, the global GS corresponds to a $p_x+ip_y$ solution with $m=1$
\begin{align}
    \ket{\phi_{2,m=1}}&=\hat{O}^\dagger_{2,m=1}\ket{\text{vac}}\nonumber\\
    &=\int_{|\bm{k}|\leq k_b}\frac{d^2\bm{k}}{(2\pi)^2Z}k_{+}e^{-\alpha\bm{k}^2}\gamma_{\bm{k}}^\dagger\gamma_{-\bm{k}}^\dagger\ket{\text{vac}}\label{eq:gs_2d_o2_main},
\end{align}
where $k_\pm=k_x\pm ik_y$ and $Z$ is a normalization factor. The solutions for general angular momenta $m>0$ have a $k_+^m$ factor instead with energy
\begin{align}
    E_{2,m}=\frac{\Gamma(1 + m) - \Gamma(1 + m, \varphi_{\text{BZ}}/\pi)}{4\varphi_{\text{BZ}} m!}Uk_b^2,
\end{align}
where we use the relation $2\alpha=2\beta=\varphi_{\text{BZ}}/A_b=\varphi_{\text{BZ}}/\pi k_b^2$ in the continuum limit $\Omega_{tot}\rightarrow\infty$.
For a negative Berry curvature $\beta<0$, the GS would instead be a $p_x-ip_y$ solution with $m=-1$. This locking of the chirality of the bound pair to the sign of the Berry curvature suggests a `ferromagnetic' coupling between the R$n$G band and the SC order parameter.

\begin{figure*}[t]
    \centering
    \includegraphics[width=1.0\linewidth]{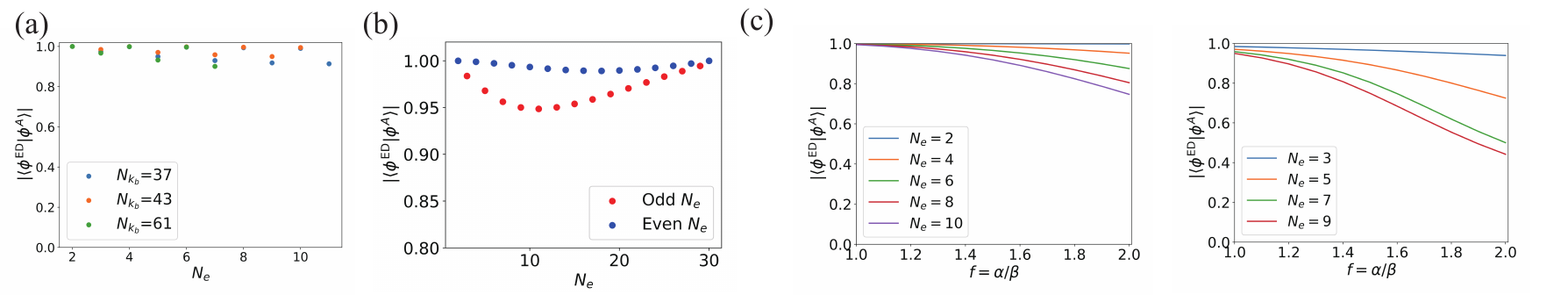}
    \caption{ Wavefunction overlap between the analytical ansatz $\ket{\phi^A}$ (Eq.~\ref{eq:2d_ansatz}) and the ED GS $\ket{\phi^{ED}}$ for the attractive 2D Berry Trashcan model with $v=\infty$. (a) Overlap at the empty filling side with $N_{k_b}=37,43$ and $61$.  (b) Overlap across all filling factors with $N_{k_b}=31$. Red (blue) dots represent the overlap for odd (even) $N_e$. (c)  Overlap with $\alpha=f\beta$ for even (odd) $N_e$ with $N_{k_b}=43$. In all plots, $\beta$ is determined by $\varphi_\text{BZ}=\frac{\pi}{2}$, and we further set $\alpha=\beta$ in (a),(b).}
    \label{fig:overlap_main_2d}
\end{figure*}

Away from the solvable limit $\alpha=|\beta|$, the interaction is no longer rank-1, but can be expressed as an infinite-rank matrix in each angular momentum channel $m$, which is amenable to a perturbative treatment when $|\alpha|, |\beta|\ll k_b^{-2}$ (see App.~\ref{sec:2d_berry_2e}). The regime $\alpha>|\beta|$ describes an exponentially decaying interaction (see Eq.~\ref{eq:2d_Vq}), which can be fitted to the gate-screened Coulomb interaction for short gate distances if repulsive~\cite{bernevig2025berrytrashcanmodelinteracting}.
We find that the GS wavefunction for each $m$ is nearly identical to the exact $\alpha=|\beta|$ solution, demonstrating the robustness of the $p_x+ip_y$ bound state away from exact solvability. For example, the overlap $|\bra{\phi_{2,m=1}}\phi_2^{ED}\rangle|$ for $\varphi_\text{BZ}=\frac{\pi}{2}$ deviates from unity by $2\times10^{-4}$ ($0.05$) for $\alpha=2\beta$ ($\alpha=5\beta$) on a $N_{k_b}=61$ momentum mesh.

In App.~\ref{appsubsec:2d_vFinf_finitep}, we also study the two-electron GS at finite momentum, which exhibits a linear dispersion at small $\bm{p}$. A finite $v$ preserves both the gapped-ness of the GS and the linear dispersion, as shown in App.~\ref{appsubsec:2d_vFfinite_p0}.

The situation of two holes on top of the fully occupied trashcan bottom for $\alpha=|\beta|$ can be solved using Weyl's inequality. In App.~\ref{app:sec:hole_doping}, we derive that the GS is gapless for each $\bm{p}$, and disperses quadratically at small momenta.
Such quadratic dispersion is contrasts with the two-electron case, and we numerically observe a crossover from linear to quadratic dispersion in the GS as $N_e$ increases towards full filling. 

Having addressed the 2-body states, we move to the many-body ones. We examine the commutator algebra. For $\alpha=\beta$ (we analyze the general $\alpha\neq|\beta|$ case in App.~\ref{app:sec:2d_many_body}), we find 
\begin{align}
\begin{aligned}
[\hat{H}^\text{int}, \hat{O}^\dagger_{2,m}]|\text{vac} \rangle &= E_{2,m} \hat{O}^\dagger_2 |\text{vac} \rangle \\
[[\hat{H}^\text{int}, \hat{O}^\dagger_{2,m}], \hat{O}^\dagger_{2,m}] &= \mathcal{O}((\alpha k_b^2)^2).
\end{aligned}\label{eq:RSGA-1_2d}
\end{align}
Compared to the 1D case (Eq.~\ref{eq:RSGA-1}), the second commutator\footnote{Higher commutators vanish because $\hat{H}^\text{int}$ only contains two annihilation operators.} here only vanishes at the lowest non-trivial (linear) order in $\alpha$, leading to an approximate RSGA-1 for small $\alpha$. Following the 1D toy model analysis, this motivates the many-body GS ansatz\footnote{We can use $\hat{O}^\dagger_{2,m}$ with higher $m$ to generate many-body ansatz wavefunctions, but they would not be related to the GS.} based on the $m=1$ pairing operator
\begin{align}
\begin{aligned}
\ket{\phi^A_{2N}} &\propto O^{\dagger N}_{2,m=1}\ket{\text{vac}}\\
\ket{\phi^A_{2N+1}} &\propto \gamma^\dagger_{\bm{0}}O^{\dagger N}_{2,m=1}\ket{\text{vac}},
\end{aligned}\label{eq:2d_ansatz}
\end{align}
which has total momentum $\bm{p}=0$ and angular momentum $N$. The even $N_e=2N$ wavefunction would be exact with energy $E_{2N}=NE_{2,m=1}$ if the second commutator vanished (i.e.~it is exact to first order in $\alpha$). Analogously to Eq.~\ref{eq:1d_ham_MM_RR}, the interaction can be rewritten
\begin{gather}
    \hat{H}^{\text{int}}\approx-\frac{\alpha U}{\Omega_{tot}}\sum_{\bm{q}}M_{\bm{q}}^\dagger M_{\bm{q}}+\frac{E_{2,m=1}}{2}\hat{N}_e=\frac{\alpha U}{\Omega_{tot}}\sum_{\bm{q}}R_{\bm{q}}^\dagger R_{\bm{q}}\\
    M_{\bm{q}}=\sum_{\bm{k}}^{\{\bm{k},\bm{k+q}\}}k_+\gamma^{\dagger}_{\bm{k}}\gamma_{\bm{k}+\bm{q}},\quad R_{q}=\sum_{k}^{\{\bm{k},\bm{k-q}\}}k_-\gamma_{\bm{q}-\bm{k}}\gamma_{\bm{k}},
\end{gather}
where $\approx$ indicates $\mathcal{O}((\alpha k_b^2)^2)$ corrections have been omitted. This demonstrates that $\ket{\phi_{2N}}$ is an exact GS at this order. Furthermore, in App.~\ref{app:sec:rsga_2d_generalization}, we show that the RSGA-1 can be extended to finite-momentum two-body operators which remains approximately solvable, in direct analogy to the 1D case~\cite{repulsive_unpub}.

The ansatz wavefunction of the GS (Eq.~\ref{eq:2d_ansatz}), exact to first order in $\alpha$, is suggestive of a $p_x+ip_y$ superconductor. 
To test the validity of the ansatz, we compute its wavefunction fidelity with the actual GS obtained numerically from ED\footnote{For numerical calculations on finite system sizes, we use a triangular momentum mesh that breaks the $SO(2)$ symmetry down to a $C_6$ subgroup, leading to weak mixing between angular momenta $m$ differing by 6.} for $\alpha=\beta,\varphi_{\text{BZ}}=\pi/2$. As shown in Fig.~\ref{fig:overlap_main_2d}, the overlap is large, and is higher for even $N_e$ than odd $N_e$. The latter is consistent with the fact that $\ket{\phi^A_{2N}}$ is exact to linear order in $\alpha$. The overlap is largest near empty and full filling of the trashcan bottom, where we recover the exact two-electron or two-hole states (see App.~\ref{app:sec:gs_ansatz_odd_2d}). In Fig.~\ref{fig:overlap_main_2d}(c), our numerics also reveal that the analytic ansatz can remain accurate even away from the limit $\alpha=|\beta|$, demonstrating its relevance for more general Hamiltonians. 

\begin{figure}[t]
    \centering
    \includegraphics[width=1.0\linewidth]{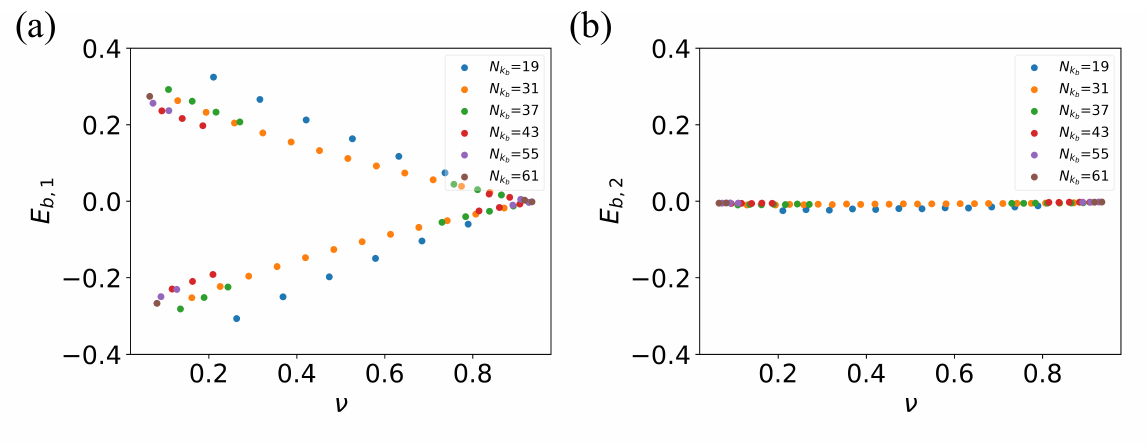}
    \caption{Binding energies (a) $E_{b,1}$ and (b) $E_{b,2}$ for the attractive 2D Berry Trashcan model with $v=\infty$, as a function of filling factor $\nu=N_e/N_{k_b}$ with $U=-2/A_b,\,\varphi_\text{BZ}=\frac{\pi}{2}$ and $\alpha=\beta$, for different $N_{k_b}$. }
    \label{fig:2D_binding_main}
\end{figure}

In Fig.~\ref{fig:2D_binding_main}, we fix $\varphi_{\text{BZ}}=\frac{\pi}{2}$ and $\alpha=\beta$, and plot the binding energies $E_{b,1}$ and $E_{b,2}$ extracted from ED as a function of filling factor $\nu$ for different system sizes $N_{k_b}$. The pair binding energy $E_{b,1}$ exhibits a clear even-odd oscillation, and both the amplitude of this oscillation and the magnitude of $E_{b,1}$ itself vanish as $\nu\to1$, indicating an energetic preference for binding electrons into pairs at any partial filling $\nu<1$. $E_{b,2}$ remains nearly zero, allowing for condensation of Cooper pairs. However, the robustness of pairing depends on the form factors of the band: the approximate RSGA-1 structure (Eq.~\ref{eq:RSGA-1_2d}) for $\alpha=|\beta|$ does not hold for sufficiently strong Berry curvature. Indeed, as we demonstrate via ED in App.~\ref{sec:app:2d_binding_energy}, for $\varphi_\text{BZ}\gtrsim2\pi$, the even/odd staggering in $E_{b,1}$ is suppressed and $E_{b,2}$ becomes significant. This suggests that the (exact) superconductivity generated by the operator $\hat{O}^\dagger_{2,m=1}$ could either give way to another phase or be modified. 

\begin{figure}
    \centering
    \includegraphics[width=1.0\linewidth]{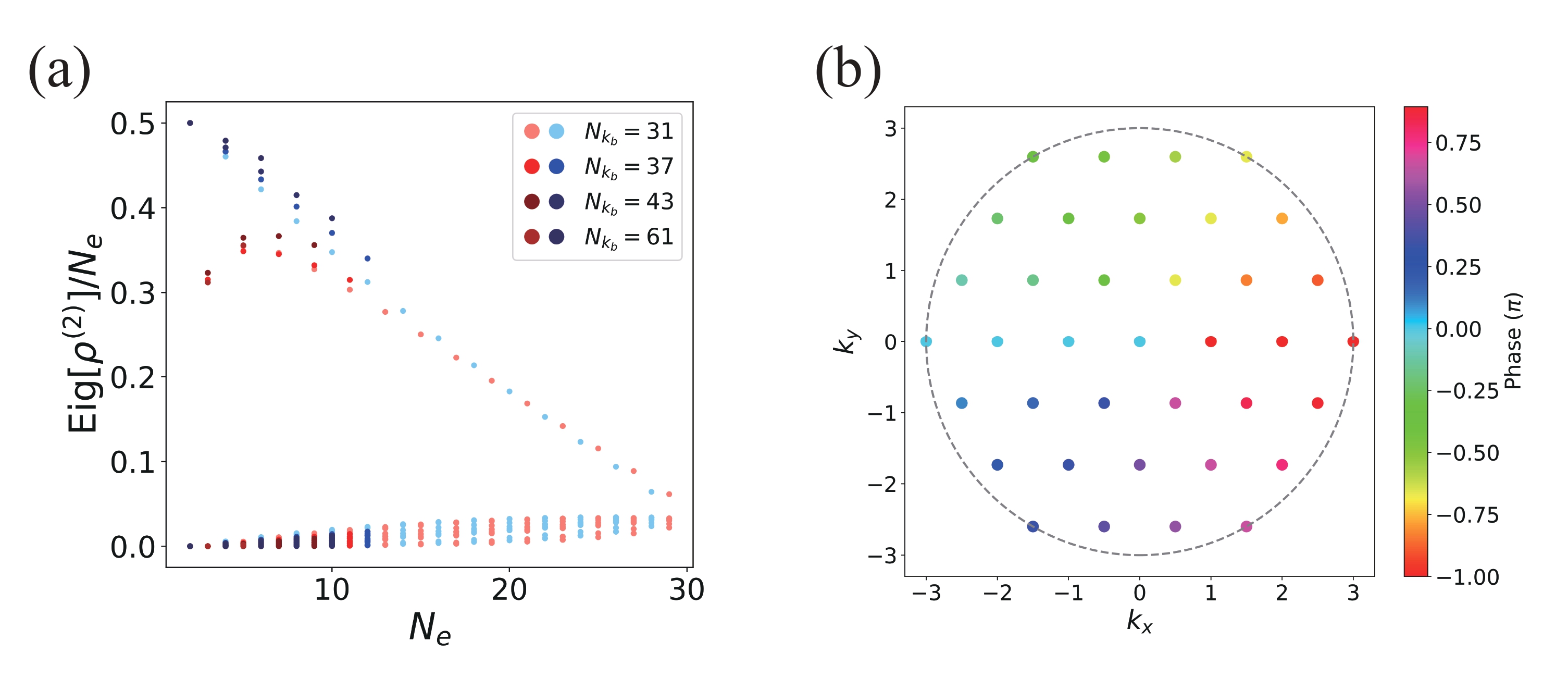}
    \caption{ODLRO in the attractive 2D Berry Trashcan model with $v=\infty$. (a) The spectrum of $\rho^{(2)}/N_e$ as a function of electron number $N_e$ for $\varphi_\text{BZ}=\frac{\pi}{2}$, $\alpha=\beta$ and different $N_{k_b}$. The red (blue) dots correspond to the odd (even) $N_e$. (b) The phase of the eigenvector corresponding to the largest eigenvalue of $\rho^{(2)}/N_e$ for $N_e=12$ and $N_{k_b}=37$. It exhibits $p_x+ip_y$ phase winding.}
    \label{fig:odlro_2d_main}
\end{figure}

Fig.~\ref{fig:odlro_2d_main} shows the eigenvalues of the two-particle density matrix $\rho^{(2)}$ (Eq.~\ref{eq:2RDM}) of the ED GS. The largest eigenvalue, when normalized by $N_e$, remains finite, indicating ODLRO. Furthermore, the dominant eigenvector of $\rho^{(2)}$ is consistent with a chiral $p_x+ip_y$ superconducting order parameter across all $N_e$, as exemplified for $N_e=12$ in Fig.~\ref{fig:odlro_2d_main}(b).
 To gain deeper real-space insight into this pairing, we analyze the pairing wavefunction of our ansatz in Eq.~\ref{eq:2d_ansatz} (App.~\ref{app:subsec:odlro_2d}). While the large $\alpha k_b^2$ limit corresponds to a strong-coupling phase with exponentially localized pairs, the small $\alpha k_b^2$ limit exhibits a long-range pairing that decays algebraically as $\sim r^{-3/2}$, distinct from the standard weak-coupling behavior $\sim r^{-1}$~\cite{PhysRevB.61.10267}.

\section{Discussion}

In this work, we have demonstrated that under short-range attraction, the 2D Berry Trashcan model, a minimal framework for moir\'e-free R$n$G, hosts a robust and (nearly) exact $p_x+ip_y$ SC whose chirality is ferromagnetically locked to the underlying Berry curvature. 
The GS, whose pairing nature is confirmed by ED calculations of binding energies and ODLRO, arises from an emergent RSGA-1~\cite{PhysRevB.102.085140,moudgalya2022review} obeyed by the Hamiltonian. The SC survives away from the solvable limit of $\alpha=|\beta|$ and exhibits unusual real-space pairing correlations that decay algebraically as $\sim r^{-3/2}$. The analytical tractability not only provides exact solutions, but also illuminates the connection between the underlying band geometry and the SC order. Our findings establish the Berry Trashcan as a powerful building block for exploring correlated phenomena in R$n$G.
The locking between the chirality of the SC order and the Berry curvature implies that the thermal Hall effect in the SC phase and the anomalous Hall effect in the normal state should exhibit the same sign. Observing this relationship experimentally would point towards pairing mediated by short-range attractive interactions, and also lend credence to the relevance of our analysis to R$n$G.

While our model captures the emergence of time-reversal-breaking SC, the experiments in R$n$G also reveal finer details in the SC phase. For example in R4G and R5G, Ref.~\cite{Han2025Signatures} observed at least two SC regions in the $n_e-D$ gatemap separated by resistive normal states. Ref.~\cite{morissette2025stripedsuperconductorrhombohedralhexalayer} also reported evidence for coexisting stripe order in R6G. Capturing these rich phenomena in the Berry Trashcan will likely require incorporating more realistic band structure details such as trigonal distortion, which represents an important direction for future study.

While our analysis in this work assumes a phenomenological short-range attraction, its microscopic origin in R$n$G remains an open question, and we discuss several possibilities below.  
In twisted bilayer graphene, the intervalley interaction between electrons and optical $K$-phonons has been identified as strong~\cite{chen2024strong,wu2018theorySC,liu2024epc,wagner2024unflattening,blason2022local,kwan2024EPC,shi2025optical}. Similarly, intravalley $\Gamma$-phonons could generate the requisite short-range attraction in our model \cite{MJL2007epc,yan2008phonon,wu2018theorySC,liu2024epc,Zan2024,wang2025epc}. Alternatively, purely electronic mechanisms could contribute, such as a Kohn-Luttinger-like mechanism~\cite{kohn1965SC,geier2024chiraltopological,yang2024incommensurate,ChouIntravalley2025,qin2024chiralfinite,jahin2025enhancedkohnluttinger,dong2025controllabletheory,maymann2025pairingmechanism,patri2025family} where over-screening of the Coulomb repulsion induces attractive components. Given the relevance of ferromagnetic states in R$n$G, pairing mediated by spin fluctuations is another scenario. Soft flavor-neutral collective modes arising from proximate (incipient) Wigner crystalline physics may also play a role. Disentangling these possibilities requires detailed microscopic modeling, which we defer to a future investigation.


\section{Acknowledgements}

We thank Jonah Herzog-Arbeitman, Minxuan Wang, Nicolas Regnault and  Andrey Chubukov for discussions, and Andreas Feuerpfeil for collaboration on related work. B.A.B. was supported by the Gordon and Betty Moore Foundation through Grant No. GBMF8685 towards the Princeton theory program, the Gordon and Betty Moore Foundation’s EPiQS Initiative (Grant No. GBMF11070), the Office of Naval Research (ONR Grant No. N00014-20-1-2303), the Global Collaborative Network Grant at Princeton University, the Simons Investigator Grant No. 404513, the NSF-MERSEC (Grant No. MERSEC DMR 2011750), Simons Collaboration on New Frontiers in Superconductivity (SFI-MPS- NFS-00006741-01), the Schmidt Foundation at Princeton University, the Princeton Catalyst Initiative. M.L. and H.Y. were supported in part by the NSFC under Grant Nos. 12347107 and 12334003 and by MOSTC under Grant No. 2021YFA1400100. H.Y. acknowledges the support in part by the New Cornerstone Science Foundation through the Xplorer Prize. M.L thanks European Research Council (ERC) under the European Union’s Horizon 2020 research and innovation program (Grant Agreement No. 101020833).

\bibliography{refs.bib}

\clearpage
\onecolumngrid
\newpage

\appendix

\startappendixlocaltoc 
\appendixlocaltoc      

\newpage

\input{sup_attractive_trashcan}
\end{document}

%% file: sup_attractive_trashcan.tex
\section{1D Toy Trashcan Model}
\subsection{Hamiltonian}\label{appsec:1d_Hamiltonian}

In this section, we discuss the Hamiltonian of the 1D trashcan model with trivial form factors. This is a toy 1D version of the 2D Berry Trashcan model~\cite{bernevig2025berrytrashcanmodelinteracting} that will be addressed in App.~\ref{appsec:2d_trashcan}. The single-particle Hilbert space of the 1D trashcan model consists of (spinless) plane waves. We quantize the momenta $k$ by imposing periodic boundary conditions with real-space system length $L$. This leads to $k=j_k\Delta k$, where $j_k$ is an integer and $\Delta k=\frac{2\pi}{L}$. We consider momentum meshes that preserve inversion symmetry, so the number of momenta $N_{k_{b}}$ is always odd (this is because we include $k=0$ which is inversion-symmetric). The creation operators are denoted by $\gamma^\dagger_k$, which satisfy canonical anticommutation relations $\{\gamma_k,\gamma_{k'}^\dagger\}=\delta_{k,k'}$. The real-space basis $\gamma^\dagger_x$ is obtained by Fourier transform
\begin{gather}
    \gamma^\dagger_x=\frac{1}{\sqrt{L}}\sum_{k}e^{-ikx}\gamma^\dagger_k\\
    \gamma^\dagger_{k}=\frac{1}{\sqrt{L}}\int_0^L dx\,e^{ikx}\gamma^\dagger_x,
\end{gather}
and satisfies $\gamma^\dagger_x\rightarrow \gamma^\dagger_{x+L}$ and $\{\gamma_x,\gamma^\dagger_{x'}\}=\delta(x-x')$. 

The kinetic energy is generically written as
\begin{equation}
    \hat{H}^\text{kin}=\sum_{k}\epsilon_k\gamma^\dagger_k\gamma_k.
\end{equation}
For the trashcan model, the dispersion is
\begin{equation}
    \epsilon_k=\theta(|k|-k_b)v(|k|-k_b),
\end{equation}
where $v$ is the velocity of the trashcan walls, $2k_b$ is the total momentum size of the flat trashcan bottom, and $\theta(k)$ is the Heaviside theta function. We parameterize $k_b=j_b\Delta k$, where $j_b$ is an integer. Hence, the number of single-particle states with zero kinetic energy is $N_{k_b}=2j_b+1$.
Note that our choice of finite-size momentum mesh and kinetic energy will always preserve inversion symmetry.

Furthermore, due to the sharp dispersion of the trashcan walls in the physical settings we are interested in, we will consider an additional hard cutoff on the momentum along the steep walls  $\Lambda=j_\Lambda\Delta k$, with $j_\Lambda$ integer. Only single-particle states with $|k|\leq k_b+\Lambda$ are allowed, the rest being prohibited by the kinetic energy. The cutoff will generally depend on the ratio of the interaction to the velocity $v$. This effectively corresponds to setting 
\begin{equation}\label{appeq:1d_ek_infty}
\epsilon_{|k|>k_b+\Lambda}\rightarrow\infty.
\end{equation}
Note that $v=\infty$ corresponds to an smaller effective hard cutoff that restricts the momenta to $|k|\leq k_b$, a limit that we will mainly focus on in this work.

The interaction is a density-density term
\begin{gather}
    \hat{H}^\text{int}=\frac{1}{2L}\sum_q V_q :\rho_q \rho_{-q}:=\frac{1}{2L}\sum^{\{k,k',k+q,k'-q\}}_{q,k,k'}V_q \gamma^\dagger_{k+q}\gamma^\dagger_{k'-q}\gamma_{k'}\gamma_{k}\label{appeq:1d_Hint}\\
    \rho_q=\sum^{\{k,k+q\}}_{k}\gamma^\dagger_{k+q}\gamma_k.
\end{gather}
We will explain the angular bracket notation in the superscript of the momentum summation in the next paragraph. The interaction is normal-ordered with respect to the vacuum state $\ket{\text{vac}}$. Note that this Hamiltonian does not explicitly include any effects of other bands (e.g.~valence bands). In App.~\ref{appsec:2d_trashcan_Hamiltonian}, in the context of the 2D Berry Trashcan model and rhombohedral graphene, we will discuss how the normal-ordering of the interaction Hamiltonian relates to the inclusion or neglect of valence band effects~\cite{kwan2023MFCI3,yu2024MFCI4}.  

Given the hard momentum cutoff (either $|k|\leq k_b+\Lambda$ for finite $v$, or $|k|\leq k_b$ for $v=\infty$), we choose to explicitly constrain the momentum summations in $\hat{H}^\text{int}$. We can do this since the occupation of states outside the cutoff is anyways energetically forbidden, so the basis states outside the cutoff do not affect the finite-energy physics that we are interested in. The summation symbol in Eq.~\ref{appeq:1d_Hint} means that the summation should be restricted so that the superscript momenta within angular brackets all lie within the hard cutoff. This notation will be used extensively below. Since our momentum cutoff and momentum mesh respect inversion symmetry, whenever $k$ lies within the cutoff, then so will $-k$. Hence, $\sum^{\{k\}}$ automatically restricts $-k$ to also lie within the cutoff.

$V_q$, which has units [energy]$\times$[length] in 1D, is the momentum-space Fourier transformation of the real-space interaction. We will refer to $V_q$ as the `interaction potential' in this work. For most of the discussion of the 1D model, we will consider a quadratic potential
\begin{equation}
    V_q=-Uq^2.
\end{equation}
This corresponds to an ultra short-range interaction $V(x)\sim \frac{d^2}{dx^2}\delta(x)$. Note that a delta function potential $V(x)\sim \delta(x)$, corresponding to a constant $V_q$, has no effect on the spinless fermions here due to the fermionic statistics. As will be demonstrated later, $U>0$ ($U<0$) corresponds to a repulsive (attractive) interaction. Also note that this potential takes the functional form of a gate-screened Coulomb interaction for very short screening lengths $\xi \ll 1/k_b$.

The total Hamiltonian is 
\begin{equation}
    \hat{H}=\hat{H}^\text{kin}+\hat{H}^\text{int}.\label{app:eq:1d_general_hamiltonian}
\end{equation}
This obeys continuous translation invariance, leading to conservation of total momentum. Hence, we can work within symmetry sectors of fixed total momentum $p$. This will be exploited in all exact diagonalization (ED) calculations to reduce the Hilbert space dimension. $\hat{H}^{\text{int}}$ also satisfies inversion symmetry $\gamma^\dagger_k\rightarrow \gamma^\dagger_{-k}$. For $p=0$, the total momentum eigenstates can be further labelled by their inversion eigenvalue. 

\subsection{Two-Body Spectrum}\label{secapp:1d_Hamiltonian_2e_spectrum}

For two electrons ($N_e=2$), the Hilbert space for total momentum $p$ is spanned by $\ket{\frac{p}{2}+k,\frac{p}{2}-k}\equiv\gamma^\dagger_{\frac{p}{2}+k}\gamma^\dagger_{\frac{p}{2}-k}\ket{\text{vac}}$ for $k>0$. Note that $\ket{\frac{p}{2}-k,\frac{p}{2}+k}$ is not independent from $\ket{\frac{p}{2}+k,\frac{p}{2}-k}$. Acting with $\hat{H}$ leads to
\begin{equation}
    \hat{H}\ket{\frac{p}{2}+k,\frac{p}{2}-k}=(\epsilon_{\frac{p}{2}+k}+\epsilon_{\frac{p}{2}-k})\ket{\frac{p}{2}+k,\frac{p}{2}-k}
    +\frac{1}{L}\sum_{k'>0}(V_{k'-k}-V_{k'+k})\ket{\frac{p}{2}+k',\frac{p}{2}-k'}.
\end{equation}

\subsubsection{$v=\infty$}

Taking $v=\infty$ restricts the allowed single-particle states to lie within the flat trashcan bottom. Consider total momentum $p\geq 0$ for concreteness. The corresponding results for $p<0$ can be found using inversion symmetry. The action of the Hamiltonian reads
\begin{equation}
    \hat{H}\ket{\frac{p}{2}+k,\frac{p}{2}-k}=\frac{4U}{L}k\sum_{k
    '>0}^{k_b-\frac{p}{2}}k'\ket{\frac{p}{2}+k',\frac{p}{2}-k'}.
\end{equation}
Note that if $p$ is an odd multiple of $\Delta k$, then $k$ and $k'$ are half-integer multiples of $\Delta k$. If $p$ is an even multiple of $\Delta k$, then $k$ and $k'$ are integer multiples of $\Delta k$.  

A crucial feature of the above equation is that the RHS is a product of a term that depends only on $k$, and another term that depends only on $k'$. One eigenstate and its eigenvalue can be straightforwardly extracted by multiplying both sides by $k$ and summing, leading to
\begin{gather}
    E_{2,p}=\sum_{k>0}^{k_b-\frac{p}{2}}\frac{4U}{L}k^2=\frac{16\pi^2 U}{3L^3}(j_b-\frac{j_p}{2})(j_b-\frac{j_p}{2}+\frac{1}{2})(j_b-\frac{j_p}{2}+1)\label{app:eq:2_electron_Ep_1d}\\
    \ket{\phi_{2,p}}\propto\sum_{k>0}^{k_b-\frac{p}{2}}k\ket{\frac{p}{2}+k,\frac{p}{2}-k}\propto\sum_{k_1,k_2\in [-k_b+\frac{p}{2},k_b-\frac{p}{2}]}(k_1-k_2)\delta_{k_1+k_2,p}\ket{k_1,k_2}\label{app:eq:2_electron_basis_1d}.
\end{gather}
The wavefunction coefficient in the second form of $\ket{\phi_{2,p}}$ in Eq.~\ref{app:eq:2_electron_basis_1d} is the explicitly anti-symmetrized version of the coefficient in the first form. For $L\rightarrow\infty$, this eigenstate has dispersion
\begin{equation}
    E_{2,p}=\frac{2U}{3\pi}(k_b-\frac{|p|}{2})^3,\label{eq:dispersion_1d_inf_vf}
\end{equation}
which is linear as $p\rightarrow0$ with velocity $-\frac{Uk_b^2}{\pi}$. Such linear behavior is consistent with the numerical results as shown in Fig.~\ref{fig:1d_energy_spectrum_vf_5}(a).

In fact, this is the only non-zero energy branch. Indeed, the matrix representation of the Hamiltonian in the symmetry sector with momentum $p$ is
\begin{gather}
    H_{k,k'}(p)\equiv \bra{\frac{p}{2}+k,\frac{p}{2}-k}\hat{H}\ket{\frac{p}{2}+k',\frac{p}{2}-k'}=E_{2,p}\phi_{k,p}\phi_{k',p}\\
    \phi_{k,p}=\frac{k}{\sqrt{\sum_{k'>0}^{k_b-\frac{p}{2}}k'^2}},
\end{gather}
which is separable and has rank 1, so all other eigenvalues for 2 electrons are zero.

\subsubsection{$v=\infty$, Even-order Polynomial $V_q$ interaction}\label{appsubsec:1d_q2gamma}

We briefly discuss the two-electron problem where the interaction $V_q\propto q^{2\gamma}$, with $\gamma$ integer, is a higher monomial of the momentum transfer $q$. We take $p=0$ and $v=\infty$ in the following. For concreteness, we consider $V_q=Uq^4$, which corresponds to a real-space interaction of the form $V(x)\sim \frac{d^4}{dx^4}\delta(x)$. The action of the Hamiltonian on the basis states reads
\begin{equation}
    \hat{H}\ket{k,-k}=-\frac{8U}{L}\sum_{k
    '>0}^{k_b}\left(k^3k'+kk'^3\right)\ket{k',-k'}.
\end{equation}
In this case, the space of non-zero energy states is spanned by $\phi^{(1)}_k=k$ and $\phi^{(2)}_k=k^3$, such that the interaction is rank 2. By multiplying both sides by $\alpha k +\beta k^3$ and summing, we obtain the coupled equations for the coefficients $\alpha,\beta$
\begin{gather}
    E\alpha=-\frac{8U}{L}(S^{(4)}\alpha+S^{(6)}\beta)\\
    E\beta=-\frac{8U}{L}(S^{(2)}\alpha+S^{(4)}\beta)\\
    S^{(n)}\equiv\sum_{k>0}^{\{k\}}k^n.
\end{gather}
The eigenvalues determine the non-zero energies, while all other states are zero modes. In the continuum limit $L\rightarrow\infty$, we have
\begin{gather}
    \frac{S^{(n)}}{L}\stackrel{L\rightarrow\infty}{\rightarrow}\frac{k_b^{n+1}}{2\pi(n+1)},
\end{gather}
so that the non-zero energies are
\begin{equation}
    E=-\frac{4Uk_b^5}{\pi}\left(\frac{1}{5}\pm\frac{1}{\sqrt{21}}\right).
\end{equation}
This takes positive and negative values, indicating that $V_q=Uq^4$ has both attractive and repulsive components.  

We now generalize to a polynomial interaction $V_q=U(-q^2+aq^4)$. The action of the Hamiltonian on the basis states reads
\begin{equation}
    \hat{H}\ket{k,-k}=\frac{U}{L}\sum_{k
    '>0}^{\{k'\}}\left((4k-8ak^3)k'-8akk'^3\right)\ket{k',-k'}.
\end{equation}
Again, the space of non-zero energy states is spanned by $\phi^{(1)}_k=k$ and $\phi^{(2)}_k=k^3$. By multiplying both sides by $\alpha k +\beta k^3$ and summing, we obtain the coupled equations for the coefficients $\alpha,\beta$
\begin{gather}
    E\alpha=\frac{4U}{L}\left[\left(S^{(2)}-2aS^{(4)}\right)\alpha+\left(S^{(4)}-2aS^{(6)}\right)\beta\right]\\
    E\beta=\frac{4U}{L}\left[-2aS^{(2)}\alpha-2aS^{(4)}\beta\right].
\end{gather}
The eigenvalues are 
\begin{align}
    E&=\frac{2U}{L}\left[S^{(2)}-4aS^{(4)}\pm\sqrt{S^{(2)}\left(S^{(2)}-8aS^{(4)}+16a^2S^{(6)}\right)}\right]\\
    &\stackrel{L\rightarrow\infty}{\rightarrow}\frac{Uk_b^3}{3\pi L}\left[
   1-\frac{12(ak_b^2)}{5}\pm\sqrt{1-\frac{24(ak_b^2)}{5}+\frac{48(ak_b^2)^2}{7}}
    \right].
\end{align}
For any non-zero value of $ak_b^2$, there is one positive and one negative eigenvalue.

For a general even-order polynomial interaction with highest power $2\gamma$, the above analysis can be generalized to show that the interaction has rank $\gamma$.

\subsubsection{Finite
$v$}\label{appsubsec:1d_2e_finitevF}

We now reintroduce a non-trivial kinetic energy by taking a finite $v$. In this case, the two-electron problem cannot  be solved analytically. However, we can derive conditions on the spectrum by leveraging Weyl's theorem, which we now review. Consider $n$-dimensional diagonalizable matrices $A,B$. For a generic diagonalizable matrix $M$, let $\lambda_i(M)$ denote the spectrum of $M$ in ascending order $\lambda_1(M)\leq \ldots \leq \lambda_n(M)$. Weyl's theorem states that
\begin{equation}
    \lambda_{i+j-n}(A+B)\leq \lambda_i(A)+\lambda_j(B)\leq \lambda_{i+j-1}(A+B).
\end{equation}
This is very useful in the situation where $A$ has low rank (e.g.~for the quadratic interaction $V_q\propto q^2$ with rank 1), where an interlacing theorem on the spectrum of $C=A+B$ can be proved. Consider the scenario where $\lambda_1(A)<0$ and $\lambda_{j>1}(A)=0$. 
Setting $i=j=1$, we find
\begin{equation}
    \lambda_1(A)+\lambda_1(B)\leq \lambda_1(C).
\end{equation}
Setting $i=n$, we find
\begin{equation}
    \lambda_j(C)\leq \lambda_j(B).
\end{equation}
Setting $i=2$, we find
\begin{equation}
    \lambda_j(B)\leq \lambda_{j+1}(C).
\end{equation}
Putting the above together, we find
\begin{equation}
    \lambda_1(A)+\lambda_1(B)\leq \lambda_1(C)\leq\lambda_1(B)\leq \lambda_2(C)\leq\ldots\leq \lambda_n(C)\leq\lambda_n(B).
\end{equation}
Hence all but one of the eigenvalues of $C$ are interlaced between the eigenvalues of $B$. The lowest eigenvalue $\lambda_1(C)$ is itself lower-bounded by $\lambda_1(A)+\lambda_1(B)$.
Note that we can also consider setting $i=1,j=n$, yielding
\begin{equation}
    \lambda_1(C)\leq \lambda_1(A)+\lambda_n(B),
\end{equation}
which is a tighter bound on $\lambda_1(C)$ if $\lambda_1(A)+\lambda_n(B)<\lambda_1(B)$.

\begin{figure}[h]
    \centering
    \includegraphics[width=0.9\linewidth]{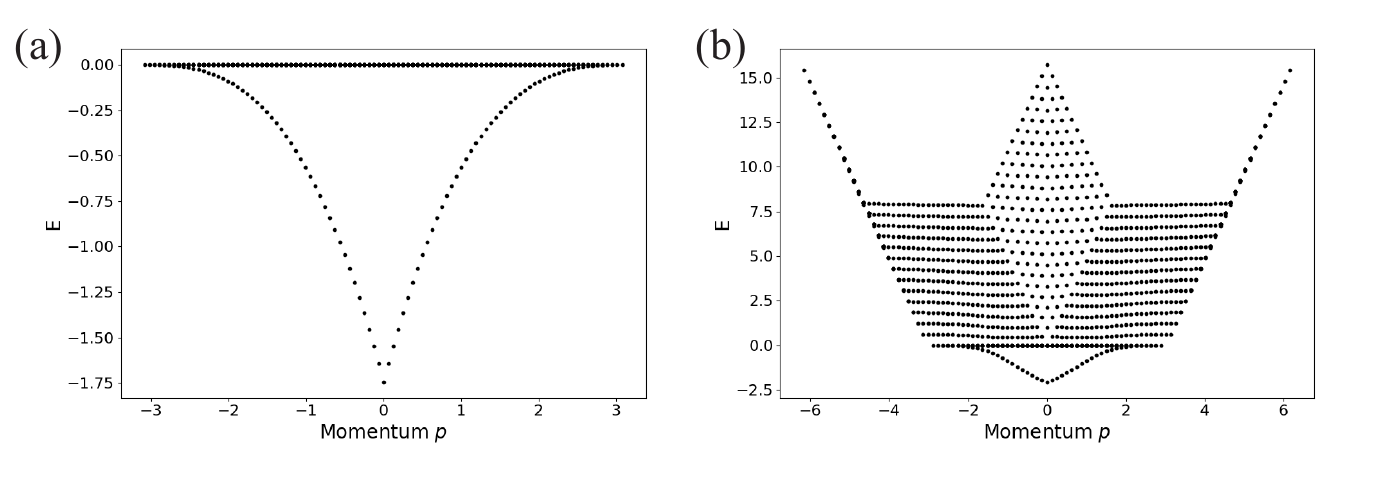}
    \caption {Spectrum for two electrons across all total momentum sectors $p$ for the attractive 1D toy model Hamiltonian (Eq.~\ref{app:eq:1d_general_hamiltonian}) with $U=-1$ and $N_{k_b}=51$. The parameters for the dispersion are $v=\infty$, $k_b=\frac{\pi}{2}$, $L=100$ for (a), and $v=5$, $\Lambda=k_b=\frac{\pi}{2}$, $L=50$ for (b).}
    \label{fig:1d_energy_spectrum_vf_5}
\end{figure}

In the case where the finite eigenvalue of A is positive instead [$\lambda_n(A)>0$ and $\lambda_{j<n}(A)=0$], we would have
\begin{equation}
    \lambda_1(B)\leq \lambda_1(C)\leq\lambda_2(B)\leq \lambda_2(C)\leq\ldots\leq \lambda_{n}(B)\leq \lambda_n(C)\leq\lambda_n(A)+\lambda_n(B).
\end{equation}
Setting $i=n,j=1$ leads to $\lambda_n(A)+\lambda_1(B)\leq\lambda_n(C)$, which is a tighter bound on $\lambda_n(C)$ if $\lambda_n(A)+\lambda_1(B)>\lambda_n(B)$.

We apply this to the two-electron problem with Hamiltonian
\begin{equation}
    \hat{H}\ket{k,-k}=(\epsilon_{k}+\epsilon_{-k})\ket{k,-k}
    +\frac{1}{L}\sum^{\{k'\}}_{k'>0}(V_{k'-k}-V_{k'+k})\ket{k',\-k'},
\end{equation}
by letting $A$ be the interaction part of the Hamiltonian. For simplicity, we consider $p=0$. Noting that the single-particle cutoff is now $|k|\leq k_b+\Lambda$, we observe that $A$ is rank 1 with eigenvalue $\frac{2U}{3\pi}(k_b+\Lambda)^3$. $B$ is then the kinetic part of the Hamiltonian, which consists of $j_b$ zero energy eigenvalues (whose wavefunctions are confined to the trashcan bottom), with the rest being positive (corresponding to the walls of the trashcan). 


We will analyze the repulsive case in Ref.~\cite{repulsive_unpub}. For the attractive case $U<0$, we again are guaranteed to have $j_b-1$ zero modes of the Hamiltonian $H=A+B$. The ground state energy satisfies $\frac{2U}{3\pi}(k_b+\Lambda)^3\leq\lambda_1(H)\leq \text{min}\left(\frac{2U}{3\pi}(k_b+\Lambda)^3+2v\Lambda,\frac{2U}{3\pi}k_b^3\right)$. The RHS of the inequality can be derived by considering the variational state consisting of the $v=\infty$ finite-energy eigenstate (which is restricted to $|k|\leq k_b$). Hence, the ground state always maintains a gap to excited states at $p=0$. The rest of the energies are interlaced with the kinetic energy eigenvalues of $B$ in the interval $[0,2v\Lambda]$. This behavior is consistent with the numerical results as shown in Fig.~\ref{fig:1d_energy_spectrum_vf_5}(b), which is computed with $U=-1$, $v=5$ and $k_b=\Lambda=\frac{\pi}{2}$. We observe the existence of a gapped (in the sense that the ground state and first excited state is separated by a finite gap within a momentum sector as $L\rightarrow\infty$) two-electron ground state at $p=0$, which persists for a finite range of $p$. The energies of the finite-energy excited states are constrained by the kinetic energy eigenvalues of the trashcan wall, which for two electrons with $p=0$ lie in the range $[0, 5\pi]$. The ground state energy at $p=0$ is $-1.75$ in Fig.~\ref{fig:1d_energy_spectrum_vf_5}(a) for $v=\infty$, and $-2.06$ in Fig.~\ref{fig:1d_energy_spectrum_vf_5}(b) for $v=5$. For small $p$, the ground state branch shows a linear dispersion which is similar to the case of $v\to\infty$.

\subsubsection{Two-hole spectrum for $v=\infty$}\label{appsec:1d_2h}

In this section, we discuss the problem of adding two holes to the fully filled trashcan bottom for $v=\infty$, which we show below can be constrained analytically. The analysis here will be useful for developing intuition for the physics near full-filling of the trashcan bottom, and comparing with the many-body ansatz developed later in App.~\ref{sec:1d_manybody_gs}.
To understand this hole-doped regime, we first rewrite the many-body Hamiltonian $\hat{H}^\text{int}$ so that it is normal-ordered with respect to the fully filled trashcan bottom $\ket{\text{full}}=\prod_{|k|\leq k_b}\gamma^\dagger_k\ket{\text{vac}}$. In other words, we bring all the creation operators to the right of all annihilation operators, which is achieved with the identity
\begin{equation}
    \gamma_{{k}+ {q}}^\dagger \gamma_{{k}'- {q}}^\dagger \gamma_{{k}'} \gamma_{{k}}= \gamma_{{k}'} \gamma_{{k}}
         \gamma_{{k}+ {q}}^\dagger \gamma_{{k}'- {q}}^\dagger-\delta_{{q},0}(\gamma_{{k}}\gamma_{{k}}^\dagger+\gamma_{{k}'}\gamma_{{k}'}^\dagger) +\delta_{{k}',{{k}+{q}}}(\gamma_{{k}+{q}}\gamma_{{k}+{q}}^\dagger+\gamma_{{k}}\gamma_{{k}}^\dagger)+\delta_{{q},0}-\delta_{{k}',{k}+{q}}.
\end{equation}
The interaction Hamiltonian becomes
\begin{align}\label{appeq:1d_Hint_hole}
    \hat{H}^\text{int}&=\frac{1}{2L}\sum^{\{k,k'k+q,k'-q\}}_{q,k,k'}V_q \gamma_{{k}'} \gamma_{{k}}
         \gamma_{{k}+ {q}}^\dagger \gamma_{{k}'- {q}}^\dagger+\frac{1}{L}\sum^{\{k,k+q\}}_{k,q}V_q\gamma_{k}\gamma^\dagger_{k}-\frac{N_{k_b}V_0}{L}\sum_{k}^{\{k\}}\gamma_k\gamma^\dagger_k+\frac{1}{2L}\left(V_0N_{k_b}^2-\sum^{\{k,k'\}}_{k,k'}V_{k-k'}\right),
\end{align}
where $N_{k_b}=2j_b+1=\frac{k_bL}{\pi}+1$ is the number of states within the trashcan bottom.

\begin{figure}[h]
    \centering
    \includegraphics[width=0.6\linewidth]{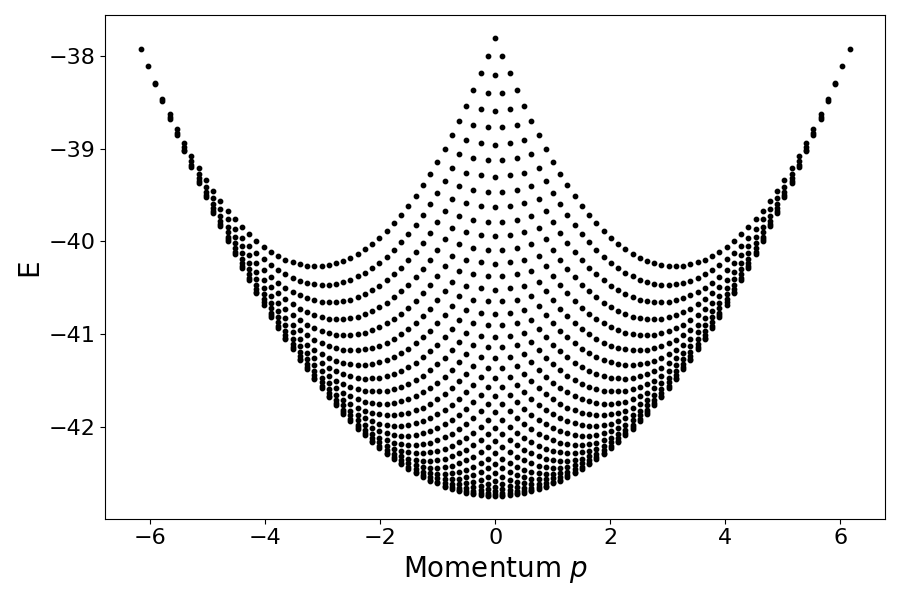}
    \caption {Spectrum across all total momentum sectors $p$ for two holes on top of fully filled trashcan bottom for the attractive 1D toy model with $U=-1,\,k_b=\frac{\pi}{2},L+1=N_{k_b}=51$ and $v=\infty$.
    }
    \label{fig:1d_energy_spectrum_hole}
\end{figure}

Eq.~\ref{appeq:1d_Hint_hole} can be viewed as an effective Hamiltonian for holes on top of the fully filled trashcan bottom.  The first term is the interaction between holes, which takes the same sign as the interaction between electrons. The second and third terms represent the interaction-induced hole dispersion. These are equivalent to the negative (arising from the particle-hole transformation) of the Fock and Hartree potentials generated by the fully-filled trashcan bottom. 
For the quadratic interaction $V_{q}=-Uq^2$, we can obtain the interaction-induced hole dispersion $\epsilon^\text{hole}_k$ defined as
\begin{equation}
    \epsilon^\text{hole}_k=\frac{1}{L}\sum_{q}^{\{k+q\}}V_q\stackrel{L\rightarrow\infty}{\rightarrow}-\frac{U}{2\pi}\int_{-k_b-k}^{k_b-k}dq\,q^2=-\frac{Uk_b}{3\pi}(k_b^2+3k^2),
\end{equation}
which consists purely of the Fock contribution (the second term of Eq.~\ref{appeq:1d_Hint_hole}). 
For the repulsive case $U>0$, the hole dispersion is minimal at $|k|=k_b$. This is because the Fock self-energy for electrons is lowest at $k=0$, which is consistent with the tendency of Fock exchange to cluster electrons in momentum space \cite{bernevig2025berrytrashcanmodelinteracting}.
On the other hand, for the attractive case $U<0$, the hole dispersion is minimal at $k=0$. 
The last term of Eq.~\ref{appeq:1d_Hint_hole} is the total energy of $\ket{\text{full}}$. 

While the four-fermion interaction term in Eq.~\ref{appeq:1d_Hint_hole} is still rank-1, the interaction-induced hole dispersion prevents an exact solution. Despite this, we can constrain the two-hole spectrum with total momentum $p$ using Weyl's theorem
in a similar fashion as in App.~\ref{appsubsec:1d_2e_finitevF}. Consider $p>0$. We let $A$ be the first term of Eq.~\ref{appeq:1d_Hint_hole}. Just like in the two-electron case, for the quadratic interaction, $A$ is a rank 1 matrix whose finite eigenvalue is $\frac{2U}{3\pi}(k_b-\frac{p}{2})^3$. $B$ is then a diagonal matrix whose diagonal entries are $\epsilon^{\text{hole}}_{\frac{p}{2}+k}+\epsilon^{\text{hole}}_{\frac{p}{2}-k}$ for $0<k\leq k_b-\frac{p}{2}$ (we have dropped the constant terms in Eq.~\ref{appeq:1d_Hint_hole}, but they can be trivially incorporated).

We can further constrain the eigenvalues of $H=A+B$ by using the Sherman-Morrison formula. This exploits the fact that the Hamiltonian matrix (for two holes) is a symmetric Diagonal-Plus-Rank-1 matrix (DPR1), i.e.~$A$ is symmetric rank-1 while $B$ is a diagonal matrix. The eigenvalues $\lambda$ of $H$ satisfy the secular equation
\begin{equation}
    1=\sum_k \frac{A_k^2}{
\lambda-B_k
    }.
\end{equation}
Applying this to the two-hole problem yields
\begin{align}
    1=&\frac{4U}{L}\sum_{k>0}^{k_b-\frac{p}{2}} \frac{k^2}{\lambda-\epsilon^{\text{hole}}_{\frac{p}{2}+k}-\epsilon^{\text{hole}}_{\frac{p}{2}-k}}\\
    \stackrel{L\rightarrow\infty}{\rightarrow}&\frac{2U}{\pi}\int_0^{k_b-\frac{p}{2}} dk\,\frac{k^2}{\lambda+\frac{Uk_b}{3\pi}\left(2k_b^2+3(\frac{p}{2}+k)^2+3(\frac{p}{2}-k)^2\right)}\\
    =&\frac{2|U|}{\pi}\int_0^{k_b-\frac{p}{2}} dk\,\frac{k^2}{\frac{|U|k_b}{6\pi}\left(4k_b^2+3p^2+12k^2\right)-\lambda},
\end{align}
where we have taken the attractive case $U=-|U|<0$. We now search for a solution $\lambda<\frac{|U|k_b}{6\pi}\left(4k_b^2+3p^2\right)=E_\text{thresh}$, where $E_\text{thresh}=2\epsilon^\text{hole}_{\frac{p}{2}}$ is equal to $\epsilon^{\text{hole}}_{\frac{p}{2}+k}+\epsilon^{\text{hole}}_{\frac{p}{2}-k}$ minimized over $k$. Define $\Delta=\sqrt{E_\text{thresh}-\lambda}>0$. Then we have
\begin{align}
    1=&\frac{2|U|}{\pi}\left[
    \frac{k_b-\frac{p}{2}}{2|U|k_b/\pi}
    -\frac{\Delta\tan^{-1}\left[\frac{({2|U|k_b/\pi})^{1/2}(k_b-\frac{p}{2})}{\Delta}\right]}
    {(2|U|k_b/\pi)^{3/2}}
    \right].
\end{align}
The RHS is smaller than 1 since the first term is less that or equal to 1, while the second term is negative. Hence, there is no eigenvalue that is less than the threshold $E_\text{thresh}$, the latter of which is constructed from the sum of hole dispersions. Combining this result with interlacing, we therefore prove that the two-hole spectrum for $U<0$ is fully gapless, in the sense that the ground state and first excited state within each momentum sector has vanishing energy separation as $L\rightarrow\infty$. The two-hole spectrum for $U=-1$ is shown in Fig.~\ref{fig:1d_energy_spectrum_hole}. The system is gapless (within each momentum sector) and exhibits a quadratic dispersion as expected. Notice that the above discussion based on analytics applies for the continuum limit $L\rightarrow\infty$. For a finite system with finite $N_{k_b}$, we cannot determine the exact energies without numerically solving the secular equation.

\subsection{Many-body Ground State For 1D Attractive Quadratic Potential}\label{sec:1d_manybody_gs}

We now consider many-body states with more general electron numbers $N_e$. We define $\nu=N_e/N_{k_b}$ to be the `filling factor' of the trashcan bottom. We focus on the many-body ground state for the attractive $(U<0)$ quadratic interaction potential.

\subsubsection{Ground State For Even $N_e$, $v=\infty$, $p=0$}\label{appsubsec:evenNe_vFinfty}
Remarkably, for even $N_e=2N$ and $v=\infty$, we can obtain the ground state analytically for $p=0$. We repeat the interaction Hamiltonian with general interaction potential $V_q$ for convenience
\begin{align}
    \hat{H}^\text{int}=\frac{1}{2L}\sum^{\{k,k',k+q,k'-q\}}_{k,k',q}V_q\gamma_{k+q}^\dagger\gamma_{k'-q}^\dagger\gamma_{k'}\gamma_{k}\label{eq:1_d_interacting_ham}.
\end{align}
Since $v\to \infty$, the momentum arguments of the creation/annihilation operators are restricted to the region $[-k_b, k_b]$. As a reminder, this is reflected in the superscripts within angular brackets of the summation symbol, namely $\{k\}$ indicates that $k$ cannot take values outside $[-k_b, k_b]$. We define a two-particle operator with zero total momentum
\begin{align}\label{appeq:1d_O2_GS_even_quadratic}
\hat{O}_2^\dagger=\frac{1}{2}\sum^{\{k\}}_{k}f_{k}\gamma_{k}^\dagger\gamma_{-k}^\dagger=\sum^{\{k\}}_{k>0}f_k\gamma^\dagger_k\gamma_{-k}^\dagger,  
\end{align}
where we have utilized fermionic statistics to set $f_k=-f_{-k}$. Note that inversion symmetry imposes $V_q=V_{-q}$. We first calculate the commutator
\begin{align}
\begin{split}
    [\hat{H}^\text{int}, \hat{O}_2^{\dagger}]&=\frac{1}{4L}\sum^{\{k,k',k_1,k+q,k'-q\}}_{k,k',q,k_1}V_qf_{k_1}\gamma_{k+q}^\dagger\gamma_{k'-q}^\dagger(\delta_{k',k_1}\gamma_{-k_1}^\dagger\gamma_{k}-\delta_{k,k_1}\gamma_{-k_1}^\dagger\gamma_{k'}\\
    &\quad\quad\quad\quad-\delta_{k',-k_1}\gamma_{k_1}^\dagger\gamma_{k}+\delta_{k,-k_1}\gamma_{k_1}^\dagger\gamma_{k'}+\delta_{k,k_1}\delta_{k',-k_1}-\delta_{k,-k_1}\delta_{k',k_1}).
\end{split}
\end{align}
From the above equation, we see that generally $[\hat{H}^\text{int},\hat{O}^\dagger_2]\ne 0$. Acting on the vacuum, we find
\begin{align}
    [\hat{H}^\text{int}, \hat{O}_2^\dagger] |\text{vac} \rangle = \frac{1}{2L}\sum^{\{k_1,k_1+q\}}_{k_1,q}V_qf_{k_1}\gamma_{k_1+q}^\dagger\gamma_{-(k_1+q)}^\dagger |\text{vac} \rangle,
\end{align}
where we used $f_{k} = -f_{-k}$. With the replacement $k_1+q=k_2$, the above reduces to
\begin{align}
    [\hat{H}^\text{int}, \hat{O}_2^\dagger] |\text{vac}  \rangle &=\frac{1}{2L} \sum^{\{k_1,k_2\}}_{k_1,k_2}V_{k_2-k_1}f_{k_1}\gamma_{k_2}^\dagger\gamma_{-k_2}^\dagger |\text{vac} \rangle\nonumber\\
    &=\frac{1}{2L}\sum^{\{k_1\}}_{k_1>0}\sum^{\{k_2\}}_{k_2}(V_{k_2-k_1}-V_{k_2+k_1})f_{k_1}\gamma_{k_2}^\dagger\gamma_{-k_2}^\dagger |\text{vac} \rangle.
\end{align}
If and only if $\frac{1}{L}\sum^{\{k_1\}}_{k_1>0}(V_{k_2-k_1}-V_{k_2+k_1})f_{k_1}=E_{2} f_{k_2}$ for some $E_{2}$, then we have 
\begin{align}
    [\hat{H}^\text{int}, \hat{O}_2^\dagger] |\text{vac}  \rangle = E_{2}\hat{O}_2^\dagger |\text{vac}  \rangle.
\end{align}
In particular, if we take $V_q=-Uq^2$ (an overall constant offset in $V_q$ does not affect the Hamiltonian due to fermion antisymmetry), then
\begin{align}
    \frac{1}{L}\sum_{k_1>0}^{\{k_1\}}(V_{k_2-k_1}-V_{k_2+k_1})f_{k_1}=\left(\sum_{k_1>0}^{\{k_1\}}\frac{4U}{L}k_1f_{k_1}\right)k_2=E_{2} f_{k_2}.
\end{align}
We therefore identify
\begin{align}
    f_k=\frac{k}{\mathcal{N}},\quad \text{with }\mathcal{N}=\sqrt{\sum_{k>0}^{\{k\}}k^2} \text{ and }E_{2}=\frac{4U}{L}\sum_{k>0}^{\{k\}}k^2=\frac{16\pi^2 U}{3L^3}j_b(j_b+\frac{1}{2})(j_b+1)\label{eq:sol_1d_energy},
\end{align}
where $\mathcal{N}$ is a normalization factor, and $j_b=k_b/\Delta_k$  with $\Delta k=2\pi/L$ is an integer parameterizing the boundary of the trashcan bottom. 

We notice that $\hat{H}^\text{int}$ is a 2-body operator while $\hat{O}^\dagger_2$ is constructed with $\gamma_{k}^\dagger\gamma_{-k}^\dagger$ terms. Therefore, we trivially have
\begin{align}
    \left[\left[\left[\hat{H}^\text{int},\hat{O}_2^\dagger\right],\hat{O}_2^\dagger\right],\gamma^\dagger_{k}\right]=0 \Rightarrow  \left[\left[\left[\hat{H}^\text{int},\hat{O}_2^\dagger\right],\hat{O}_2^\dagger\right],\hat{O}_2^\dagger\right]=0.
\end{align}
We further calculate
\begin{align}   
\left[\left[\hat{H}^\text{int},\hat{O}_2^\dagger\right],\hat{O}_2^\dagger\right]=\sum^{\{k_1,k_2,k_1-q,k_2+q\}}_{q,k_1,k_2}\frac{V_q}{L}f_{k_1}f_{k_2}\gamma_{k_2+q}^\dagger\gamma_{k_1-q}^\dagger\gamma_{-k_1}^\dagger\gamma_{-k_2}^\dagger.\label{eq:HOOcommutator}
\end{align}
In general, Eq.~\ref{eq:HOOcommutator} does not vanish. Later in App.~\ref{appsubsec:RSGA_generalization} (see Tab.~\ref{tab:W_zero_table}), we will study different choices of $f_k$ to understand the conditions for $\left[\left[\hat{H}^\text{int},\hat{O}_2^\dagger\right],\hat{O}_2^\dagger\right]$ to vanish.

For now, we point out that Eq.~\ref{eq:HOOcommutator} vanishes under the following  sufficient conditions 
\begin{itemize}
    \item $V_q=-Uq^2$ (up to an overall constant), 
    \item $f_k\sim k$.
\end{itemize}
To prove this, we manipulate Eq.~\ref{eq:HOOcommutator}
\begin{align}
    \left[\left[\hat{H}^\text{int},\hat{O}_2^\dagger\right],\hat{O}_2^\dagger\right]&=\frac{1}{L}\sum^{\{k_1,k_2,k_3,k_4\}}_{k_{1},k_2,k_3,k_4}V_{k_1+k_4}f_{k_1}f_{k_2}\gamma_{k_4}^\dagger\gamma_{k_3}^\dagger\gamma_{k_2}^\dagger\gamma_{k_1}^\dagger\delta_{k_1+k_2+k_3+k_4=0}\nonumber\\
    &=\frac{1}{24L}\sum^{\{k_1,k_2,k_3,k_4\}}_{k_{1},k_2,k_3,k_4}W_{k_1,k_2,k_3,k_4}\gamma_{k_4}^\dagger\gamma_{k_3}^\dagger\gamma_{k_2}^\dagger\gamma_{k_1}^\dagger\delta_{k_1+k_2+k_3+k_4=0},
\end{align}
where
\begin{align}
    W_{k_1,k_2,k_3,k_4}=\sum_{\sigma\in S_4}\text{sgn}(\sigma)V_{k_{\sigma(1)}+k_{\sigma(4)}}f_{k_{\sigma(1)}}f_{k_{\sigma(2)}}
\end{align}
contains a summation that runs over elements $\sigma$ the permutations group $S_4$ of four objects. For the separable interaction $V_q=-Uq^2$ with $f_k\sim k$, we obtain
\begin{align}
    W_{k_1,k_2,k_3,k_4}=-U\sum_{\sigma\in S_4}\text{sgn}(\sigma)[k_{\sigma(1)}^3k_{\sigma(2)}+k_{\sigma(4)}^2k_{\sigma(1)}k_{\sigma(2)}+2k_{\sigma(1)}^2k_{\sigma(2)}k_{\sigma(4)}]=0,
\end{align}
which vanishes because every summand in the square brackets has one pair of indices $\sigma(i),\sigma(j)$ that appear symmetrically. Swapping $i$ and $j$ leads to a cancelling contribution due to the $\text{sgn}(\sigma)$ factor.

In summary, for $f_k\sim k$ and $V_q=-Uq^2$, we have found 
\begin{gather}
[\hat{H}^\text{int}, \hat{O}^\dagger_2]|\text{vac} \rangle = E_{2} \hat{O}^\dagger_2 |\text{vac} \rangle\label{eq:commutator_HO}\\
[[\hat{H}^\text{int}, \hat{O}^\dagger_2],\hat{O}^\dagger_2]= 0.\label{eq:commutator_HOO}    
\end{gather}
For the attractive case $U<0$, $E_{2}$ is the ground state (with negative energy) for the two-electron problem, and all other energies for $p=0$ are zero as the interaction is rank-1. Furthermore, Eq.~\ref{eq:commutator_HOO}  demonstrates that this Hamiltonian exhibits a Restricted Spectrum Generating Algebra of order 1 (RSGA-1), a notion first introduced in the context of quantum scars~\cite{PhysRevB.102.085140,moudgalya2022review}. The existence of this RSGA-1 means that
\begin{align}
\ket{\phi_{2N}}=(\hat{O}_2^\dagger)^N\ket{\text{vac}}\label{app:eq:ground_state_1d_even}    
\end{align}
 for integer $N$ is an eigenstate with $N_e=2N$ particles and energy $E=NE_{2}$.

Given the conditions above, we now further prove that $\ket{\phi_{2N}}$ is the \emph{ground state} for $p=0$ in the $N_e=2N$ particle sector. To see this, we recast the interaction Hamiltonian into the form
\begin{align}
    \hat{H}^\text{int}&=\frac{1}{2L}\sum^{\{k_1,k_2,k_3,k_4\}}_{k_1,k_2,k_3,k_4}V_{k_4-k_1}\gamma_{k_4}^\dagger\gamma_{k_3}^\dagger\gamma_{k_2}\gamma_{k_1}\delta_{k_1+k_2,k_3+k_4}\\
    &=-\frac{U}{2L}\sum^{\{k_1,k_2,k_3,k_4\}}_{k_1,k_2,k_3,k_4}(k_4^2+k_1^2-2k_4k_1)\gamma_{k_4}^\dagger\gamma_{k_3}^\dagger\gamma_{k_2}\gamma_{k_1}\delta_{k_1+k_2,k_3+k_4}\label{eq:1d_decomp_step_1}\\
    &=\frac{U}{L}\sum^{\{k_1,k_2,k_3,k_4\}}_{k_1,k_2,k_3,k_4}k_4k_1\gamma_{k_4}^\dagger\gamma_{k_3}^\dagger\gamma_{k_2}\gamma_{k_1}\delta_{k_1+k_2,k_3+k_4}\\
    &=\frac{U}{L}\sum_{k_3,k_4}^{\{k_3,k_4\}}k_4\gamma_{k_4}^\dagger\gamma_{k_3}^\dagger\sum_{k_1,k_2}^{\{k_1,k_2\}}k_1\gamma_{k_2}\gamma_{k_1}\delta_{k_1+k_2,k_3+k_4}\label{eq:1d_ham_k1234}.
\end{align}
We have discarded the first two terms in Eq.~\ref{eq:1d_decomp_step_1} since they either vanish under the interchange of $k_1$ and $k_2$, or $k_3$ and $k_4$. 
We define a pairing operator
\begin{align}
    R_q=\sum^{\{k,q-k\}}_{k}k \gamma_{q-k}\gamma_{k},
\end{align}
in terms of which the Hamiltonian reduces to
\begin{align}
    \hat{H}^\text{int}=\frac{U}{L}\sum_{q}R_q^\dagger R_q.\label{eq:ham_1d_rqrq}
\end{align}
For an attractive interaction ($U<0$), this form of the Hamiltonian ensures that the spectrum of $\hat{H}^\text{int}$ is bounded from above by zero. Notably, the operator $R_q^\dagger$ creates the exact two-electron ground state with total momentum $q$, which will be discussed in App.~\ref{appsubsec:RSGA_generalization}. Interestingly, as we will briefly comment on in App.~\ref{appsubsec:RSGA_generalization} and demonstrate in a future paper~\cite{repulsive_unpub}, products of these operators can also be used to construct an approximate tower of low-energy, finite-momentum excited states.

To bound the ground state energy, we instead rewrite the Hamiltonian as
\begin{align}
    \hat{H}^\text{int}&=\frac{U}{L}\sum^{\{k_1,k_2,k_3,k_4\}}_{k_1,k_2,k_3,k_4}k_4k_1\gamma_{k_4}^\dagger\gamma_{k_3}^\dagger\gamma_{k_2}\gamma_{k_1}\delta_{k_1+k_2,k_3+k_4}\\
    &=\frac{U}{L}\sum^{\{k_1,k_2,k_3,k_4\}}_{k_1,k_2,k_3,k_4}\left(
    -k_4k_1\gamma_{k_3}^\dagger\gamma_{k_1}\gamma_{k_4}^\dagger\gamma_{k_2}\delta_{k_1+k_2,k_3+k_4}
    +k_4k_1\gamma_{k_3}^\dagger\gamma_{k_2}\delta_{k_1+k_2,k_3+k_4}\delta_{k_4,k_1}\right)\\
    &=-\frac{U}{L}\sum^{\{k_1,k_4,k_1+q,k_4+q\}}_{k_{1},k_{4},q}k_4k_1\gamma_{k_1+q}^\dagger\gamma_{k_1}\gamma_{k_4}^\dagger\gamma_{k_4+q}+\frac{U}{L}\sum^{\{k_1\}}_{k_1}k_1^2\sum^{\{k_2\}}_{k_2}\gamma_{k_2}^\dagger\gamma_{k_2}.
\end{align}
In terms of the fermion bilinear
\begin{equation}
    M_q=\sum^{\{k,k+q\}}_{k}k\gamma^{\dagger}_{k}\gamma_{k+q},
\end{equation}
we find
\begin{align}
    \hat{H}^\text{int}
    &=-\frac{U}{L}\sum_{q}M_q^\dagger M_q+\frac{2U}{L}\sum_{k_1>0}^{\{k_1\}}k_1^2\sum^{\{k_2\}}_{k_2}\gamma_{k_2}^\dagger\gamma_{k_2}\nonumber\\
    &=-\frac{U}{L}\sum_{q}M_q^\dagger M_q+\frac{E_{2}}{2} N_e,\label{appeq:1d_H_Mq}
\end{align}
where in the last line, we assume that we work in a symmetry sector of fixed particle number $N_e$. Note that unlike the density operator $\rho_q=\sum_{k}\gamma_{k+q}^\dagger\gamma_{k}$ which satisfies $\rho_q^\dagger=\rho_{-q}$, here we have $R_{-q}\ne R_q^\dagger$ and $M_{-q}\ne M_q^\dagger$. For $U<0$, Eq.~\ref{appeq:1d_H_Mq} shows that the ground state energy satisfies $E\geq N_e\frac{E_{2}}{2}$. Since we have previously shown that $\ket{\phi_{2N}}$ is an eigenstate of the Hamiltonian with $N_e=2N$ particles and energy $N E_{2}$, then the above analysis demonstrates that it is also a ground state across all momentum sectors. 

The above analysis also implies that $M_{q}\ket{\phi_{2N}}=0$ for all $q$. To see this, we calculate the commutator
\begin{align}
    \left[M_q,\hat{O}_2^\dagger\right]&=\frac{1}{2}\sum^{\{k,k+q\}}_{k}\sum^{\{p\}}_{p}kp\left[\gamma_{k}^\dagger\gamma_{k+q},\gamma_{p}^\dagger\gamma_{-p}^\dagger\right]\nonumber\\
    &=\sum^{\{k,k+q\}}_{k}k(k+q)\gamma_{k}^\dagger\gamma_{-k-q}^\dagger\nonumber\\
    &=-\sum^{\{k_1,k_2\}}_{k_1,k_2}k_1k_2\gamma_{k_1}^\dagger\gamma_{k_2}^\dagger\delta_{k_1+k_2,-q}\nonumber\\
    &=0.
\end{align}
This implies
\begin{align}
    M_{q}\ket{\phi_{2N}}=M_q(\hat{O}_2^\dagger)^N\ket{\text{vac}}=(\hat{O}_2^\dagger)^NM_q\ket{\text{vac}}=0.
\end{align}

\subsubsection{Ground State Ansatz for odd $N_e$, $v=\infty$, $p=0$}\label{appsubsec:oddNe_vFinfty}
Here, our aim is to construct many-body eigenstates with an odd number $N_e$ of electrons. 
We first compute the commutator of the Hamiltonian with the creation operator $\gamma_k^\dagger$ 
\begin{align}
    \left[\hat{H}^\text{int},\gamma_{k}^\dagger\right]&=-\frac{U}{L}\sum_{q}\left(M_q^\dagger[M_q,\gamma_{k}^\dagger]+[M_q^\dagger,\gamma_k^\dagger]M_q\right)+\frac{E_{2}}{2}[\sum^{\{k'\}}_{k'}\gamma_{k'}^\dagger\gamma_{k'},\gamma_{k}^\dagger]\nonumber\\
    &=-\frac{U}{L}\left[\sum^{\{k-q\}}_{q}(k-q)M^\dagger_q\gamma_{k-q}^\dagger+\sum^{\{k+q\}}_{q}k\gamma_{k+q}^\dagger M_q\right]+\frac{E_{2}}{2}\gamma_k^\dagger.
\end{align}
We act this on the even-particle ground state $\ket{\phi_{2N}}$ and obtain
\begin{align}
    \left[\hat{H}^\text{int},\gamma_{k}^\dagger\right]\ket{\phi_{2N}}=\left[-\frac{U}{L}\sum^{\{k-q\}}_{q}(k-q)M^\dagger_q\gamma_{k-q}^\dagger+\frac{E_{2}}{2}\gamma_k^\dagger\right]\ket{\phi_{2N}},\label{eq:H_gamma_commutator_1d}
\end{align}
where we have used the property $M_q\ket{\phi_{2N}}=0$. The second term is  $\gamma_k^\dagger\ket{\phi_{2N}}$ multiplied by a constant $\frac{E_{2}}{2}$. 
If the commutator acting on $\ket{\phi_{2N}}$ only contained this term, then the equation above would imply that $\gamma^\dagger_k\ket{\phi_{2N}}$ is a $N_e=2N+1$ eigenstate with energy $N_e\frac{E_{2}}{2}$ independent of $k$, suggesting that there is a `flat' (in momentum space) energy change associated with adding a single particle to $\ket{\phi_{2N}}$.
However, the first term is a three-fermion operator which involves non-trivial electron scattering processes and hinders an exact solution. 

 Since a simple and exact solution is lacking for this many-body scattering problem for odd-$N_e$, 
 we propose an approximate ansatz for the total momentum $p=0$ ground state that is based on the even-$N_e$ exact solution. For attractive $U<0$, we will demonstrate that our construction corresponds to an approximation to the true odd-$N_e$ ground state.
We first pick one momentum $k_0$ that will correspond to a single `unpaired electron'. 
Analogously to the analysis of the Richardson-Gaudin model~\cite{RevModPhys.76.643}, we consider that this unpaired electron prevents the momenta $\pm k_0$ from being involved in the pairing of the remaining $N_e-1=2N$ electrons (of which there are an even number). We repeat the analysis of App.~\ref{appsubsec:evenNe_vFinfty} for these remaining $2N$ electrons, except that the set of single-particle momenta that can participate in pairing no longer includes $\pm k_0$. The resulting energy of the paired electrons (after excluding $\pm k_0$) is
\begin{gather}
    E_{2N}'=N E_{2}'\\
    E_{2}'=\frac{4U}{L}\sum^{\{k\}}_{k>0; k\neq k_0}k^2.\label{appeq:1d_E2prime}
\end{gather}
To minimize $E_{2N}'$, we should set $k_0=0$ such that $E_{2}'=E_{2}$, since $k=0$ cannot contribute to pairing anyway. Based on these observations, we propose the following ansatz for odd particle numbers with total momentum $p=0$
\begin{align}
    \ket{\phi_{2N+1}^{A}}=\gamma_0^\dagger(\hat{O}_2^\dagger)^N\ket{\text{vac}}.\label{eq:odd_ansatz_1d}
\end{align}

\begin{figure}[h]
    \centering
    \includegraphics[width=1.0\linewidth]{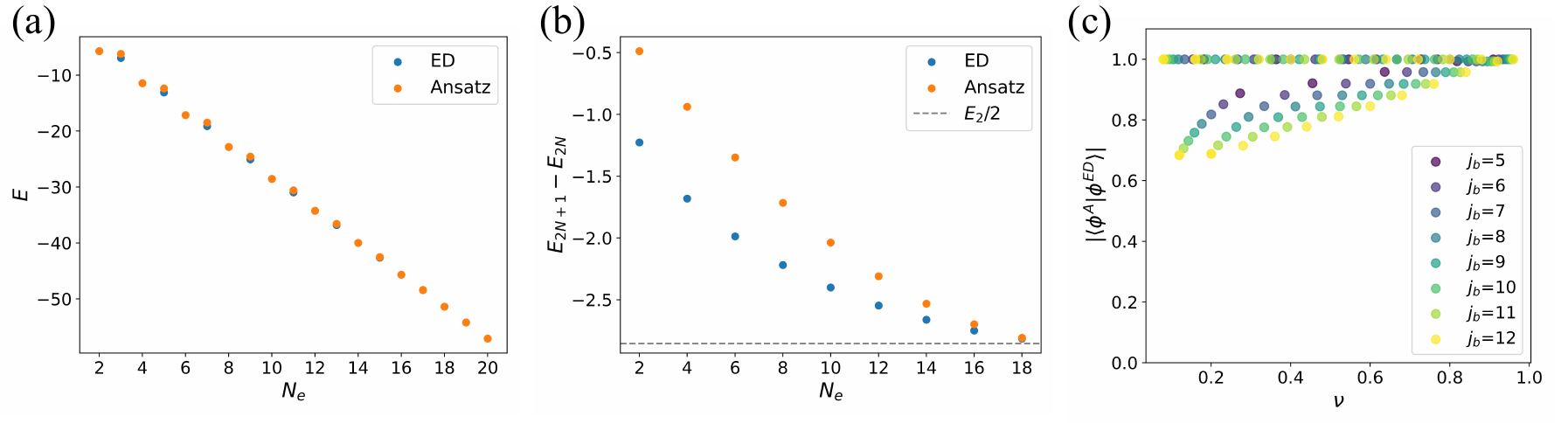}
    \caption{(a) The comparison of the ground state energy from ED with the energy expectation value of the ansatz $\bra{\phi^A }\hat{H}^{\text{int}}\ket{\phi^A}$.  (b) The comparison between the single-charge excitation energy $E_{2N+1}-E_{2N}$ extracted using the ED ground state energy, and using the expectation value of the ansatz $\bra{\phi^A }\hat{H}^{\text{int}}\ket{\phi^A}$. Near the full-filling limit, the excitation energy approaches the value $E_{2}/2$, which is indicated by the dashed line. The calculations in (a) and (b) are performed with $U=-1,\, L+1=N_{k_b}=21$. (c) The wavefunction overlap $|\bra{\phi^{A}}\phi^\text{ED}\rangle|$ between the ansatz and the ED ground state, plotted as a function of electron density $\nu$ with $L+1=N_{k_b}$ from 11 to 25.}
    \label{fig:overlap_1d_all_length}
\end{figure}

To test the validity of our proposed odd-particle ansatz, we numerically calculate the expectation value of the Hamiltonian $E_{g,_{2N+1}}^{A}=\bra{\phi^A_{2N+1} }\hat{H}^\text{int}\ket{\phi^A_{2N+1}}$ in the ansatz, and compare it with the actual ground state energy computed using ED. We also use the ansatz to estimate the single-particle excitation energy $E_{g,2N+1}^{A}-E_{g,2N}$ on top of the $2N$-particle state (for which we have the exact analytic solution), and compare it with the corresponding exact result from ED. We take $U=-1, L+1=N_{k_b}=21$, and the results are shown in Figs.~\ref{fig:overlap_1d_all_length}(a) and (b). The energy and the single-particle excitation energy estimated using the ansatz are close to the ED values, and the deviations even vanish when the system approaches full filling $\nu\rightarrow 1$. 

In addition, we also calculated the overlap $|\bra{\phi^A_{N_e}}\phi^{ED}_{N_e} \rangle|$ between our ansatz wavefunction $\ket{\phi_{N_e}^{A}}$ and the true GS $\ket{\phi_{N_e}^{ED}}$ obtained from ED. Fig.~\ref{fig:overlap_1d_all_length}(c) shows the results for $U=-1$ and a range of $L+1=N_{k_b}$ from 11 to 25. For the even-$N_e$ wavefunction  Eq.~\ref{app:eq:ground_state_1d_even}, the overlap is exactly 1 as expected. The odd-$N_e$ ED wavefunctions also exhibit high overlaps with the corresponding ansatz (Eq.~\ref{eq:odd_ansatz_1d}), especially near full filling of the trashcan bottom.

To understand this high fidelity, we now show that $ \ket{\phi_{2N+1}^{A}}$ is an approximate ground state near full filling. Using the commutator from Eq.~\ref{eq:H_gamma_commutator_1d}, we can express the action of the Hamiltonian on the ansatz state as
\begin{align}
    \hat{H}^\text{int}\ket{\phi_{2N+1}^{A}}=\frac{E_{2}}{2}(2N+1)\ket{\phi_{2N+1}^A}+\frac{U}{L}\sum_{q}qM_{q}^\dagger\gamma_{-q}^\dagger\ket{\phi_{2N}}.
\end{align}
If the second term is vanished, then $\ket{\phi_{2N+1}^{A}}$ would be an eigenstate with energy $\frac{E_{2}}{2}(2N+1)$. Such behavior would be consistent with our numerical observations at the full filling side, where the single-particle excitation energy $E_{2N+1}-E_{2N}$ approaches $E_{2}/2$ as indicated by the dashed line in Fig.~\ref{fig:overlap_1d_all_length}(b).
 To quantify the deviation from a true eigenstate, we evaluate the energy difference $\Delta E(2N+1)$, defined as
\begin{align}
    \Delta E(2N+1) &= \frac{\bra{\phi_{2N+1}^{A}}\hat{H}^\text{int}\ket{\phi_{2N+1}^{A}}}{\bra{\phi_{2N+1}^{A}}\phi_{2N+1}^{A}\rangle}-\frac{E_{2}}{2}(2N+1)=-\frac{U}{L}\frac{\bra{\phi_{2N+1}^{A}}\sum_q M_q^\dagger M_q\ket{\phi_{2N+1}^{A}}}{\bra{\phi_{2N+1}^{A}}\phi_{2N+1}^{A}\rangle}=-\frac{U}{L}\sum_{q}\frac{||q\gamma_{-q}^\dagger\ket{\phi_{2N}}||^2}{\bra{\phi_{2N+1}^{A}}\phi_{2N+1}^{A}\rangle},\label{app:eq:delta_e_1d_odd}
\end{align}
where we have used the property
\begin{align}
    M_q\ket{\phi_{2N+1}^{A}}&=[M_q,\gamma_0^\dagger]\ket{\phi_{2N}}=-q\gamma_{-q}^\dagger\ket{\phi_{2N}}.
\end{align}
Near full filling of the trashcan bottom, most momenta within the trashcan bottom are occupied. Consequently, the sum over $q$ in the numerator of $\Delta E$ is restricted to a few unoccupied momenta, which suggests that the energy deviation is small.
To justify this, we now compute $\Delta E$ exactly. We consider the ground state $\ket{\phi_{2\frac{N_{k_b}-2n-1}{2}}}$ of an even number $N_{k_b}-2n-1$ of electrons (in other words, $2n+1$ holes), and express the wavefunction in the many-body Fock basis
\begin{align}
    \ket{\phi_{2\frac{N_{k_b}-2n-1}{2}}}=\frac{1}{Z}\sum_{k_1,k_2,\cdots ,k_{(N_{k_b}-2n-1)/2}}^{\{k_1,k_2,\cdots ,k_{(N_{k_b}-2n-1)/2}\}}k_1k_2\cdots k_{(N_{k_b}-2n-1)/2}\ket{\pm k_1,\pm k_2,\cdots,\pm k_{(N_{k_b}-2n-1)/2}},\label{app:eq:occupation_ele_1d}
\end{align}
where $Z$ is a normalization factor. The Fock basis state $\ket{\pm k_1,\pm k_2,\cdots,\pm k_{(N_{k_b}-2n-1)/2}}$, which is described by the occupied single-particle momenta $\{\pm k_1,\pm k_2,\cdots,\pm k_{(N_{k_b}-2n-1)/2}\}$, can be equivalently described by hole-occupying the remaining single-particle momenta on top of $\ket{\text{full}}$. The latter perspective is described by the hole-occupied momenta $\{0,\pm k'_1,\cdots \pm k'_n\}$, which is just the complement of $\{\pm k_1,\pm k_2,\cdots,\pm k_{(N_{k_b}-2n-1)/2}\}$. Eq.~\ref{app:eq:occupation_ele_1d} can then be reduced to
\begin{align}\label{appeq:1d_manybody_e_to_h}
        \ket{\phi_{2\frac{N_{k_b}-2n-1}{2}}}=&\frac{1}{Z}\sum_{k'_1,k'_2,\cdots,k'_{n}}^{\{k'_1,k'_2,\cdots,k'_{n}\}}\frac{j_b!\Delta k^{j_b}}{k'_1k'_2\cdots k'_{n}}\ket{\pm k_1,\pm k_2,\cdots,\pm k_{(N_{k_b}-2n-1)/2}}\nonumber\\
        =&\frac{1}{Z'}\sum_{k'_1,k'_2,\cdots,k'_{n}}^{\{k'_1,k'_2,\cdots,k'_{n}\}}\frac{1}{k'_1k'_2\cdots k'_{n}}\ket{0,\pm k'_1,\pm k'_2,\cdots,\pm k'_{n}}_h.
\end{align}
We emphasize that $\ket{\pm k_1,\pm k_2,\cdots,\pm k_{(N_{k_b}-2n-1)/2}}$ and $\ket{0,\pm k'_1,\pm k'_2,\cdots,\pm k'_{n}}_h$ are different ways of writing the same Fock basis state: the first one indicates explicitly the electron occupations on top of $\ket{\text{vac}}$, while the second one indicates explicitly the corresponding hole occupations on top of $\ket{\text{full}}$. The first line of Eq.~\ref{appeq:1d_manybody_e_to_h} is obtained by recognizing that $k_1k_2\cdots k_{(N_{k_b}-2n-1)/2}=\frac{j_b!\Delta k^{j_b}}{k'_1k'_2\cdots k'_{n}}$, where the numerator of the RHS is simply the product of all positive momenta ($k=j\Delta k$ for $j=1,\ldots,j_b$) that lie within the trashcan bottom. In the second line of Eq.~\ref{appeq:1d_manybody_e_to_h}, we absorbed constants into a new normalization constant $Z'$ of the wavefunction.

Eq.~\ref{appeq:1d_manybody_e_to_h} allows us to write $\Delta E$ as
\begin{align}
    \Delta E(N_b-2n)=-\frac{U}{L}\frac{(N_{k_b}-2n+1)\sum_{k'_1,k'_2,\cdots,k'_{n-1}}^{\{k'_1,k'_2,\cdots,k'_{n-1}\}}(\frac{1}{k'_1k'_2\cdots k'_{n-1}})^2}{\sum_{k'_1,k'_2,\cdots,k'_{n}}^{\{k'_1,k'_2,\cdots,k'_{n}\}}(\frac{1}{k'_1k'_2\cdots k'_n})^2}.\label{eq:delta_e_1d}
\end{align}
In the limit where $N_{k_b}\gg 1$ and the number of holes is small $n\ll N_{k_b}$, the denominator scales as $\sim N_{k_b}^{n}k_b^{-2n}$ while the numerator scales as $\sim N_{k_b}^{n}k_b^{-2(n-1)}$. The energy deviation therefore scales as:
\begin{align}
     \Delta E(N_b-2n) \sim -\frac{U}{L}k_b^2.
\end{align}
In comparison, $E_{2}\sim U k_b^3\sim \frac{U}{L}k_b^2\cdot\pi N_{k_b}$. Because $\Delta E$ is smaller that $E_{2}$ by a factor of the system size $N_{k_b}$, we conclude that the true ground state deviates only slightly from the ansatz in Eq.~\ref{eq:odd_ansatz_1d} in the regime near full-filling, which is consistent with the numerical results.

\subsubsection{Generalization of the RSGA to Toy Model Hamiltonians}\label{appsubsec:RSGA_generalization}

More general (but unphysical) Hamiltonians can satisfy a RSGA-1 and have an exact superconducting ground state. The discussion below is valid in any dimension, and the momentum $k$ should be interpreted as a vector. Similar types of Hamiltonians were first considered in Ref.~\cite{herzogarbeitman2022manybodysuperconductivity} (and expanded on in  e.g.~Refs.~\cite{han2024nesting,han2025exactmodels}).

A density-density Hamiltonian (i.e.~one that takes the form $\hat{H}^\text{int}=\frac{1}{2L}\sum^{\{k_1,k_2,k_3,k_4\}}_{k_1,k_2,k_3,k_4}V_{k_4-k_1}\gamma_{k_4}^\dagger\gamma_{k_3}^\dagger\gamma_{k_2}\gamma_{k_1}\delta_{k_1+k_2,k_3+k_4}$ as in Eq.~\ref{eq:1d_ham_k1234}) is very constraining if we seek to satisfy the RSGA-1 condition. We have not yet been able to find a interaction potential beyond $V(q) \propto q^2 $ where the RSGA-1 is valid. However, if we are willing to go beyond density-density interactions, and construct more general toy interactions such as
\begin{eqnarray}
    \hat{H}=\frac{1}{2L}\sum^{\{k_1,k_2,k_3,k_4\}}_{k_1,k_2,k_3,k_4}V_{k_4,k_3,k_2,k_1} \gamma_{k_4}^\dagger\gamma_{k_3}^\dagger\gamma_{k_2}\gamma_{k_1}\delta_{k_1+k_2,k_3+k_4},
\end{eqnarray} 
we can find more Hamiltonians that obey a RSGA-1. Note that $V_{k_4,k_3,k_2,k_1}$ can be chosen to be antisymmetric under interchange of first two arguments separately, and the last two arguments separately, due to fermion antisymmetry.

To proceed, we first consider a zero-momentum operator parameterized as
\begin{equation}
    \hat{O}_2^\dagger=\sum^{\{k\}}_{k}f_{k}\gamma_{k}^\dagger\gamma_{-k}^\dagger,
\end{equation}
where $f_k$ is chosen to be antisymmetric. (Note that we have absorbed an unimportant factor of $1/2$ in $f_k$ compared to Eq.~\ref{appeq:1d_O2_GS_even_quadratic} for convenience).
We compute the first-order commutator with the Hamiltonian acting on the vacuum
\begin{eqnarray}
    [\hat{H},O_2^\dagger]\ket{\text{vac}} = \frac{1}{L} \sum^{\{k_4\}}_{k_4} \sum^{\{k\}}_k V_{k_4, -k_4, -k, k}  f_k \gamma_{k_4}^\dagger\gamma_{-k_4}^\dagger \ket{\text{vac}}, 
\end{eqnarray} where we have used the fermionic antisymmetry of  $V_{k_4,k_3,k_2,k_1}$. This leads to the following condition for $\hat{O}_2^\dagger$ to create a two-body eigenstate on top of the vacuum
\begin{equation}
   [\hat{H},\hat{O}_2^\dagger]\ket{\text{vac}} = E_2 O_2^\dagger\ket{\text{vac}} \implies \frac{1}{L}\sum^{\{k\}}_k V_{k_4, -k_4, -k, k}  f_k = E_2 f_{k_4}.
\end{equation}

We now want to find the functions $V_{k_4,k_3,k_2,k_1}$ that lead to a solution above.  Functions that have the separable form $V_{k_4,k_3,k_2,k_1} = T_{1(k_4, k_3)} T_{2 (k_2, k_1)} $ are sufficient for a solution. Hermiticity of $\hat{H}$  (i.e.~$V_{k_4,k_3,k_2,k_1} = V^\star_{k_1,k_2,k_3, k_4}$) implies $ T_{2 (k_2, k_1)}/T_{1(k_1, k_2)}^\star =  T_{2 (k_3, k_4)}^\star/T_{1(k_4, k_3)} = c \in \mathcal{R}$, since two  functions depending on different arguments are equal only if the functions are a constant. The reality of the constant $c$ can be established by setting $k_1=k_4$ and $k_2=k_3$. We hence have $V_{k_4,k_3,k_2,k_1} = c T_{1(k_4, k_3)} T_{1(k_1,k_2)}^\star $. We define $T_{k_1,k_2}=T_{1,(k_1,k_2)}$ so that the interaction function can be written
\begin{equation}\label{appeq:H_1d_genRSGA_separableV}
    V_{k_4,k_3,k_2,k_1}=cT_{k_4,k_3}T^*_{k_1,k_2}.
\end{equation}
Due to antisymmetry of $V_{k_4,k_3,k_2,k_1}$, we have $T_{k_1, k_2} = -T_{k_2, k_1}$.  Note that density-density interactions, which are necessarily expressible as $V_{k_4,k_3,k_2,k_1}  = V_{k_4-k_1}$, \emph{do not} generally satisfy the separable form of Eq.~\ref{appeq:H_1d_genRSGA_separableV}. However, the quadratic interaction potential $V_q \propto q^2$ that we have used previously does satisfy this separable form, because the $k_4^2 + k_1^2$ terms vanish in the Hamiltonian due to antisymmetry of the fermionic operators. Only the $\propto k_4 k_1$ term remains, which satisfies the separability condition. 
For the two body state $\hat{O}_2^\dagger \ket{\text{vac}}$ to be an eigenstate, we require 
\begin{eqnarray}
   &  \frac{c}{L}\sum^{\{k\}}_k  T_{k_4, -k_4} T_{k, -k}^\star      f_k = E_2 f_{k_4} \implies \nonumber \\ & f_{k}= A  T_{k, -k},\;\;\; E_2=  \frac{c}{L}\sum^{\{k\}}_k  |T_{k, -k}|^2 
   \end{eqnarray} 
with $A$ a constant determined by wavefunction normalization. While any separable interaction of this form will have an exact $2$-particle  state $\hat{O}^\dagger_2\ket{\text{vac}}$, we emphasize that most  realistic interactions (besides $V_q \propto q^2$) do not.  

We now check the second condition for a RSGA-1, i.e.~$ [[\hat{H},\hat{O}_2^\dagger],\hat{O}_2^\dagger]=0$, which will give rise to an exact tower of states. We find (recalling that the allowed set of momenta respects inversion symmetry, so if $k$ is allowed, then so is $-k$)
\begin{align}
    [[\hat{H},\hat{O}_2^\dagger],\hat{O}_2^\dagger]&= \frac{4c}{L}\sum^{\{k_1,k_2,k_3,k_4\}}_{k_1,k_2,k_3,k_4}V_{k_4,k_3,k_2,k_1} f_{k_1} f_{k_2}  \gamma_{k_4}^\dagger\gamma_{k_3}^\dagger\gamma^\dagger_{-k_2}\gamma^\dagger_{-k_1}\delta_{k_1+k_2,k_3+k_4} \nonumber \\ & = \frac{4c}{L}\sum^{\{k_1,k_2,k_3,k_4\}}_{k_1,k_2,k_3,k_4}V_{k_4,k_3,-k_2,-k_1} f_{k_1} f_{k_2}  \gamma_{k_4}^\dagger\gamma_{k_3}^\dagger\gamma^\dagger_{k_2}\gamma^\dagger_{k_1}\delta_{k_1+k_2+ k_3+k_4,0} 
    \nonumber \\ & = \frac{c}{6 L}\sum^{\{k_1,k_2,k_3,k_4\}}_{k_1,k_2,k_3,k_4}W_{k_4,k_3,k_2,k_1}  \gamma_{k_4}^\dagger\gamma_{k_3}^\dagger\gamma^\dagger_{k_2}\gamma^\dagger_{k_1}\delta_{k_1+k_2+ k_3+k_4,0}\label{eq:HOO_general_mom_0}  
\end{align}
\begin{align}
    W_{k_1,k_2,k_3,k_4}&=\sum_{\sigma\in S_4}\text{sgn}(\sigma)V_{k_{\sigma(4)},  k_{\sigma(3)}, -k_{\sigma(2)},  - k_{\sigma(1)}}f_{k_{\sigma(1)}}f_{k_{\sigma(2)}} \nonumber \\ &=A^2 \sum_{\sigma\in S_4}\text{sgn}(\sigma)T_{k_{\sigma(4)},  k_{\sigma(3)}}T^\star_{-k_{\sigma(1)},  - k_{\sigma(2)}}T_{k_{\sigma(1)},-k_{\sigma(1)}}  T_{k_{\sigma(2)},-k_{\sigma(2)}}.
\end{align} 
Without further restriction to the form of $T_{k_1, k_2}$, we generally have $  W_{k_1,k_2,k_3,k_4} \ne 0$ and hence the tower of states stops at the $2$-particle state. In Tab.~\ref{tab:W_zero_table}, we investigate sufficient conditions to achieve $W_{k_1,k_2,k_3,k_4}=0$. In particular, the first column of Tab.~\ref{tab:W_zero_table} denotes the parameterization of $T_{k_1,k_2}$ (in terms of generic odd functions $g_o(k),h_o(k)$ and even functions $g_e(k),h_e(k)$). The second and fifth columns indicate whether $W_{k_1,k_2,k_3,k_4}$ vanishes for an infinite and finite momentum cutoff $k_b$ respectively. Note that (antisymmetrized) $f_k\propto T_{k,-k}$ needs to be non-zero for a non-vanishing operator $\hat{O}^\dagger_2$. For instance, this precludes the parameterization $T_{k_1,k_2}=g_e(k_1)-g_e(k_2)$ from being a valid solution for a RSGA-1. On the other hand, for example,  $T_{k_1,k_2}=g_o(k_1)-g_o(k_2)$ for a generic odd function $g_o(k)$ is a valid solution for a RSGA-1.


\begin{table}[h!] 
\centering 

\begin{tabular}{|c|c|c|c|c|c|c|}
\hline
 $T_{k_1,k_2}$& \makecell{$k_b= \infty$ \\ $q_1=q_2=0$} & 
\makecell{$k_b= \infty$, \\ $q_{1(2)}=0, q_{2(1)}\ne 0$} & 
\makecell{$k_b= \infty$, \\ $q_1\ne0, q_2\ne0$} & \makecell{$k_b$ finite \\ $q_1=q_2=0$} & 
\makecell{$k_b$ finite, \\ $q_{1(2)}=0, q_{2(1)}\ne 0$} & 
\makecell{$k_b$ finite, \\ $q_1\ne0, q_2\ne0$} \\
\hline
$g_o(k_1)-g_o(k_2)$ & 0 &X & X & 0& X & X \\
\hline
$g_e(k_1)-g_e(k_2)$ & 0 & 0 & X & 0& 0 & X \\
\hline
$g_o(k_1)\cdot h_o(k_2) - h_o(k_1)\cdot g_o(k_2)$ & 0 & 0 & X  & 0& 0 & X \\
\hline
$g_o(k_1)\cdot h_e(k_2) - h_e(k_1)\cdot g_o(k_2)$ & X & X & X & X& X & X \\
\hline
$g_e(k_1)\cdot h_e(k_2) - h_e(k_1)\cdot g_e(k_2)$ & 0 & 0 & X  & 0& 0 & X\\
\hline
$k_1 - k_2$ & 0 & 0 & 0  & 0& X & X\\
\hline
\end{tabular}\caption{
Table shows whether $W^{q_1,q_2}_{k_1,k_2,k_3,k_4}$ (Eq.~\ref{appeq:1d_W_qq_kkkk}) vanishes for different parameterizations of the interaction coefficients. The latter are constrained to satisfy the separable form in Eq.~\ref{appeq:H_1d_genRSGA_separableV}, in terms of $T_{k_1,k_2}$. 
$g_o(k),h_o(k)$ are general odd functions, and $g_e(k),h_e(k)$ are general even functions. 
An entry `0' (`X') indicates that  $W^{q_1,q_2}_{k_1,k_2,k_3,k_4}$ is zero (non-zero) based on the specified conditions for $k_b$ and $q_1$, $q_2$ in the first row.\label{tab:W_zero_table} }
\end{table}

Further exact many-body eigenstates with non-zero momenta can constructed by similar methods.
To see this, we rewrite the toy model Hamiltonian with separable $V_{k_4,k_3,k_2,k_1}=-T_{k_4,k_3}T_{k_1,k_2}^*$ as (we ignore factors of $c$, $L$, and other constants for simplicity)
\begin{align}\label{appeq:1d_H_PqPq}
    \hat{H}=-\sum_{q}P_{q}^\dagger P_q,
\end{align}
where
\begin{align}
    P_q=\sum^{\{k_1,k_2\}}_{k_1,k_2}T_{k_1,k_2}^* \gamma_{k_2}\gamma_{k_1}\delta_{q,k_1+k_2}.
\end{align}
Similarly to the zero momentum case, we define a two-particle operator with finite momentum $p$
\begin{align}
    \hat{O}_{2,p}^\dagger=\sum^{\{k,k'\}}_{k,k'}f_{k,k'}\gamma_{k}^\dagger\gamma_{k'}^\dagger\delta_{p,k+k'},
\end{align}
where $f_{k,k'}$ is antisymmetric under interchange of its arguments. We evaluate
\begin{align}
    P_q\hat{O}_{2,p}^\dagger\ket{\text{vac}}&=\sum^{\{{k_1,k_2,k,k'}\}}_{k_1,k_2,k,k'}T_{k_1,k_2}^* f_{k,k'}\gamma_{k_2}\gamma_{k_1}\gamma_{k}^\dagger\gamma_{k'}^\dagger\delta_{p,k+k'}\delta_{q,k_1+k_2}\ket{\text{vac}}\nonumber\\
    &=\sum^{\{{k_1,k_2,k,k'}\}}_{k_1,k_2,k,k'}T_{k_1,k_2}^* f_{k,k'}\delta_{p,k+k'}\delta_{q,k_1+k_2}(\delta_{k,k_1}\delta_{k',k_2}-\delta_{k_1,k'}\delta_{k_2,k})\ket{\text{vac}}\nonumber\\
    &=\sum^{\{k_1,k_2\}}_{k_1,k_2}T_{k_1,k_2}^* \delta_{p,q}\delta_{q,k_1+k_2}(f_{k_1,k_2}-f_{k_2,k_1})\ket{\text{vac}}\nonumber\\
    &=2\sum^{\{k_1,k_2\}}_{k_1,k_2}T_{k_1,k_2}^* \delta_{p,q}\delta_{q,k_1+k_2}f_{k_1,k_2}\ket{\text{vac}}.
\end{align}
With the above relation, we act the Hamiltonian on $\hat{O}_{2,p}^\dagger\ket{\text{vac}}$
\begin{align}
    \hat{H}\hat{O}_{2,p}^\dagger\ket{\text{vac}}&=-\sum_{q}P_q^\dagger P_q\hat{O}_{2,p}^\dagger\ket{\text{vac}}=-2\sum^{\{k_1,k_2\}}_{k_1,k_2}T_{k_1,k_2}^* f_{k_1,k_2}\delta_{p,k_1+k_2}P_p^\dagger\ket{\text{vac}}.
\end{align}
Note that if and only if $f_{k,k'}=A T_{k,k'}$ for some constant $A$, then $P_p^\dagger\ket{\text{vac}}\propto \hat{O}_{2,p}^\dagger\ket{\text{vac}}$, which leads to
\begin{align}
    \hat{H}\hat{O}_{2,p}^\dagger\ket{\text{vac}}=-2\sum^{\{k_1,k_2\}}_{k_1,k_2}|T_{k_1,k_2}|^2 \delta_{p,k_1+k_2}\hat{O}_{2,p}^\dagger\ket{\text{vac}}.
\end{align}
Note that within the 2-electron Hilbert space, the Hamiltonian is rank-1 due to its separability. Hence, $\hat{P}_{p}^\dagger\ket{\text{vac}}$ is the ground state (given our choice of negative semi-definite $\hat{H}$ in Eq.~\ref{appeq:1d_H_PqPq}) within momentum sector $p$, with energy
\begin{equation}
    E_{2,p}=-2\sum^{\{k_1,k_2\}}_{k_1,k_2}|T_{k_1,k_2}|^2 \delta_{p,k_1+k_2},
\end{equation}
while all other energies are $0$.

To construct higher-body states, we consider higher-order commutators. We trivially have
\begin{align}
    \left[\left[\left[\hat{H},P_{q_1}^\dagger\right],P_{q_2}^\dagger\right],\gamma_{k}^\dagger\right]=0, \forall q_1,q_2.
\end{align}
Thus, 
\begin{align}
     \left[\left[\left[\hat{H},P_{q_1}^\dagger\right],P_{q_2}^\dagger\right],P_{q_3}^\dagger\right]=0.
\end{align}
To find a RSGA-1, we can therefore focus on $\left[\left[\hat{H},P_{q_1}^\dagger\right],P_{q_2}^\dagger\right]$. Since $\hat{H}=-\sum_{q}P_{q}^\dagger P_q$, we study the commutators between $P_q$ and $P^\dagger_{q'}$. It is clear that $[P_{q_1}^\dagger,P_{q_2}^\dagger]=0$. Combining
\begin{align}
\left[P_q,P_{q_1}^\dagger\right]&=\sum^{\{k_1,k_2\}}_{k_1,k_2}\sum^{\{k_3,k_4\}}_{k_3,k_4}T_{k_1,k_2}^*T_{k_3,k_4}\delta_{k_1+k_2,q}\delta_{k_3+k_4,q_1}\left[\gamma_{k_2}\gamma_{k_1},\gamma_{k_3}^\dagger\gamma_{k_4}^\dagger\right]
\end{align}
with the identity
\begin{align}
\left[\gamma_{k_2}\gamma_{k_1},\gamma_{k_3}^\dagger\gamma_{k_4}^\dagger\right]&=\delta_{k_1 ,k_3}\gamma_{k_2}\gamma_{k_4}^\dagger - \delta_{k_1, k_4}\gamma_{k_2}\gamma_{k_3}^\dagger + \delta_{k_2, k_3}\gamma_{k_4}^\dagger\gamma_{k_1} - \delta_{k_2, k_4}\gamma_{k_3}^\dagger\gamma_{k_1}\nonumber\\
&=-\delta_{k_1, k_3}\gamma_{k_4}^\dagger\gamma_{k_2} + \delta_{k_1, k_4}\gamma_{k_3}^\dagger\gamma_{k_2} + \delta_{k_2 ,k_3}\gamma_{k_4}^\dagger\gamma_{k_1} - \delta_{k_2, k_4}\gamma_{k_3}^\dagger\gamma_{k_1}+\delta_{k_1, k_3}\delta_{k_2,k_4}-\delta_{k_1 k_4}\delta_{k_2, k_3},
\end{align}
leads to
\begin{align}
    \left[\left[P_q,P_{q_1}^\dagger\right],P_{q_2}^\dagger\right]&=\sum^{\{k_1,k_2,k_3,k_4,k_5,k_6\}}_{k_1,k_2,k_3,k_4,k_5,k_6}T_{k_1,k_2}^*T_{k_3,k_4}T_{k_5,k_6}\delta_{k_1+k_2,q}\delta_{k_3+k_4,q_1}\delta_{k_5+k_6,q_2}\Big[(\delta_{k_2, k_3}\delta_{k_1,k_5}-\delta_{k_1, k_3}\delta_{k_2,k_5})\gamma_{k_4}^\dagger\gamma_{k_6}^\dagger\nonumber\\
    &+(\delta_{k_1, k_3}\delta_{k_2,k_6}-\delta_{k_2, k_3}\delta_{k_1,k_6})\gamma_{k_4}^\dagger\gamma_{k_5}^\dagger \nonumber+ (\delta_{k_1, k_4}\delta_{k_2,k_5}-\delta_{k_2, k_4}\delta_{k_1,k_5})\gamma_{k_3}^\dagger\gamma_{k_6}^\dagger\\
    &+(\delta_{k_2, k_4}\delta_{k_1,k_6}-\delta_{k_1, k_4}\delta_{k_2,k_6})\gamma_{k_3}^\dagger\gamma_{k_5}^\dagger\Big]\nonumber\\
    &=8\sum^{\{k_1,\cdots,k_6\}}_{k_{1},\cdots,k_{6}}T_{k_1,k_2}^*T_{k_3,k_4}T_{k_5,k_6}\delta_{k_1+k_2,q}\delta_{k_3+k_4,q_1}\delta_{k_5+k_6,q_2}\delta_{k_1,k_5}\delta_{k_2,k_3}\gamma_{k_4}^\dagger\gamma_{k_6}^\dagger.
\end{align}
With the above results, we find
\begin{align}
    \left[\left[\hat{H},P_{q_1}^\dagger\right],P_{q_2}^\dagger\right]&=-8\sum^{\{k_1,\ldots,k_6\}}_{k_{1},\cdots,k_{6},q}T_{k_1,k_2}^*T_{k_3,k_4}T_{k_5,k_6}\delta_{k_1+k_2,q}\delta_{k_3+k_4,q_1}\delta_{k_5+k_6,q_2}\delta_{k_1,k_5}\delta_{k_2,k_3}P_{q}^\dagger\gamma_{k_4}^\dagger\gamma_{k_6}^\dagger\nonumber\\
    &=-8\sum^{\{k_1,\ldots,k_8\}}_{k_{1},\cdots,k_{8}}T_{k_7,k_8}T_{k_1,k_2}^*T_{k_3,k_4}T_{k_5,k_6}\delta_{k_7+k_8,q}\delta_{k_1+k_2,q}\delta_{k_3+k_4,q_1}\delta_{k_5+k_6,q_2}\delta_{k_1,k_5}\delta_{k_2,k_3}\gamma_{k_7}^\dagger\gamma_{k_8}^\dagger\gamma_{k_4}^\dagger\gamma_{k_6}^\dagger\nonumber\\
    &=-8\sum^{\{k_7,k_8,k_4,k_6\}}_{k_7,k_8,k_4,k_6}W_{k_7,k_8,k_4,k_6}'^{q_1,q_2}\gamma_{k_7}^\dagger\gamma_{k_8}^\dagger\gamma_{k_4}^\dagger\gamma_{k_6}^\dagger.
\end{align}
Here, $W_{k_7,k_8,k_4,k_6}'^{q_1,q_2}$ is defined as
\begin{align}
    W_{k_7,k_8,k_4,k_6}'^{q_1,q_2}&=T_{k_7,k_8}\sum^{\{k_5,k_3\}}_{k_{3},k_{5}}T_{k_5,k_3}^*T_{k_3,k_4}T_{k_5,k_6}\delta_{k_7+k_8,k_5+k_3}\delta_{k_3+k_4,q_1}\delta_{k_5+k_6,q_2}\nonumber\\
    &=T_{k_7,k_8}T_{q_2-k_6,q_1-k_4}^*T_{q_1-k_4,k_4}T_{q_2-k_6,k_6}\delta_{k_7+k_8+k_4+k_6,q_1+q_2}\delta_{q_1-k_4\in\mathcal{H}}\delta_{q_2-k_6\in\mathcal{H}},
\end{align}
where the symbol $\delta_{k\in\mathcal{H}}$ is 1 if $k$ lies in the momentum cutoff, and 0 otherwise. For clarity, we relabel the momenta as 
\begin{align}
    W_{k_4,k_3,k_2,k_1}'^{q_1,q_2}    &=T_{k_4,k_3}T_{q_2-k_1,q_1-k_2}^*T_{q_1-k_2,k_2}T_{q_2-k_1,k_1}\delta_{k_1+k_2+k_3+k_4,q_1+q_2}\delta_{q_1-k_2\in\mathcal{H}}\delta_{q_2-k_1\in\mathcal{H}}.
\end{align}
Utilizing the full anti-symmetry of $\gamma_{k_7}^\dagger\gamma_{k_8}^\dagger\gamma_{k_4}^\dagger\gamma_{k_6}^\dagger$ ($\gamma_{k_4}^\dagger\gamma_{k_3}^\dagger\gamma_{k_2}^\dagger\gamma_{k_1}^\dagger$ with the relabeled momenta), we consider the fully antisymmetrized version of $W'$
\begin{align}\label{appeq:1d_W_qq_kkkk}
    W_{k_4,k_3,k_2,k_1}^{q_1,q_2}=\sum_{\sigma\in S_4}\text{sign}(\sigma)W_{k_{\sigma_4},k_{\sigma_3},k_{\sigma_2},k_{\sigma_1}}'^{q_1,q_2},
\end{align}
in terms of which we have
\begin{align}
    \left[\left[\hat{H},P_{q_1}^\dagger\right],P_{q_2}^\dagger\right]=-\frac{1}{3}\sum^{\{k_1,k_2,k_3,k_4\}}_{k_1,k_2,k_3,k_4} W_{k_4,k_3,k_2,k_1}^{q_1,q_2}\gamma_{k_4}^\dagger\gamma_{k_3}^\dagger\gamma_{k_2}^\dagger\gamma_{k_1}^\dagger.
\end{align}
This is the generalization of Eq.~\ref{eq:HOO_general_mom_0} to double commutators of non-zero momentum operators. 

In Tab.~\ref{tab:W_zero_table}, we test several parameterizations of (antisymmetric) $T_{k_1,k_2}$ in search of the solution of $W_{k_4,k_3,k_2,k_1}^{q_1,q_2}=0$. Interestingly, we find that $T_{k_1,k_2}=k_1-k_2$ leads to vanishing $W_{k_4,k_3,k_2,k_1}^{q_1,q_2}=0$ in the absence of a momentum cutoff. 

If $W_{k_4,k_3,k_2,k_1}^{q_1,q_2}=0$, we can generate exact towers of states with finite momentum. To see this, we first use the fact that $P^\dagger_q\ket{\text{vac}}$ is the 2-electron ground state in the momentum sector $q$
\begin{align}
    \hat{H}P^\dagger_q\ket{\text{vac}}=E_{2,q}P^\dagger_q\ket{\text{vac}}\Rightarrow [\hat{H},P^\dagger_q]\ket{\text{vac}}=E_{2,q}P^\dagger_q\ket{\text{vac}}.
\end{align}
With the property that $\left[\left[\hat{H},P_{q_1}^\dagger\right],P_{q_2}^\dagger\right]=0$, we find
\begin{align}
    \hat{H}P_{q_1}^\dagger P_{q_2}^\dagger\ket{\text{vac}}&=-P_{q_1}^\dagger P_{q_2}^\dagger \hat{H}\ket{\text{vac}}+P_{q_1}^\dagger \hat{H}P_{q_2}^\dagger\ket{\text{vac}}+P_{q_2}^\dagger \hat{H}P_{q_1}^\dagger\ket{\text{vac}}\nonumber\\
    &=(E_{2,q_1}+E_{2,q_2})P_{q_1}^\dagger P_{q_2}^\dagger\ket{\text{vac}}.
\end{align}
We have used the fact that $[P_{q_1}^\dagger,P_{q_2}^\dagger]=0$ in the last line of the above equation. Hence, using this RSGA-1, we can construct the eigenstates with finite momentum $q=q_1+\cdots+ q_N$ by the action of pair-creation operators $P_{q_i}^\dagger$ on the vacuum state
\begin{align}
    \hat{H}\prod_{i}P_{q_i}^\dagger\ket{\text{vac}}=\left(\sum_{j}E_{2,q_j}\right)\prod_{i}P_{q_i}^\dagger\ket{\text{vac}}.\label{app:eq:towerofstate_finite}
\end{align}
 The resulting energy eigenvalue is exactly the sum of the pair energies $\sum_j E_{2,q_j}$. This perfect additivity of the energy spectrum signifies that the electron pairs are effectively non-interacting, rendering the tower of states exactly solvable. In this way, we can construct exact many-body eigenstates for any total momentum sector.

For the density-density interaction $V_q\propto q^2$, we find that introducing a sharp momentum cutoff $k_b$ (induced by $v=\infty$) yields  non-zero values for $W_{k_4,k_3,k_2,k_1}^{q_1,q_2}$ (see Tab.~\ref{tab:W_zero_table}). Nevertheless, if the non-zero corrections to $W_{k_4,k_3,k_2,k_1}^{q_1,q_2}$ are small compared the energy spacings (related to differences in $E_{2,p}$) in the limit of an exact RSGA-1, it is possible to perform a perturbation theory to determine the perturbed many-body ground states at finite momentum. A detailed treatment of this will be the subject of a future paper~\cite{repulsive_unpub}.

\subsubsection{Dispersion at $p\rightarrow 0$}\label{appsubsec:1d_2e_general_dispersion}
\begin{figure}[H]
    \centering
    \includegraphics[width=1.0\linewidth]{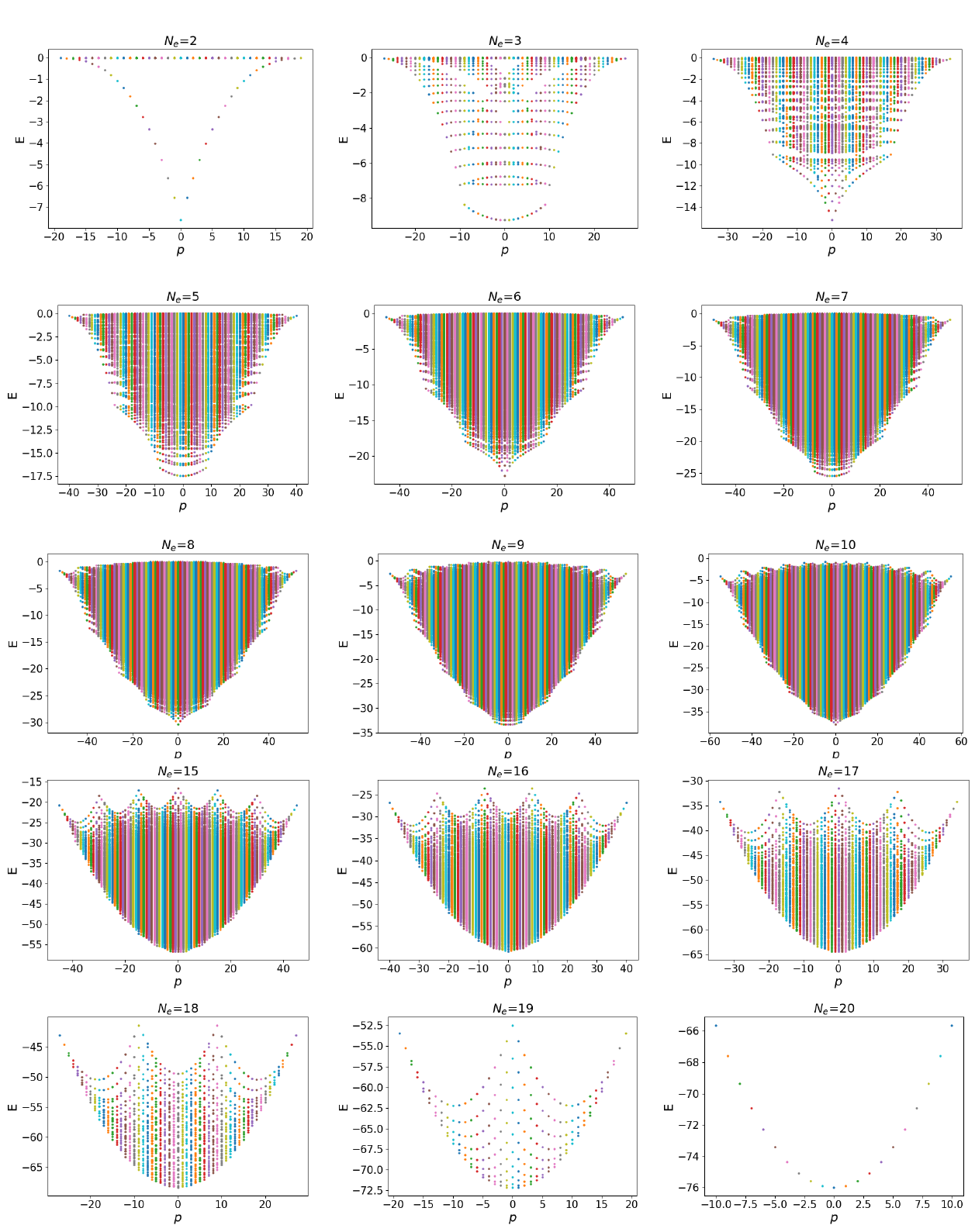}
    \caption{The full energy spectrum of the attractive 1D trashcan model with $U=-1, L+1=N_{k_b}=21$ and varying particle number $N_e$. }
    \label{fig:dispersion_1d_varying_N}
\end{figure}

\begin{figure}[H]
    \centering
    \includegraphics[width=1.0\linewidth]{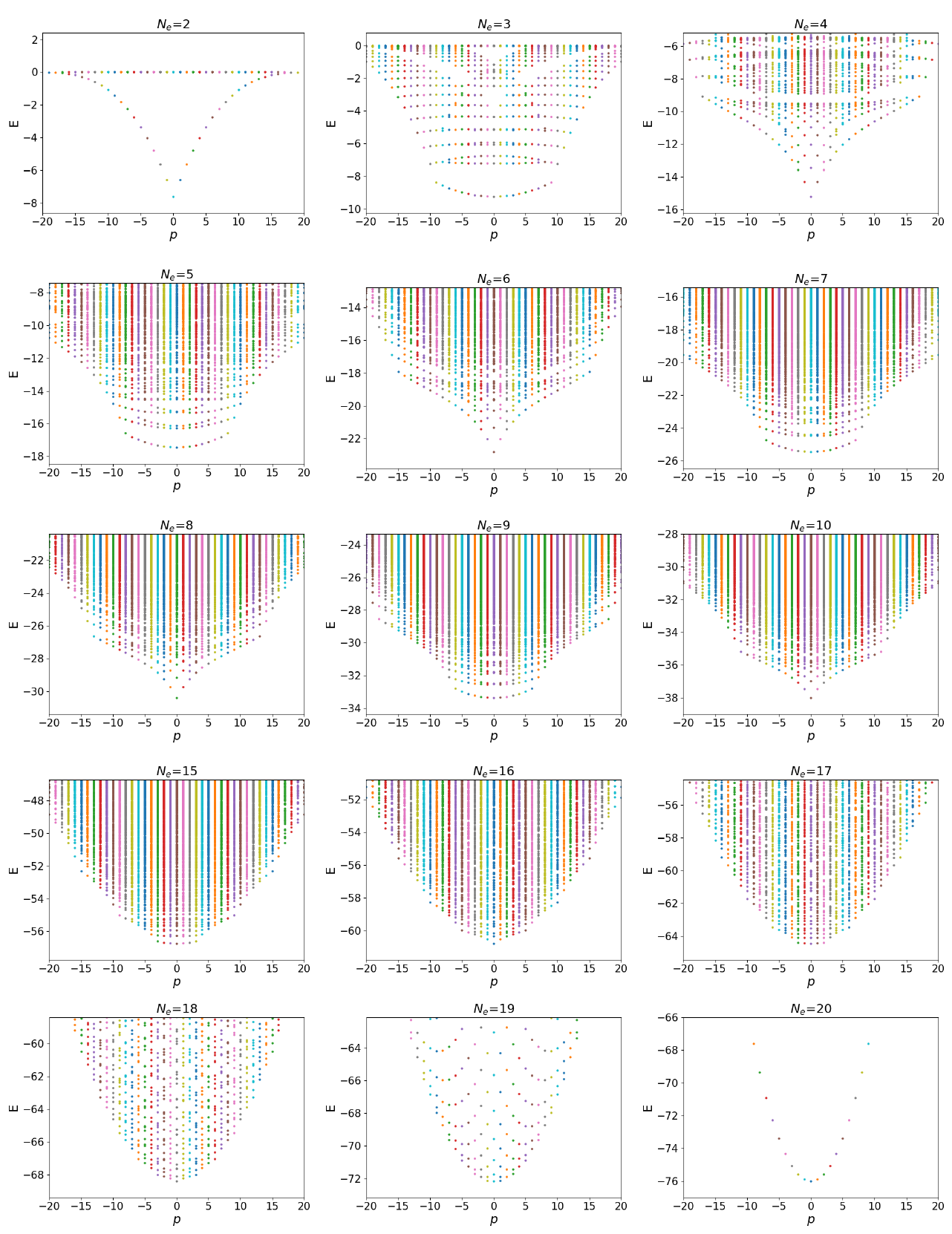}
    \caption{Zoomed in view of the full energy spectrum of the attractive 1D trashcan model with $U=-1, L+1=N_{k_b}=21$ and varying particle number $N_e$. }
    \label{fig:dispersion_1d_varying_N_zoom}
\end{figure}

In this section, we discuss the energy of the ground state as a function of total momentum $p$, i.e.~the dispersion. The dispersion of the 2-electron state is given in Eq.~\ref{eq:dispersion_1d_inf_vf} and exhibits linear behavior at small $p$, while the dispersion of the 2-hole state is quadratic. We numerically observe a crossover between linear and quadratic dispersion for increasing even $N_e$ (see Fig.~\ref{fig:dispersion_1d_varying_N}, and a zoomed-in view of the low-energy states in Fig.~\ref{fig:dispersion_1d_varying_N_zoom}). On the other hand, the derivative of the dispersion at $p=0$ appears to vanish for all odd $N_e$.

\begin{figure}[H]
    \centering
    \includegraphics[width=0.7\linewidth]{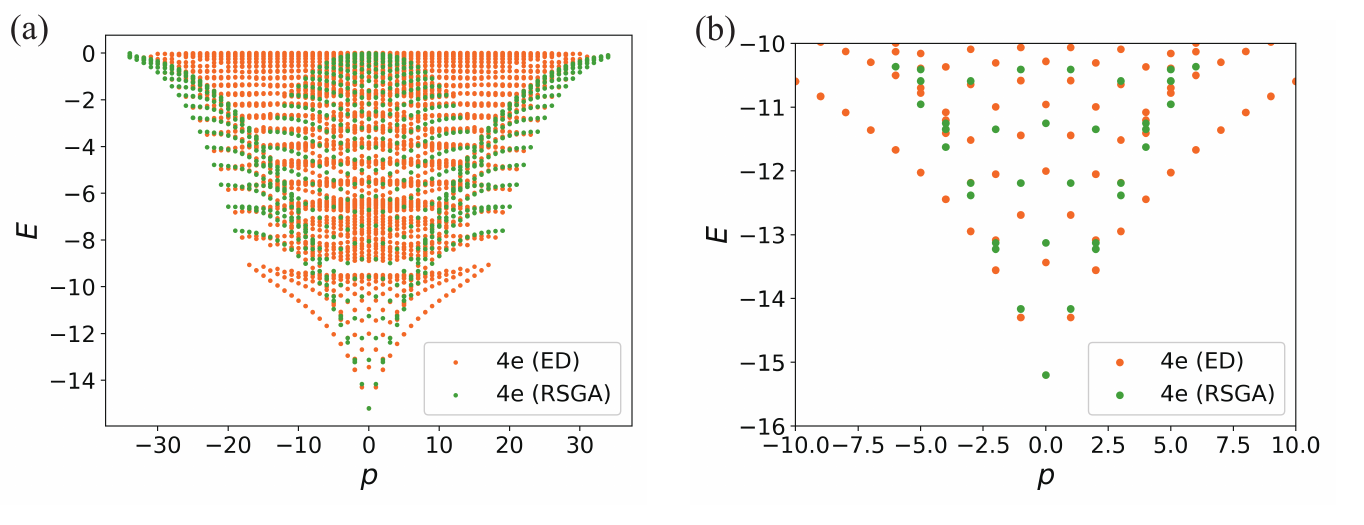}
    \caption{(a) Comparison between the ED spectrum and the spectrum generated by the RSGA-1 for the attractive 1D trashcan model with $N_e=4,U=-1, L+1=N_{k_b}=21$. The ED spectrum consists all the states. The RSGA-1 energies consist of a sum of two 2-electron energies $E_{2,p_1}+E_{2,p_2}$  (see Eq.~\ref{app:eq:2_electron_Ep_1d}) for all combinations of $p_1,p_2$. (b) Zoom in view of (a) near the ground state at $p=0$.}
    \label{fig:rsga_spectrum_4e}
\end{figure}

To motivate the linear dispersion for even $N_e$, we compare in Fig.~\ref{fig:rsga_spectrum_4e} the four-electron ($N_e=4$) ED energy spectrum, with the spectrum generated by the RSGA-1 assuming that it is exact for finite momentum (see discussion in App.~\ref{appsubsec:RSGA_generalization}). The latter corresponds to states $P_{p_1}^\dagger P_{p_2}^\dagger\ket{\text{vac}}$ constructed from the 2-electron pairing operators, with corresponding energies $E_{2,p_1}+E_{2,p_2}$ (see Eq.~\ref{app:eq:2_electron_Ep_1d}).
We find that the spectrum constructed from the RSGA-1 approximately captures the low-energy ED spectrum for $N_e=4$. This points to the validity of an approximate, finite-momentum RSGA-1 structure at low energies, as discussed in App.~\ref{appsubsec:RSGA_generalization}. A zoomed-in view of the spectrum near zero total momentum, shown in Fig.~\ref{fig:rsga_spectrum_4e}(b), reveals a linear dispersion, which is a direct consequence of the RSGA-1 framework. This linear dispersion is a general feature for even-electron states in this model. Since the two-electron ground state itself exhibits a linear dispersion $E_{2,p}\propto |p|$ (up to a constant), the total ground state energy of a low-momentum $2N$-electron state composed of $N$ such pairs is also linear in $|p|$. This follows from the energy additivity (Eq.~\ref{app:eq:towerofstate_finite}) inherent to the RSGA-1 structure.

We also provide a possible reason for the non-linear dispersion for odd $N_e$ based on the ansatz in Eq.~\ref{eq:odd_ansatz_1d}. The ansatz for $N_e=2N+1$ at $p=0$ consisted of creating a single unpaired electron at $k=0$ on top of the exact $N_e=2N$ wavefunction. In App.~\ref{appsubsec:oddNe_vFinfty}, we argued that the momentum of the single unpaired electron $\gamma^\dagger_0$ should be at $k=0$ since that minimizes the pairing energy $E_2'$ of the remaining electron pairs (see Eq.~\ref{appeq:1d_E2prime}). In particular, having the unpaired electron at $k=0$ does not `block' the binding of the remaining electrons, which form $\pm k$ pairs with non-zero $k$. For finite total momentum $p$, we could either (i) keep the unpaired electron at $k=0$ and rearrange the pairing of the remaining electrons to obtain momentum $p$, or (ii) simply shift the momentum of the unpaired electron to $k=p$. In the former scenario, we expect an energy change that is linear in $p$, since the dispersion of the even-particle ground state is itself linear as discussed in the previous paragraph. In the latter scenario, the pair-blocking induced by having an unpaired electron at $k$ would lead to a quadratic-in-$k$ change in the pairing energy $E_2'$ (Eq.~\ref{appeq:1d_E2prime}). Hence for sufficiently small $p$, we anticipate that the dispersion for odd $N_e$ is quadratic.



\subsubsection{Binding Energies}\label{app:sec:1d_binding_energy}

To investigate the presence of superconductivity in this system, we numerically compute the binding energy $E_{b,m}$, defined as
\begin{align}
  E_{b,m}(N_e) \;=\; -2E(N_e)\;+\;E(N_e-m)\;+\;E(N_e+m)\,,
\end{align}
where $E(N_e)$ is the ground‐state energy of a system with $N_e$ particles.  We focus on the cases $m=1$ and $m=2$, corresponding respectively to the pair binding energy $E_{b,1}$ and the quartet binding energy $E_{b,2}$.

A primary signature of Cooper pairing is an even–odd staggering in $E_{b,1}$, for example a positive (negative) pair binding energy $E_{b,1}>0$ ($E_{b,1}<0$) for even (odd) $N_e$. This signifies that the ground state is energetically stable for even $N_e$, with an energy cost to break a pair. The negative binding energy for odd $N_e$ indicates that it is energetically favorable for two systems with an odd number of particles to instead form two systems with an even number of particles. This energetic preference for paired, even-particle ground states is a hallmark of a pairing instability. 
Furthermore, a superconducting ground state composed of Cooper pairs is expected to exhibit a small quartet binding energy, $|E_{b,2}|$. This condition is required for the spontaneous breaking of the global charge-$U(1)$ symmetry, which enables the coherent superposition of states with different particle numbers.

\begin{figure}[t]
    \centering
    \includegraphics[width=1.0\linewidth]{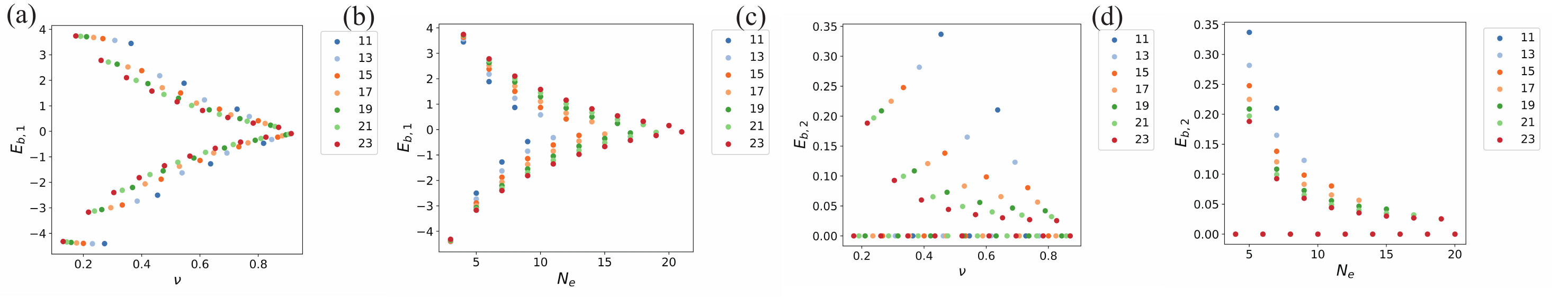}
    \caption{Binding energy (a)-(b) $E_{b,1}$ and (c)-(d) $E_{b,2}$ for the attractive 1D toy model, as a function of electron number $N_e$ and filling factor $\nu=N_e/N_{k_b}$. Different colors indicate different system sizes $N_{k_b}$ (see legend), while $U=-1$ and $L+1=N_{k_b}$.}
    \label{fig:binding_energy_1d}
\end{figure}

Fig.~\ref{fig:binding_energy_1d} shows the binding energies calculated using ED. 
The pair binding energy $E_{b,1}$ exhibits clear even-odd staggering, with its amplitude decaying to zero as the electron number increases to full filling.
In particular, $E_{b,1}$ is positive for even $N_e$ and negative for odd $N_e$,  confirming the binding of electrons into pairs. This behavior is captured by our ansatz (see App.~\ref{appsubsec:oddNe_vFinfty}), for which the binding energies are
\begin{gather}
    E_{b,1}^A(2N)=\Delta E(2N+1)+\Delta E(2N-1)\\
    E_{b,1}^A(2N+1)=-2\Delta E(2N+1),
\end{gather}
where $\Delta E$ (defined in Eq.~\ref{app:eq:delta_e_1d_odd}) is positive for an attractive interaction and grows with the number of holes.

For the quartet binding energy, $E_{b,2}$ is exactly zero for even $N_e$. This is a direct consequence of the exact linearity of the even-particle ground-state energy as a function of the number of pairs. For odd $N$, we find that $E_{b,2}$ is an order of magnitude smaller than $E_{b,1}$, providing another strong indicator of superconductivity in the system.

\subsubsection{Off-Diagonal-Long-Range-Order}\label{secapp:1d_Hamiltonian_ODLRO}
Off-diagonal long-range order (ODLRO)~\cite{PhysRevLett.63.2144,yang1962ODLRO} provides another diagnostic for superconductivity. We begin with the 4-point correlator
\begin{align}
        \rho^{(2)}_{(k_1,k_2),(k_3,k_4)}=\bra{\text{GS}}\gamma^\dagger_{k_1}\gamma^\dagger_{k_2}\gamma_{k_4}\gamma_{k_3}\ket{\text{GS}},
\end{align}
where $\ket{\text{GS}}$ is the ground state wavefunction under consideration.
Performing a Fourier transformation, we obtain 
\begin{align}
    \rho^{(2)}_{(r_1,r_2),(r_3,r_4)}=\int\frac{dk_1dk_2dk_3 dk_4}{(2\pi)^4} e^{-i(k_1r_1 + k_2r_2 - k_3r_3 - k_4r_4)} \bra{\text{GS}}\gamma^\dagger_{k_1}\gamma^\dagger_{k_2}\gamma_{k_4}\gamma_{k_3}\ket{\text{GS}}.\label{eq:2RDM_1d_real}
\end{align}
This can be interpreted as the correlation function for destroying an electron pair at $r_3,r_4$, then creating an electron pair at $r_1,r_2$. ODLRO manifests as a non-vanishing correlator in the situation where the positions $r_3,r_4$ of the destroyed pair are infinitely far away from the positions $r_1,r_2$ of the created pair.

The ground wavefunction for an even number $N_e=2N$ of particles 
\begin{align}
    \ket{\phi_{2N}}=(\hat{O}_2^\dagger)^N\ket{\text{vac}}
\end{align}
is constructed by repeated application of $\hat{O}^\dagger_2$, a two-particle operator carrying zero momentum, on the vacuum. This implies that in $\ket{\phi_{2N}}$, if the single-particle momentum $k$ is occupied, then $-k$ is also necessarily occupied. Thus, $\rho^{(2)}$ is only nonzero in 3 cases: (i) $k_1+k_2=k_3+k_4=0$, (ii) $k_1=k_4,\,k_2=k_3$, and (iii) $k_1=k_3,\,k_2=k_4$. Therefore, we can rewrite Eq.~\ref{eq:2RDM_1d_real} as 
\begin{align}
    \rho^{(2)}_{(r_1,r_2),(r_3,r_4)}&=\int\frac{dk_1dk_3}{(2\pi)^2} e^{-i(k_1(r_1 - r_2) - k_3(r_3 - r_4))} \bra{\text{GS}}\gamma^\dagger_{k_1}\gamma^\dagger_{-k_1}\gamma_{-k_3}\gamma_{k_3}\ket{\text{GS}}\nonumber\\
    &+\int\frac{dk_1dk_2}{(2\pi)^2} e^{-i(k_1(r_1 - r_4) + k_2(r_2 - r_3))} \bra{\text{GS}}\gamma^\dagger_{k_1}\gamma^\dagger_{k_2}\gamma_{k_1}\gamma_{k_2}\ket{\text{GS}}\nonumber\\
    &+\int\frac{dk_1dk_2}{(2\pi)^2} e^{-i(k_1(r_1 - r_3) + k_2(r_2 - r_4))} \bra{\text{GS}}\gamma^\dagger_{k_1}\gamma^\dagger_{k_2}\gamma_{k_2}\gamma_{k_1}\ket{\text{GS}}.\label{eq:odlro_fourier}
\end{align}
In the limit where the intra-pair separations $|r_1 - r_2|$ and $|r_3 - r_4|$ are finite, while the inter-pair separation tends to infinity so that
\begin{align}
    k_b|r_1 -r_3|,k_b|r_2-r_4|\to\infty,
\end{align}
the second and third terms in Eq.~\ref{eq:odlro_fourier} become negligible due to rapid oscillation of their exponential factors. 

For odd $N_e$ though, $\rho^{(2)}$ is non-zero even when the momenta do not belong to one of the 3 cases  mentioned above, since the electrons do not necessarily form $\pm k$ pairs in the ground state.
Setting $r_1 = 0$ without loss of generality, we express Eq.~\eqref{eq:2RDM_1d_real} as
\begin{align}
\rho^{(2)}_{(0,r_2),(r_3,r_4)} = \int \frac{dk_1 dk_2 dk_3 dk_4}{(2\pi)^4} e^{-i(k_2 r_2 - k_3 r_3 - k_4 r_4)} \bra{\mathrm{GS}} \gamma^\dagger_{k_1} \gamma^\dagger_{k_2} \gamma_{k_4} \gamma_{k_3} \ket{\mathrm{GS}} \delta(k_1 + k_2 - k_3 - k_4),\label{app:eq:general_odlro_1d}
\end{align}
where momentum conservation is explicitly indicated. We consider the limit where the inter-pair separation tends to infinity. Given that $k_b r_2$ and $k_b |r_3 - r_4|$ are finite while $k_br_4\to \infty$, then $\rho^{(2)}_{(0,r_2),(r_3,r_4)}$ vanishes when $k_3 + k_4 \ne 0$ due to the rapid oscillations. When $k_1+k_2=k_3+k_4=0$, Eq.~\ref{app:eq:general_odlro_1d} reduces to the first term in Eq.~\eqref{eq:odlro_fourier}
\begin{align}
\rho^{(2)}_{(0,r_2),(r_3,r_4)} \stackrel{k_br_4\rightarrow\infty}{\approx} \int\frac{dk_1dk_3}{(2\pi)^2} e^{-i(-k_1 r_2 - k_3(r_3 - r_4))} \bra{\text{GS}}\gamma^\dagger_{k_1}\gamma^\dagger_{-k_1}\gamma_{-k_3}\gamma_{k_3}\ket{\text{GS}}.
\end{align}

\begin{figure}[h]
    \centering
    \includegraphics[width=0.8\linewidth]{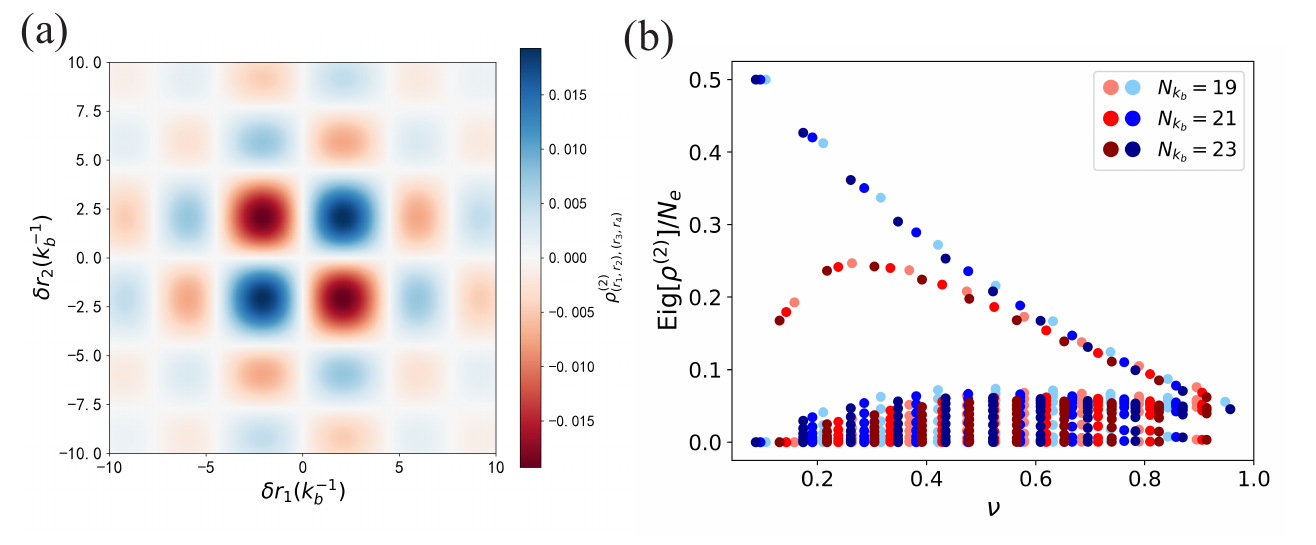}
    \caption{(a) Two-particle density matrix $\rho^{(2)}_{(r_1,r_2),(r_3,r_4)}$ of the ground state for two electrons (Eq.~\ref{app:eq:rho_2_2e}) in the attractive 1D toy trashcan model with $v=\infty$ as a function of $\delta r_1=r_1-r_2$ and $\delta r_2=r_3-r_4$ in units of $k_b^{-1}$. We have assumed that $k_b|r_1 -r_3|,k_b|r_2-r_4|\to\infty$. (b) Eigenvalues of the two-particle ground state density matrix (Eq.~\ref{app:eq:reduced_1d_rho2}) normalized by the electron number $N_e$ for  $L+1=N_{k_b}=19,21,23$, as a function of filling. The results are obtained from ED calculation. Blue (red) dots indicate even (odd) $N_e$. The presence of a finite eigenvalue for finite filling factor $\nu$ indicates ODLRO.}
    \label{app:fig:ODLRO_1d}
\end{figure}

Thus, to demonstrate the existence of ODLRO, it suffices to evaluate the expectation value 
\begin{align}
\rho_{k_1,k_2}^{(2)}=\bra{\mathrm{GS}} \gamma^\dagger_{k_1} \gamma^\dagger_{-k_1} \gamma_{-k_2} \gamma_{k_2} \ket{\mathrm{GS}}.\label{app:eq:reduced_1d_rho2}
\end{align}
For even-particle states, we begin with the exact ground state wavefunction in Eq.~\eqref{app:eq:ground_state_1d_even} and express it as
\begin{align}
    \ket{\phi_{2N}}&=\frac{1}{Z}\sum^{\{k_{1},\ldots, k_N\}}_{0<k_{1}<\ldots< k_N}k_1\cdots k_N\gamma_{k_1}^{\dagger}\gamma_{-k_1}^\dagger\cdots \gamma_{k_N}^{\dagger}\gamma_{-k_N}^\dagger\ket{\text{vac}}\nonumber\\
    &\equiv\frac{1}{Z}\sum^{\{k_{1},\ldots, k_N\}}_{0<k_{1}<\ldots< k_N}k_1\cdots k_N\ket{k_1,\cdots, k_N},
\end{align}
where $Z$ is a normalization factor, and $\ket{k_1,\cdots, k_N}$ is a Fock basis state where the momenta $\pm k_1,\ldots,\pm k_N$ are occupied. The 4-point correlator can be evaluated as (where $k_i,k_j>0$ for simplicity)
\begin{align}
    \bra{\phi_{2N}}\gamma^\dagger_{k_i}\gamma^\dagger_{-k_i}\gamma_{-k_j}\gamma_{k_j}\ket{\phi_{2N}}=\frac{1}{Z^2}k_ik_j\sum^{\{k_{1},\ldots, k_{N-1}\}}_{\stackrel{0<k_1<\ldots<k _{N-1}}{\text{such that }k_i,k_j\notin\{k_1,\ldots,k _{N-1}\}}}(k_1\cdots k_{N-1})^2.
\end{align}
In particular, for $N_e=2$, we obtain
\begin{align}
    \bra{\phi_{2}}\gamma^\dagger_{k_i}\gamma^\dagger_{-k_i}\gamma_{-k_j}\gamma_{k_j}\ket{\phi_{2}}=\frac{1}{Z^2}k_ik_j,
\end{align}
which is a rank-1 matrix with only one finite eigenvalue of 1. Letting $r_1-r_2=\delta r_1$ and $r_3-r_4=\delta r_2$, and taking $k_b|r_1-r_3|$ and $k_b|r_2-r_4|$ to infinity, we find that the real-space two-particle density matrix for the two-particle state becomes
\begin{align}
    \rho^{(2)}_{(r_1,r_2),(r_3,r_4)}&\approx\int\frac{dk_1dk_3}{(2\pi)^2} e^{-i(k_1\delta r_1 - k_3\delta r_2)} \bra{\phi_2}\gamma^\dagger_{k_1}\gamma^\dagger_{-k_1}\gamma_{-k_3}\gamma_{k_3}\ket{\phi_2}\nonumber\\
    &=\int\frac{dk_1dk_3}{(2\pi)^2} e^{-i(k_1\delta r_1 - k_3\delta r_2)} \frac{1}{Z^2}k_1k_3\nonumber\\
    &=\frac{1}{\pi^2 Z^2} \left[ \frac{\sin(k_b\delta r_1) - k_b\delta r_1\cos(k_b\delta r_1)}{\delta r_1^2} \right] \left[ \frac{\sin(k_b\delta r_2) - k_b\delta r_2\cos(k_b\delta r_2)}{\delta r_2^2} \right].\label{app:eq:rho_2_2e}
\end{align}
This remains finite for $\delta r_1, \delta r_2\sim k_b^{-1}$ as shown in Fig.~\ref{app:fig:ODLRO_1d}(a).
Note that if $\delta r_1$ and $\delta r_2 \to 0$, $\rho^{(2)}_{(r_1,r_2),(r_3,r_4)}$ also approaches zero due to the Pauli exclusion principle. To generalize the above result to higher electron numbers, we numerically compute the normalized spectrum of the two-particle reduced density matrix $\rho^{(2)}/N_e$ for three different sizes $L+1=N_{k_b}=19,21,23$. Fig.~\ref{app:fig:ODLRO_1d}(b) shows that  $\rho^{(2)}/N_e$ has a single dominant eigenvalue at finite fillings. A pronounced even-odd effect is also clear at low electron densities,
where the dominant eigenvalue of $\rho^{(2)}/N_e$ systematically oscillates between even (blue) and odd (red) particle numbers. As the electron filling increases, this eigenvalue decays, a trend that is similar to the binding energy results.

\section{2D Berry Trashcan Model}\label{appsec:2d_trashcan}
\subsection{Hamiltonian}\label{appsec:2d_trashcan_Hamiltonian}

In this section, we discuss the Hamiltonian for the 2D Berry Trashcan model. The Hamiltonian is
\begin{gather}
    \hat{H}=
    \hat{H}^\text{kin}+\hat{H}^\text{int}\label{eq:2dham_trashcan_0}
\end{gather}
The kinetic term is
\begin{gather}
    \hat{H}^\text{kin}=\sum_{\bm{k}}\epsilon_{\bm{k}}\gamma^\dagger_{\bm{k}}\gamma_{\bm{k}}\\
    \epsilon_{\bm{k}}=\theta(k-k_b)v(k-k_b)
\end{gather}
where $k=|\bm{k}|$, $v$ is the velocity of the trashcan wall, and $k_b$ is the radius of the flat trashcan bottom. We will also consider an additional hard cutoff $\Lambda$ so that only single-particle momenta $\bm{k}$ with $k\leq k_b+\Lambda$ are allowed. This effectively corresponds to $\epsilon_{k>k_b+\Lambda}\rightarrow\infty$. Note that setting $v=\infty$ effectively leads to a smaller hard cutoff that restricts $k\leq k_b$.

The interaction term is
\begin{gather}
    \hat{H}^\text{int}=\frac{1}{2\Omega_{tot}}  \sum^{\{\bm{k},\bm{k}',\bm{k}+\bm{q},\bm{k}'-\bm{q}\}}_{\mbf{k}, \mbf{k'}, \mbf{q}}   V_{\bm q} \mathcal{M}_{\bm k,\bm q}\mathcal{M}^*_{\bm k',\bm q}     \gamma_{\mbf{k}+ \mbf{q}}^\dagger \gamma_{\mbf{k}'- \mbf{q}}^\dagger \gamma_{\mbf{k}'} \gamma_{\mbf{k}}.
    \label{appeq:2d_Hint}
\end{gather}
where $\Omega_\text{tot}$ is the total real-space area of the system. 
$V_{\bm{q}}$, which has units [energy]$\times$[length]$^2$ in 2D, is the momentum-space Fourier transformation of the real-space interaction. We will refer to $V_{\bm{q}}$ as the `interaction potential' in this work.
The form factor $\mathcal{M}_{\bm{k},\bm{q}}$, as well as the angular brackets in the superscript on the momentum summations, will be explained in the next paragraphs. The interaction is normal-ordered with respect to the vacuum state $\ket{\text{vac}}$. Note that here we are not considering the effect of the valence bands, which are not included in this work. This neglect of the valence bands is not valid when considering hBN-aligned samples of R$n$G, such as in the case of the experiments of Refs.~\cite{Lu2024fractional,lu2025extended,choi2024electricfieldcontrolsuperconductivity,xie2024even}. In the latter situation, Refs.~\cite{kwan2023MFCI3, yu2024MFCI4} demonstrated that in the moir\'e-distant regime (where the displacement field drives the doped conduction electrons away from the moir\'e interface), incorporating valence bands is crucial for inducing moir\'e effects in the conduction bands. This is achieved by using an interaction scheme, such as the `average scheme', that enables the occupied valence bands to impart a moir\'e-modulated potential onto the conduction electrons. In the current case, since we are not developing a microscopic theory of the origin of the interaction, and since we have no moir\'e pattern (as we are considering R$n$G without hBN-alignment), we discard the valence bands.
We leave potential effects of the valence band (such as interband polarizability) for a future publication.

Given the existence of a hard momentum cutoff (either at $k=k_b+\Lambda$ for finite $v$, or $k=k_b$ for $v=\infty$), we choose to explicitly constrain the momentum summations in $\hat{H}^\text{int}$. We can do this since the occupation of states outside the cutoff is anyways energetically forbidden, so the basis states outside the cutoff do not affect the finite-energy physics that we are interested in. The summation symbol in Eq.~\ref{appeq:2d_Hint} means that the summation should be restricted so that the superscript momenta with angular brackets all lie within the hard cutoff. This notation will be used extensively below. Since our momentum cutoff and momentum mesh respect inversion symmetry, then whenever $\bm{k}$ lies within the cutoff, then so will $-\bm{k}$. Hence, $\sum^{\{\bm{k}\}}$ automatically restricts $-\bm{k}$ to also lie within the cutoff.

$\mathcal{M}_{\bm k,\bm q}$ is the form factor of the Berry Trashcan continuum band, which takes the form~\cite{bernevig2025berrytrashcanmodelinteracting}
\begin{align}\label{appeq:2d_Mkq}
    \mathcal{M}_{\bm{k},\bm{q}}=e^{-\frac{|\beta| q^2}{2}}e^{-i\beta\bm{q}\times\bm{k}}.
\end{align}
The corresponding Berry curvature is $2\beta$. The Berry flux enclosed by the flat bottom is $\varphi_\text{BZ}=2\beta A_{b}$, where $A_{b}$ is the momentum area of the trashcan bottom. 
The above form factor is extracted from the Bloch wavefunctions of rhombohedral $n$-layer graphene (R$n$G) in the vicinity of the valley $K$ Dirac momentum. In this context, $\beta$ is related to the square of the ratio of the graphene Dirac velocity and the nearest-neighbor interlayer hopping. A derivation of Eq.~\ref{appeq:2d_Mkq} is provided in Ref.~\cite{bernevig2025berrytrashcanmodelinteracting}. Similarly, the appropriate values of the parameters $k_b$ and $v$ for R$n$G (as a function of the number  of layers $n$) are derived in Ref.~\cite{bernevig2025berrytrashcanmodelinteracting}. For R5G, Ref.~\cite{bernevig2025berrytrashcanmodelinteracting} finds that $\varphi_\text{BZ}\simeq \pi/2$. Hence, most of the numerical calculations in this work will use $\varphi_\text{BZ}=\pi/2$.

The interacting Hamiltonian can be written
\begin{equation}\label{appeq:2d_Ham_Uq}
     \hat{H}^{\text{int}}= \frac{1}{2\Omega_{tot}}  \sum^{\{\bm{k},\bm{k}',\bm{k}+\bm{q},\bm{k}'-\bm{q}\}}_{\bm{q},\bm{k},\bm{k'}}  U_{\bm{q}}e^{-i \beta (  \mbf{q}\times (\mbf{k}- \mbf{k'})) }     \gamma_{\mbf{k}+ \mbf{q}}^\dagger \gamma_{\mbf{k}'- \mbf{q}}^\dagger \gamma_{\mbf{k}'} \gamma_{\mbf{k}}
\end{equation}
where we have absorbed all the real parts into $U_{\bm{q}}$ for convenience
\begin{align}
    U_{\bm{q}}=V_{\bm{q}}e^{-|\beta| q^2}.
\end{align}
Note that $U_{\bm{q}}$ only depends on $q$. In this paper, unless otherwise specified, we consider a Gaussian-type interaction 
\begin{equation}
V_{\bm{q}}=Ue^{-(\alpha-|\beta|)q^2},
\end{equation}
so that $U_{\bm{q}}=Ue^{-\alpha q^2}$. The resulting interaction Hamiltonian is 
\begin{equation}
    \hat{H}^{\text{int}}= \frac{U}{2\Omega_{tot}}  \sum^{\{\bm{k},\bm{k}',\bm{k}+\bm{q},\bm{k}'-\bm{q}\}}_{\bm{q},\bm{k},\bm{k'}}  e^{-\alpha q^2}e^{-i \beta (  \mbf{q}\times (\mbf{k}- \mbf{k'})) }     \gamma_{\mbf{k}+ \mbf{q}}^\dagger \gamma_{\mbf{k}'- \mbf{q}}^\dagger \gamma_{\mbf{k}'} \gamma_{\mbf{k}}.\label{appeq:Hint_2d_alpha_beta}
\end{equation}

Note that for $U>0$ ($U<0$), the interaction term is positive semi-definite and hence purely repulsive (negative semi-definitive and hence purely attractive) when $\alpha\geq |\beta|$. For $\alpha=|\beta|$, the interaction potential $V_{\bm{q}}$ is constant in momentum space, which corresponds to a delta function interaction in real space. For $\alpha=\beta=0$, the interaction Hamiltonian vanishes due to fermionic statistics. 


The Hamiltonian $\hat{H}$ satisfies continuous translation invariance, leading to a conserved total momentum $\bm{p}$. In the infinite size limit $\Omega_{tot}\rightarrow\infty$, there is also $SO(2)$ rotation symmetry, which enables $\bm{p}=0$ eigenstates to be labelled by an angular momentum quantum number. Otherwise, there is a discrete rotational symmetry (such as $C_6$) depending on the momentum mesh, which is determined by the choice of periodic boundary conditions on the finite-size real-space torus. $\hat{H}$ also satisfies an antiunitary symmetry $M_1\mathcal{T}$ which takes $(k_x,k_y)\rightarrow (k_x,-k_y)$.

\subsection{Density operator and GMP algebra}\label{app:subsec:gmp}

In this subsection, we consider the density-density commutator of the 2D Berry Trashcan model, and compare it to the Girvin-MacDonald-Platzman (GMP) algebra~\cite{girvin1986magnetoroton} of the Lowest Landau Level (LLL). The density operator for the Berry Trashcan model in the absence of a cutoff is given as
\begin{gather}
    \rho_{\bm{q}}=\sum_{\bm{k}}\mathcal{M}_{\bm{k},\bm{q}}\gamma^\dagger_{\bm{k}+\bm{q}}\gamma_{\bm{k}},
\end{gather}
with the form factor
\begin{align}
    \mathcal{M}_{\bm{k},\bm{q}}=e^{-\frac{\alpha'}{2}q^2}e^{-i\beta\bm{q}\times \bm{k}},
\end{align}
where for generality, we have allowed for independent $\alpha'$ and $\beta$.
Note that $\rho_{\bm{q}}$ in first quantization is just the projection of $e^{i\bm{q}\cdot\hat{\bm{r}}}$ into the continuum band of the Berry Trashcan.

The commutator of the density operator can be evaluated as
\begin{align}
    [\rho_{\bm{q}},\rho_{\bm{q}'}]&=\sum_{\bm{k}}(\mathcal{M}_{\bm{k}+\bm{q}',\bm{q}}\mathcal{M}_{\bm{k},\bm{q}'}-\mathcal{M}_{\bm{k},\bm{q}}\mathcal{M}_{\bm{k}+\bm{q},\bm{q}'})\gamma^\dagger_{\bm{k}+\bm{q}+\bm{q}'}\gamma_{\bm{k}}\nonumber\\
    &=e^{-\frac{\alpha'}{2}q^2}e^{-\frac{\alpha'}{2}q'^2}\sum_{\bm{k}}e^{-i\beta(\bm{q}+\bm{q}')\times \bm{k}}\times (-2i\sin\left(\beta\bm{q}\times\bm{q}'\right) )\gamma^\dagger_{\bm{k}+\bm{q}+\bm{q}'}\gamma_{\bm{k}}\nonumber\\
    &=\left(e^{\alpha'(\bm{q}\cdot\bm{q'})-i\beta(\bm{q}\times\bm{q}')}-e^{\alpha'(\bm{q}\cdot\bm{q'})+i\beta(\bm{q}\times\bm{q}')}\right)\rho_{\bm{q}+\bm{q'}}.\label{app:eq:GMP_trashcan}
\end{align}

We now demonstrate that the above density algebra maps exactly onto the GMP algebra of the LLL. We follow Appendix A in Ref.~\cite{girvin1999quantumhalleffectnovel}. Consider the symmetric gauge so that the LLL wavefunctions are spanned by
\begin{equation}
    \phi_m(\bm{r})=\frac{1}{\sqrt{2\pi 2^m m!}}z_\tau ^m e^{-\frac{|z_\tau|^2}{4}},
\end{equation}
where $z_\tau=x+i\tau y$ and we have set $\ell=1$.  The parameter $\tau$ determines the sign of the effective magnetic field; for instance, $\tau=+$ yields wavefunctions analytic in $z=x+iy$, corresponding to a negative field $-B\hat{z}$.
The resulting GMP algebra for the projected density operators for $\tau=+$ is
\begin{equation}
 [\bar\rho_{\bm{q}},\bar\rho_{\bm{q}'}]=\left(e^{\frac{\ell^2}{2}q_{+}'q_{-}}-e^{\frac{\ell^2}{2}q_{+}q_{-}'}\right)\bar\rho_{\bm{q}+\bm{q}'},
\end{equation}
where $q_\pm=q_x\pm iq_y$. We establish a correspondence by comparing to Eq.~\ref{app:eq:GMP_trashcan}. When $\alpha'=-\beta=\frac{\ell^2}{2}$, the density algebra of the Berry Trashcan matches the GMP algebra in a negative magnetic field. Conversely, when for the case $\alpha'=\beta=\frac{\ell^2}{2}$, the case of primary interest in our work, the algebra matches that of a positive magnetic field. This is consistent with the Berry curvature of our model, which is proportional to $2\beta$ \cite{bernevig2025berrytrashcanmodelinteracting}, thus fixing the sign of the effective magnetic field experienced by the electrons in the Berry Trashcan. 

\subsection{Two-Body Spectrum}\label{sec:2d_berry_2e}


\subsubsection{$v=\infty$, $\bm{p}=0$}\label{app:sec:vf_inf_p_0}

Here, we consider the simplest case of two electrons with zero total momentum $\bm{p}=0$ and an infinite trashcan wall dispersion $v=\infty$. The latter means that the allowed single-particle momenta lie on a disk $|\bm{k}|\leq k_b$.

The most general two-electron wavefunction in this symmetry sector can be written as
\begin{equation}
    \ket{\Psi}=\frac{1}{\Omega_{tot}}\sum^{\{\bm{k}\}}_{\bm{k}}f_{\bm{k}}\gamma^\dagger_{\bm{k}}\gamma^\dagger_{-\bm{k}}\ket{\text{vac}}=\frac{1}{\Omega_{tot}}\sum^{\{\bm{k}\}}_{\bm{k}}f_{\bm{k}}\ket{\bm{k}},\label{app:eq:definiton_of_psi}
\end{equation}
where we can impose $f_{\bm{k}}=-f_{-\bm{k}}$ due to fermionic statistics, and we have defined $\ket{\bm{k}}\equiv \gamma^\dagger_{\bm{k}}\gamma^\dagger_{-\bm{k}}\ket{\text{vac}}=-\ket{-\bm{k}}$. The action of the interaction Hamiltonian is
\begin{align}
    \hat{H}^\text{int}\ket{\Psi}&=\frac{1}{2\Omega_{tot}}\sum^{\{\bm{k},\bm{k}+\bm{q}\}}_{\bm{k},\bm{q}}\left[f_{\bm{k}}U_{\bm{q}}e^{-2i\beta\bm{q}\times \bm{k}}-f_{-\bm{k}}U_{\bm{q}}e^{-2i\beta\bm{q}\times \bm{k}}\right]\gamma^\dagger_{\bm{k}+\bm{q}}\gamma^\dagger_{-\bm{k}-\bm{q}}\ket{\text{vac}}\nonumber\\
    &=\frac{1}{\Omega_{tot}}\sum^{\{\bm{k},\bm{k}+\bm{q}\}}_{\bm{k},\bm{q}}f_{\bm{k}}U_{\bm{q}}e^{-2i\beta\bm{q}\times \bm{k}}\gamma^\dagger_{\bm{k}+\bm{q}}\gamma^\dagger_{-\bm{k}-\bm{q}}\ket{\text{vac}}\nonumber\\
    &=\frac{1}{\Omega_{tot}}\sum^{\{\bm{k},\bm{k}'\}}_{\bm{k},\bm{k}'}f_{\bm{k}}U_{\bm{k}'-\bm{k}}e^{-2i\beta\bm{k}'\times \bm{k}}\gamma^\dagger_{\bm{k}'}\gamma^\dagger_{-\bm{k}'}\ket{\text{vac}}=\frac{1}{\Omega_{tot}}\sum^{\{\bm{k},\bm{k}'\}}_{\bm{k},\bm{k}'}f_{\bm{k}}U_{\bm{k}'-\bm{k}}e^{-2i\beta\bm{k}'\times \bm{k}}\ket{\bm{k}'}.
\end{align}
We remind the reader that the function $U_{\bm{q}}=U_q$ only depends on the modulus of $\bm{q}$.

In this subsection, we consider the infinite size limit $\Omega_{tot}\rightarrow \infty$. We also refer to this as the `continuum limit', though we emphasize that even for finite $\Omega_{tot}$, the model is defined on the real-space continuum. We can therefore replace summations with integrals
\begin{gather}
    \ket{\Psi}=\int_{|\bm{k}|\leq k_b} \frac{d^2\bm{k} }{(2\pi)^2}f_{\bm{k}}\ket{\bm{k}},\\
    \hat{H}^\text{int}\ket{\Psi}=\int_{|\bm{k}|\leq k_b} \frac{d^2\bm{k} }{(2\pi)^2}\int_{|\bm{k}'|\leq k_b} \frac{d^2\bm{k}' }{(2\pi)^2}f_{\bm{k}}U_{\bm{k}'-\bm{k}}e^{-2i\beta\bm{k}'\times \bm{k}}\ket{\bm{k}'}.\label{app:eq:h_int_uk'k}
\end{gather}
Within this continuum limit, the system has full $SO(2)$ rotational symmetry. Thus, we can decompose the Hamiltonian and the wave functions into angular momentum channels labeled by angular momentum $m$. The basis states with definite relative angular momentum $m$ can be defined as
\begin{gather}
    \ket{k,m}\equiv \sqrt{k}\int \frac{d\varphi_{\bm{k}}}{2\pi}e^{im\varphi_{\bm{k}}}\ket{\bm{k}}
\end{gather}
where $\bm{k}=k(\cos\varphi_{\bm{k}},\sin\varphi_{\bm{k}})$. Only states with $m$ odd are non-vanishing, since due to fermionic statistics we have $\ket{k,m} = \frac{1}{2}( 1- e^{i m \pi}) \ket{k,m}  $. 
Because the total linear momentum is zero, the total angular momentum 
\begin{align}
\hat{\mathbf{L}}_{tot}=\hat{\br}_1\times\hat{\bk}_1+\hat{\br}_2\times\hat{\bk}_2=(\hat{\br}_1-\hat{\br}_2)\times\hat{\bk}_1=\hat{\br}\times\hat{\bk}=\hat{\mathbf{L}}_{rel}
\end{align}
equals to the relative angular momentum. Thus, from now on, we will not distinguish between relative and total angular momentum, and will refer to both simply as angular momentum for convenience.
We also justify the normalization of the states $\ket{k, m}$ introduced above. The plane-wave basis states obey the normalization $\bra{\bm{k}}\bm{k}'\rangle=(2\pi)^2\delta^{(2)}(\bm{k}-\bm{k}')$ in the continuum limit. In polar coordinates, this becomes 
\begin{align}
\bra{k,\varphi}k',\varphi'\rangle=(2\pi)^2\frac{1}{k}\delta(k-k')\delta(\varphi-\varphi').    
\end{align}
 Accordingly, the normalization of the angular momentum eigenstates $\ket{k, m}$ is given by
 \begin{align}
 \bra{k,m}k',m'\rangle=\sqrt{kk'}\int \frac{d\varphi_{\bm{k}}}{2\pi}e^{-im\varphi_{\bm{k}}}\int \frac{d\varphi_{\bm{k}'}}{2\pi}e^{im'\varphi_{\bm{k}}}\bra{\bm{k}}\bm{k}'\rangle=2\pi\delta(k-k')\delta_{m,m'}.    
 \end{align} 
Therefore, within each angular momentum sector labeled by $m$, the problem reduces to an effective one-dimensional continuum system with the standard 1D plane-wave normalization.

We now decompose the interaction onto the angular momentum basis as
\begin{align}
        U_{\bm{k}'-\bm{k}}&=\sum_{n=-\infty}^{\infty}u_n(k',k)e^{in(\varphi_{\bm{k}'}-\varphi_{\bm{k}})}\\
    u_n(k',k)&=\int_0^{2\pi}\frac{d\theta}{2\pi}e^{-in\theta}U_{\sqrt{k^2+k'^2-2kk'\cos\theta}}=u_{-n}^*(k',k)=u_{-n}(k',k)=u_n(k,k').
\end{align}

For the remainder of this section, we specialize to
\begin{equation}\label{appeq:2d_Uq_exp}
    U_{\bm{q}}=Ue^{-\alpha q^2},
\end{equation}
with $U$ being negative which corresponds to a purely attractive interaction if $\alpha\geq|\beta|$.
With this choice, the kernel $u_m$ is symmetric with
\begin{equation}
    u_m(k',k)=\int_0^{2\pi}\frac{d\theta}{2\pi}e^{-im\theta}U_{\sqrt{k^2+k'^2-2kk'\cos\theta}}=Ue^{-\alpha(k^2+k'^2)}I_m(2\alpha k k'),\label{eq:interaction_only_alpha}
\end{equation}
where $I_n(x)$ is a modified Bessel function of the first kind. Then, the integral equation of the Hamiltonian is reduced to 
\begin{align}
    \hat{H}^{\text{int}}\ket{k,m}
    &=\frac{\sqrt{k}}{\Omega_{tot}}\sum^{\{\bm{k}'\}}_{\bm{k}'}\int \frac{d\varphi_{\bm{k}}}{2\pi}e^{im\varphi_{\bm{k}}}U_{\bm{k}'-\bm{k}}e^{-2i\beta \bm{k}'\times \bm{k}}\ket{\bm{k}'}\nonumber\\
    &={\sqrt{k}}\int_0^{k_b} \frac{dk'}{2\pi}k'\int\frac{d\varphi_{\bm{k}'}}{2\pi}\int \frac{d\varphi_{\bm{k}}}{2\pi}e^{im\varphi_{\bm{k}}}U_{\bm{k}'-\bm{k}}e^{-2i\beta \bm{k}'\times \bm{k}}\ket{\bm{k}'}\nonumber\\
    &={\sqrt{k}}\int_0^{k_b} \frac{dk'}{2\pi}k'\int \frac{d\varphi_{\bm{k}'}}{2\pi}\int \frac{d\varphi_{\bm{k}}}{2\pi}\sum_{n=-\infty}^{\infty}u_n(k',k)e^{in(\varphi_{\bm{k}'}-\varphi_{\bm{k}})+im\varphi_{\bm{k}}-2i\beta \bm{k}'\times \bm{k}}\ket{\bm{k}'}\label{eq:hamiltonian_angular_momentum}.
\end{align}
We can also decompose the phase factor $G(\bm{k}',\bm{k})=e^{-2i\beta \bm{k}'\times \bm{k}}=e^{2i\beta kk'\sin(\varphi_{\bm{k}'}-\varphi_{\bm{k}})}$ into angular momentum components
\begin{align}
    G(\bm{k}',\bm{k})&=\sum_{n=-\infty}^{\infty}g_n(k',k)e^{in(\varphi_{\bm{k}'}-\varphi_{\bm{k}})}\label{eq:phase_angular_com},
\end{align}
where $g_n(k',k)$ can be evaluated 
\begin{align}
    g_n(k',k)
    &=\int_0^{2\pi}\frac{d\theta}{2\pi}e^{-in\theta}e^{2i\beta kk'\sin \theta}\nonumber\\
    &=\int_0^\pi\frac{d\theta}{\pi}\cos(n\theta-2\beta kk'\sin\theta)\nonumber\\
    &=J_n(2\beta kk').
\end{align}
$J_n(x)$ is the Bessel function of the first kind. 
Substituting Eq.~\ref{eq:phase_angular_com} into Eq.~\ref{eq:hamiltonian_angular_momentum}, we find
\begin{align}
    \hat{H}^\text{int}\ket{k,m}&={\sqrt{k}}\int_0^{k_b} \frac{dk'}{2\pi}k'\int \frac{d\varphi_{\bm{k}'}}{2\pi}\int \frac{d\varphi_{\bm{k}}}{2\pi}\sum_{n=-\infty}^{\infty}\sum_{n'=-\infty}^{\infty}u_n(k',k)g_{n'}(k',k)e^{i(n+n')(\varphi_{\bm{k}'}-\varphi_{\bm{k}})+im\varphi_{\bm{k}}}\ket{\bm{k}'}\nonumber\\
    &={\sqrt{k}}\int_0^{k_b} \frac{dk'}{2\pi}k'\int \frac{d\varphi_{\bm{k}'}}{2\pi}\sum_{n=-\infty}^{\infty}u_{n}(k',k)g_{m-n}(k',k)e^{im\varphi_{\bm{k}'}}\ket{\bm{k}'}\nonumber\\
    &={\sqrt{k}}\int_0^{k_b} \frac{dk'}{2\pi}\sqrt{k'}\sum_{n=-\infty}^{\infty}u_{n}(k',k)g_{m-n}(k',k)\ket{k',m}\nonumber\\
    &=\int_0^{k_b} \frac{dk'}{2\pi}{\sqrt{kk'}}\sum_{n=-\infty}^{\infty}Ue^{-\alpha(k^2+k'^2)}I_n(2\alpha k k')J_{m-n}(2\beta kk')\ket{k',m}.\label{eq:angularmomentumgmp}
\end{align}
Note that $J_n(x)=(-1)^nJ_{-n}(x)$, which introduces an asymmetry in the energies between angular momenta $m$ and $-m$. This is a consequence of explicit time-reversal symmetry breaking induced by the finite Berry curvature.

Using the identities
\begin{gather}
I_n(z)=\sum_{k=0}^{\infty}\frac{z^k}{k!}J_{n+k}(z), \\
\sum_{\nu=-\infty}^{\infty}J_{\nu}(x)J_{n-\nu}(y)=J_{n}(x+y),
\end{gather}
Eq.~\ref{eq:angularmomentumgmp} can be reduced to
\begin{align}
    \hat{H}^\text{int}\ket{k,m} &= U\int_0^{k_b} \frac{dk'}{2\pi}{\sqrt{kk'}} e^{-\alpha(k^2+k'^2)} \sum_{n=-\infty}^{\infty} \sum_{j=0}^{\infty} \frac{(2\alpha k k')^j}{j!} J_{n+j}(2\alpha k k') J_{m-n}(2\beta kk') \ket{k',m} \nonumber \\
    &= U\int_0^{k_b} \frac{dk'}{2\pi}{\sqrt{kk'}} e^{-\alpha(k^2+k'^2)} \sum_{n=-\infty}^{\infty} \sum_{j=0}^{\infty} \frac{(2\alpha k k')^j}{j!} J_{n}(2\alpha k k') J_{m-n+j}(2\beta kk') \ket{k',m} \nonumber \\
    &= U\int_0^{k_b} \frac{dk'}{2\pi}{\sqrt{kk'}} e^{-\alpha(k^2+k'^2)} \sum_{j=0}^{\infty} \frac{(2\alpha k k')^j}{j!} J_{m+j}(2(\alpha + \beta) k k') \ket{k',m}. \label{eq:h_int_general_exp}
\end{align}



From Eq.~\ref{eq:h_int_general_exp}, we can understand the limit where $\alpha=-\beta>0$, where we find that the interaction Hamiltonian vanishes for all channels with angular momentum $m>0$. To see this, we note that $J_{m+j}$ in Eq.~\ref{eq:h_int_general_exp} is only nonzero when $m+j=0$ (because $\alpha+\beta=0$, the argument of the Bessel function vanishes), so the Hamiltonian vanishes for any positive $m$. On the other hand, if $m<0$,
\begin{align}
    \hat{H}^{\text{int}}\ket{k,m}=U\int_0^{k_b} \frac{dk'}{2\pi}\sqrt{kk'}e^{-\alpha(k^2+k'^2)}\frac{(2\alpha kk')^{-m}}{(-m)!}\ket{k',m},\label{eq:interacting_ham_a_-b}
\end{align}
which is separable, i.e.~it is a product of a factor that depends solely on $k$, and another factor that depends solely on $k'$. Thus, the interaction matrix is a rank-1 matrix for each angular momentum. This means that there is one finite energy ground state for each odd $m<0$ with energy
\begin{align}
    E_{2,m}&=\frac{U(2\alpha)^{-m}}{2\pi(-m)!}\int_0^{k_b} e^{-2\alpha k^2}k^{-2m+1}dk\nonumber\\
    &=U\frac{\Gamma(1 - m) - \Gamma(1 - m, 2\alpha k_b^2)}{8\pi\alpha(-m)!}\nonumber \\
    &\approx\begin{cases}
   \frac{U}{8\pi\alpha}=-\frac{Uk_b^2}{4\varphi_{\text{BZ}}} & \quad \text{if } \alpha k_b^2\to \infty, \\
    \\
    \frac{U(2\alpha)^{-m}k_b^{2-2m}}{4\pi(1-m)(-m)!}=-\frac{U(\varphi_{\text{BZ}}/\pi)^{-m}k_b^2}{4\pi(1-m)(-m)!} & \quad \text{if } \alpha k_b^2\to 0
\end{cases},
\end{align}
where we recall that we have an effective cutoff at $k_b$ since $v=\infty$. In the last equation we use the relation $2\alpha=2\beta=\frac{\varphi_{\text{BZ}}}{A_b}$ where $A_b$ is the momentum area of the trashcan bottom which equals to $\pi k_b^2$ in the continuum limit. Note that the interaction strength parameter $U$ takes the unit of $[\text{Energy}]\cdot[\text{length}]^{2}$ in our convention (see Eq.~\ref{appeq:2d_Uq_exp}).
The eigenfunction is
\begin{align}
    |\psi_m\rangle&=\int_0^{k_b}\frac{dk}{2\pi}k^{-m+\frac{1}{2}}e^{-\alpha k^2}|k,m\rangle\nonumber\\
    &=\int_0^{k_b}\frac{dk}{2\pi}k\int \frac{d\varphi_{\bm{k}}}{2\pi}k^{-m}e^{-\alpha k^2+im\varphi_{\bm{k}}}|\bm{k}\rangle\nonumber\\
    &=\int_{|\bm{k}|\leq k_b}\frac{d^2\bm{k}}{4\pi^2}k_-^{-m}e^{-\alpha \bm{k}^2}\ket{\bm{k}},
\end{align}
where $k_\pm=k_x\pm ik_y$. All other energies for angular momenta $m<0$ are zero.

To obtain the solution for more general values of $\alpha$ and $\beta$, we go back to Eq.~\ref{eq:h_int_general_exp} and expand the Bessel function \( J_{m + j} \) as a series in powers of \( (\alpha + \beta) \)
\begin{align}
    \hat{H}^\text{int}\ket{k,m} &= U\int_0^{k_b} \frac{dk'}{2\pi}{\sqrt{kk'}} e^{-\alpha(k^2+k'^2)} \sum_{j=0}^{\infty} \sum_{\nu=0}^{\infty} \frac{(-1)^{\nu}}{j! \nu! (\nu+m+j)!} (2\alpha kk')^j ((\alpha + \beta) kk')^{2\nu + m + j} \ket{k',m} \nonumber \\
    &= U\int_0^{k_b} \frac{dk'}{2\pi}{\sqrt{kk'}} e^{-\alpha(k^2+k'^2)} \sum_{j=0}^{\infty} \sum_{\nu=0}^{\infty} \frac{(-1)^{\nu}}{j! \nu! (\nu+m+j)!} (2\alpha kk')^{2\nu + 2j + m} \left( \frac{\alpha + \beta}{2\alpha} \right)^{2\nu + m + j} \ket{k',m} \nonumber \\
    &= U\int_0^{k_b} \frac{dk'}{2\pi}{\sqrt{kk'}} e^{-\alpha(k^2+k'^2)} \sum_{n=0}^{\infty} \sum_{\nu=0}^{n} \frac{(-1)^{\nu}}{(n-\nu)! \nu! (n+m)!} (2\alpha kk')^{2n+m} \left( \frac{\alpha + \beta}{2\alpha} \right)^{n+m+\nu} \ket{k',m} \nonumber \\
    &= U\int_0^{k_b} \frac{dk'}{2\pi}{\sqrt{kk'}} e^{-\alpha(k^2+k'^2)} \sum_{n=0}^{\infty} \frac{1}{n!(n+m)!} (\sqrt{\alpha^2 - \beta^2} kk')^{2n} ((\alpha + \beta) kk')^m \ket{k',m} \label{eq:interaction_alpha_beta_exp} \\
    &= U\left( \frac{\alpha + \beta}{\sqrt{\alpha^2 - \beta^2}} \right)^m \int_0^{k_b}  \frac{dk'}{2\pi} \sqrt{kk'} e^{-\alpha(k^2+k'^2)} I_m(2 \sqrt{\alpha^2 - \beta^2} kk') \ket{k',m}.\label{eq:interaction_alpha_beta}
\end{align}
Note that the transition to the fourth line is achieved by applying the binomial theorem  to the inner sum over $\nu$: $
\frac{1}{n!} \sum_{\nu=0}^{n} \binom{n}{\nu} (x)^\nu = \frac{1}{n!} (1+x)^n$, where  $x = - \frac{\alpha + \beta}{2\alpha}$.  In the final expression, the apparent singularity at $\alpha=\pm\beta$ is regularized by the small-argument limit of the modified Bessel function $I_m(z)\propto z^m$. Since its argument $z\propto\sqrt{\alpha^2-\beta^2}$, the diverging prefactor is canceled, ensuring the result remains finite.

We first consider $\alpha=\beta>0$. The Hamiltonian (see Eq.~\ref{appeq:Hint_2d_alpha_beta}) in this case is related to $\alpha=-\beta>0$ by time-reversal, so the corresponding two-electron solutions can be straightforwardly inferred. We can also explicitly examine Eq.~\ref{eq:interaction_alpha_beta_exp} directly, which for $\alpha=\beta>0$ reduces to 
\begin{align}
    \hat{H}^{\text{int}}\ket{k,m}=U\int_0^{k_b}\frac{dk'}{2\pi}{\sqrt{kk'}}e^{-\alpha(k^2+k'^2)}\frac{(2\alpha kk')^{m}}{m!}\ket{k',m}.\label{eq:PD_hamiltonian}
\end{align}
This is the same as Eq.~\ref{eq:interacting_ham_a_-b} with $-m\to m$. 

In summary, we conclude that when $\alpha=|\beta|$, only the odd angular momenta $m$ that satisfy $m\beta>0$ have a gapped finite-energy ground state with energy
\begin{align}
    E_m=U\frac{\Gamma(1 + |m|) - \Gamma(1 + |m|, 2\alpha k_b^2)}{8\pi\alpha(|m|)!}\approx\begin{cases}
   \frac{U}{8\pi\alpha}=\frac{Uk_b^2}{4|\varphi_{BZ}|} & \quad \text{if } \alpha k_b^2\to \infty \\
    \\
    \frac{U(2\alpha)^{|m|}k_b^{2+2|m|}}{4\pi(1+|m|)(|m|)!}=\frac{U(|\varphi_{\text{BZ}}|/\pi)^{|m|}k_b^2}{4\pi(1+|m|)(|m|)!} & \quad \text{if } \alpha k_b^2\to 0
\end{cases}\label{app:eq:E_m_2limit},
\end{align}
and normalized wavefunction
\begin{align}
        |\psi_m\rangle&=\int_0^{k_b}\frac{dk}{2\pi Z}k^{|m|+\frac{1}{2}}e^{-\alpha k^2}|k,m\rangle\nonumber\\
    &=\int_0^{k_b}\frac{dk}{2\pi Z}k\int \frac{d\varphi_{\bm{k}}}{2\pi}k^{|m|}e^{-\alpha k^2+im\varphi_{\bm{k}}}|\bm{k}\rangle\nonumber\\
    &=\begin{cases}
    \displaystyle \int_{|\bm{k}|\leq k_b}\frac{d^2\bm{k}}{(2\pi)^2 Z} k_+^m e^{-\alpha \bm{k}^2}\ket{\bm{k}} & \quad \text{if } m > 0, \\
    \\
    \displaystyle \int_{|\bm{k}|\leq k_b}\frac{d^2\bm{k}}{(2\pi)^2 Z} k_-^{-m} e^{-\alpha \bm{k}^2}\ket{\bm{k}} & \quad \text{if } m < 0.
\end{cases}.\label{eq:GMP_LM_WF}
\end{align}
All other eigenvalues (including those corresponding to angular momentum satisfying $m\beta<0$) are zero. Here, we introduce a normalization factor $Z$ which is
\begin{align}
Z^2&=\frac{\left(\Gamma(1 + |m|) - \Gamma(1 + |m|, 2\alpha k_b^2)\right)}{4\pi (2\alpha)^{|m|+1}}\approx\frac{ k_b^{2+2|m|}}{4\pi(1+|m|)} \quad \quad \text{for } \alpha k_b^2\to0.
\end{align}
A key physical insight is that the chirality of the electron pairs in the ground state is directly governed by the sign of $\beta$. As discussed in App.~\ref{app:subsec:gmp}, the Berry curvature and the effective magnetic field of the GMP algebra take the same sign of $\beta$. Taken together, these results imply a ``ferromagnetic'' coupling between the GS chirality and the underlying Berry curvature of the Berry Trashcan.

To analyze the (approximate) solutions of the Hamiltonian for general $\alpha,\beta$,
we begin by expanding the modified Bessel function within Eq.~\ref{eq:interaction_alpha_beta} in powers of $\alpha$ and $\beta$ 
\begin{align}
    \hat{H}^{\text{int}}\ket{k,m}&=U\int_0^{k_b}\frac{dk'}{2\pi}\sum_{j=0}^{\infty}e^{-\alpha (k^2+k'^2)}\frac{(k k')^{2j+|m|+\frac{1}{2}}(\alpha^2-\beta^2)^{j+\frac{|m|-m}{2}}(\alpha+\beta)^m}{j!(j+|m|)!}\ket{k',m}\nonumber\\
    &=U\sum_{j=0}^{\infty}(\alpha^2-\beta^2)^{j+\frac{|m|-m}{2}}(\alpha+\beta)^{m}\int_0^{k_b} dk'f_{m,j}(k)f_{m,j}(k')\ket{k',m},\label{appeq:1d_H_fmjfmj}
\end{align}
with $f_{m,j}$ defined as
\begin{align}
f_{m,j}(k)=e^{-\alpha k^2}\frac{k^{2j+|m|+\frac{1}{2}}}{\sqrt{2\pi j!(j+|m|)!}}=e^{-(1+ x)c k^2}\frac{k^{2j+|m|+\frac{1}{2}}}{\sqrt{2\pi j!(j+|m|)!}}\label{eq:f_mj(k)}.
\end{align}
In the final expression above, we adopt the parameterization 
\begin{equation}\label{eqapp:alphabeta_xc_param}
\alpha\equiv(1+x)c,\quad \beta\equiv(1- x)c,
\end{equation} 
with $|x|\leq1$ and $c>0$. 
For the subsequent analysis, we focus on $\beta>0$ (results for $\beta<0$ follow from acting with the time-reversal operator).

Next, we rewrite the Hamiltonian in a more useful basis motivated by the way $f_{m,j}(k)$ enters Eq.~\ref{appeq:1d_H_fmjfmj}. After inserting the parameterization of $\alpha,\beta$ in terms of $x,c$ in Eq.~\ref{appeq:1d_H_fmjfmj}, we express the Hamiltonian as a sum over projection operators $|\psi_{m,j}\rangle\langle\psi_{m,j}|$
\begin{align}
        \hat{H}^{\text{int}}\ket{k,m}&=U\sum_{j=0}^{\infty}x^{j+\frac{|m|-m}{2}}(2c)^{2j+|m|}\int_0^{k_b} dk'f_{m,j}(k)f_{m,j}(k')\ket{k',m}\nonumber\\
        &=U\sum_{j=0}^{\infty}x^{j+\frac{|m|-m}{2}}(2c)^{2j+|m|}|\psi_{m,j}\rangle\langle\psi_{m,j}\ket{k,m}
\end{align}
where the basis states $\ket{\psi_{m,j}}$ are defined by the wavefunctions $f_{m,j}(k)$
\begin{align}\label{appeq:2d_psimj}
    \ket{\psi_{m,j}}=\int_0^{k_b} d{k}f_{m,j}(k)\ket{k,m}.
\end{align}
A challenge arises because the basis states $\ket{\psi_{m,j}}$ are not orthogonal for different $j$. The overlaps between different basis states are quantified by the overlap matrix
\begin{align}
    \bra{\psi_{m,i}}\psi_{m,j}\rangle=S_{i,j}^{m}=\frac{\gamma(i+j+|m|+1,2(1+x)c k_b^2)}{2(2(1+x)c)^{i+j+|m|+1}\sqrt{i!j!(i+|m|)!(j+|m|)!}}\ne\delta_{ij},\label{appeq:2d_Smij}
\end{align}
where $\gamma(s,x)$ is the lower incomplete Gamma function. If $c k_b^2\ll1$, then
\begin{align}
    S_{i,j}^m\approx&\frac{(2(1+x)c k_b^2)^{i+j+|m|+1}}{2(2(1+x)c)^{i+j+|m|+1}\sqrt{i!j!(i+|m|)!(j+|m|)!}},\nonumber\\
    =&\frac{k_b^{2(i+j+|m|+1)}}{2\sqrt{i!j!(i+|m|)!(j+|m|)!}},\label{eq:app:s_ij_small_c}
\end{align}
which is a constant and independent of $c$ and $x$. 

To diagonalize the Hamiltonian, we first construct an orthonormal basis. We achieve this by applying the Gram-Schmidt orthogonalization procedure to the set $\{\ket{\psi_{m,j}}\}$, generating a new orthonormal basis $\{\ket{e_{m,j}}\}$
\begin{align}
    \ket{e_{m,j}'}&=\ket{\psi_{m,j}}-\sum_{i=0}^{j-1}\bra{e_{m,i}}\psi_{m,j}\rangle\ket{e_{m,i}}\label{eq:gs_ort_1}\\
    \ket{e_{m,j}}&=\frac{\ket{e_{m,j}'}}{\mathcal{N}_j^m},\text{ where } \mathcal{N}_j^m=\bra{e_{m,j}'}e_{m,j}'\rangle^{1/2}=\sqrt{S_{jj}-\sum_{i=0}^{j-1}|\bra{e_{m,i}}\psi_{m,j}\rangle} |^2.\label{eq:gs_ort_2}
\end{align}
In the orthonormal basis $\{\ket{e_{m,j}}\}$, we can express the Hamiltonian matrix elements as
\begin{align}
    H_{ij}^{m}=\bra{e_{m,i}}\hat{H}^\text{int}\ket{e_{m,j}}=U\sum_{n=0}^{\infty}x^{n+\frac{|m|-m}{2}}(2c)^{2n+|m|}\bra{e_{m,i}}\psi_{m,n}\rangle\bra{\psi_{m,n}}e_{m,j}\rangle.\label{eq:H_int_ortho}
\end{align}
To simplify this expression, we study the properties of $\bra{e_{m,i}}\psi_{m,j}\rangle$. By construction, $\ket{e_{m,i}}$ is orthogonal to the subspace spanned by $\{\ket{\psi_{m,0}},\cdots,\ket{\psi_{m,i-1}}\}$. This implies that the transformation matrix between the two bases is upper triangular
\begin{align}
    \bra{e_{m,i}}\psi_{m,j}\rangle=0\quad\text{when } i>j.
\end{align}
We can therefore write the expansion of $\ket{\psi_{m,j}}$ as
\begin{align}
\ket{\psi_{m,j}} = \sum_{i=0}^{j} \bra{e_{m,i}}\psi_{m,j}\rangle \ket{e_{m,i}} = \sum_{i=0}^{j} \beta_{ij}^{m} \ket{e_{m,i}},\label{appeq:2d_beta}
\end{align}
where we have defined the real transformation coefficients $\beta_{ij}^m=\langle e_{m,i} | \psi_{m,j} \rangle$ and the diagonal elements $\beta_{jj}^m\equiv\mathcal{N}_{j}^m$ are the normalization factors from Eq.~\ref{eq:gs_ort_2}. 

In the following, we will focus on the small $ck_b^2$ limit with $ck_b^2\ll1$, which is the relevant limit for the RnG system \cite{bernevig2025berrytrashcanmodelinteracting}.  Furthermore, we consider a finite $x$ to analyze the system's spectrum away from the exactly solvable point $\alpha=\beta$. In this limit, the argument of the Gaussian factor in the basis functions $f_{m,j}(k)$ (Eq.~\ref{eq:f_mj(k)}), $\alpha k^2=(1+x)ck^2$, remains small across the entire trashcan bottom $k\leq k_b$. Consequently, the exponential term can be approximated as unity, $e^{-\alpha k^2}\approx 1$. Within this approximation, the basis functions $f_{m,j}(k)$ (Eq.~\ref{eq:f_mj(k)}), and consequently the states $\ket{\psi_{m,j}}$ (Eq.~\ref{appeq:2d_psimj}), the orthonormal basis $\ket{e_{m,j}}$ (Eq.~\ref{eq:gs_ort_2}), and the transformation coefficients $\beta_{ij}$ (Eq.~\ref{appeq:2d_beta}), become independent of $c$ and $x$. In addition, we also have already shown that the overlap matrix $S_{ij}^m$ is independent of $x$ and $c$ and scales with $k_b$ as $S_{ij}^m\propto k_{b}^{2(i+j+|m|+1)}$ (see Eq.~\ref{eq:app:s_ij_small_c}). The scaling of the overlap matrix, in turn, determines the scaling of the transformation coefficients $\beta_{ij}^m$ according to the Gram-Schmidt orthogonalization procedure. As can be easily proven via mathematical induction, the coefficients $\beta_{ij}^m$ follow a scaling behavior
\begin{align}
    \beta_{ij}^{m}\propto k_b^{2j+|m|+1}.
\end{align}
To make this scaling explicit in our analysis, we can therefore write $\beta_{ij}^{m}= \overline{\beta}_{ij}^{m}k_b^{2j+|m|+1}$, where $\overline{\beta}_{ij}^{m}$ is a positive dimensionless coefficient that is independent of $k_b$, $x$, and $c$.

Substituting this expansion back into Eq.~\ref{eq:H_int_ortho}, we arrive at the final form of the Hamiltonian matrix elements
\begin{align}
    H_{ij}^m=Uk_b^2\sum_{n=\text{max}(i,j)}^{\infty}x^{n+\frac{|m|-m}{2}}(2ck_b^2)^{2n+|m|}\overline{\beta}_{in}^m\overline{\beta}_{jn}^m.\label{eq:int_h_orthogonal}
\end{align}
 The lower limit of the sum, $n=\text{max}(i,j)$, arises because $\overline{\beta}_{in}^m$ is zero if $i>n$ and $\overline{\beta}_{jn}^m$ is zero if $j>n$. The Hamiltonian matrix in Eq.~\ref{eq:int_h_orthogonal} possesses several key properties. First, since the basis functions $f_{m,j}(k)$ are real, the overlap coefficients $\overline{\beta}_{jn}^m$ are also real, ensuring the Hamiltonian is Hermitian. Second, the leading-order scaling with $ck_b^2$ of $H_{ij}^m$ is $H_{ij}^m\sim (ck_b^2)^{2\max{(i,j)}+|m|}$. Third, since $ck_b^2\ll1$, the sign of $H_{ij}^m$ is determined by the sign of the leading term which is $\text{sign}(x^{\text{max}(i,j)+\frac{|m|-m}{2}})$.


In this small $ck_b^2$ limit, we can exploit the hierarchical scaling of the Hamiltonian elements to perform a perturbative treatment in $ck_b^2$. We treat the diagonal elements $H_{ii}^m$ as the unperturbed energy levels and the off-diagonal elements $H_{ij}^m$ as the perturbation. To validate this approach, we examine the perturbed energy $E_i^p$ up to the second-order correction for the $i$-th level
\begin{align}
    E_{i}^p=H_{ii}^m+\sum_{j\ne i}\frac{H_{ij}^mH_{ji}^m}{H_{ii}^m-H_{jj}^m}.
\end{align}
Analyzing the scaling of each term in the sum reveals that it is always of higher order in $ck_b^2$ than the leading-order energy $H_{ii}^m\sim (ck_b^2)^{2i+|m|}$. Explicitly, for $j > i$, we have 
\begin{align}
H_{ii}^m \sim (ck_b^2)^{2i+|m|} \gg H_{jj}^m \simeq H_{ij}^m  \sim (ck_b^2)^{2j+|m|}
\end{align}
and the corresponding second-order energy correction, scaling as $\sim (ck_b^2)^{4j - 2i + |m|}$, constitutes a higher-order correction to $H_{ii}^m$.
Conversely, when $j < i$, we find
\begin{align}
    H_{jj}^m\sim(ck_b^2)^{2j+|m|} \gg H_{ii}^m \simeq H_{ij}^m\sim (ck_b^2)^{2i+|m|},
\end{align}
and the second-order correction scales as $\sim (ck_b^2)^{4i - 2j + |m|}$, which is again a high-order contribution to $H_{ii}^m$.
\begin{figure}[h]
    \centering
    \includegraphics[width=1.0\linewidth]{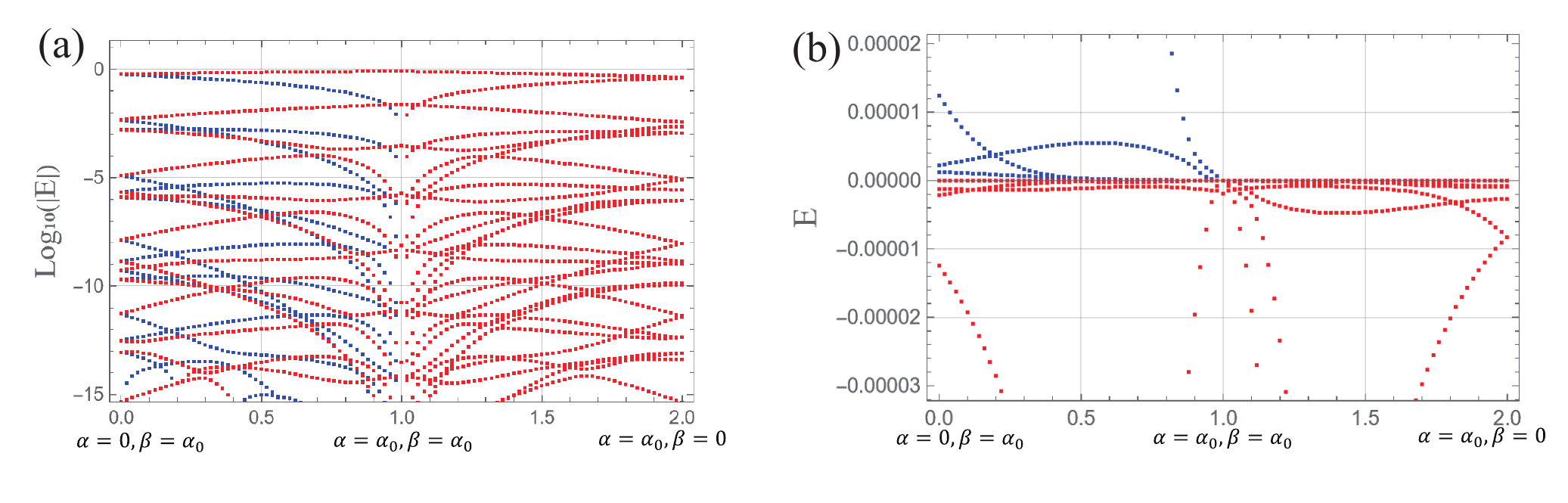}
    \caption{Two-electron energy spectrum of the attractive 2D Berry Trashcan model with $v=\infty,U=-2/A_b,N_{k_b}=61$ and total momentum $\bm{p}=0$ (we absorb a factor of $\Omega_{tot}$ in $U$). $\beta=\alpha_0$ corresponds to $\varphi_{\text{BZ}}=\pi/2$. (a) Plot showing $\log_{10}(|E|)$. The red (blue) dots represent the negative (positive) energies. On the horizontal axis, the middle corresponds to $\alpha=\beta=\alpha_0$. Moving towards the left from the middle, $\alpha$ is linearly decreased to 0.  Moving towards the right from the middle, $\beta$ is linearly decreased to 0. (b)  
    Same as (a) except that $E$ is plotted.}
    \label{fig:2d_spectrum}
\end{figure}
Since all second-order corrections are negligible at the leading order and decay exponentially as the index $j$ deviates from $i$, their sum is thus convergent and remains of higher order than the diagonal terms. The energy spectrum is therefore well-approximated by the diagonal elements of the Hamiltonian
\begin{align}
    E\approx\{E_{m,0}, E_{m,1}, \cdots\}
\end{align}
where
\begin{align}
    E_{m,i}=H_{ii}^m&=Uk_b^2\sum_{n=i}^{\infty}x^{n+\frac{|m|-m}{2}}(2ck_b^2)^{2n+|m|}(\overline{\beta}_{in}^m)^2.
\end{align}
This expression is dominated by its first term $(n=i)$, yielding the leading-order approximation for the energy levels
\begin{align}
     E_{m,i}\approx Uk_b^2x^{i+\frac{|m|-m}{2}}(2ck_b^2)^{2i+|m|}(\overline{\mathcal{N}}_{j}^m)^2,\label{eq:e_ii_approx}
\end{align}
where $\overline{\mathcal{N}_{j}^m}$ is defined in an analogous way to the $\overline{\beta_{ij}^m}$, i.e.~$\mathcal{N}_{j}^m=\overline{\mathcal{N}_{j}^m}k_b^{2j+|m|+1}$.


Our analytical model qualitatively reproduces the key features of the energy spectrum numerically obtained via ED (Fig.~\ref{fig:2d_spectrum}):
\begin{enumerate}
    \item \textbf{Energy Clustering:} The numerical spectrum organizes into distinct clusters near the two ends of the spectrum ($|x|\approx1$), whose energies are separated by orders of magnitude. The approximate energy of the $i$-th cluster ($i=1, 2, \dots$) is given by the leading-order scaling relation
    \begin{equation}
        E_i \sim (ck_b^2)^{2i-1},
        \label{eq:cluster_energy_scaling}
    \end{equation}
    which is captured by our analytics in Eq.~\ref{eq:e_ii_approx}. This power-law dependence on the cluster index $i$ is responsible for the large energy gaps observed on a logarithmic scale between clusters (Fig.~\ref{fig:2d_spectrum}(a)). Furthermore, our analysis correctly predicts the detailed structure of these energy clusters. The $i$-th cluster consists of $2i$ states, with angular momenta $m = \pm 1, \pm 3, \dots, \pm (2i-1)$. The sign for the energy of each of these states is also accurately captured by Eq.~\ref{eq:e_ii_approx}; for a state in the $i$-th cluster with angular momentum $m$, its sign is determined by $\text{sign}(x^{i-(m+1)/2})$.

    \item \textbf{$\alpha-\beta$ Symmetry:} The spectrum shown in Figs.~\ref{fig:2d_spectrum}(a) and (b) exhibits an approximate symmetry under the interchange of parameters, $\alpha \leftrightarrow \beta$. In particular, this transformation leaves  $|E|$ approximately invariant while flipping the signs of the positive energies in the region $\beta>\alpha$. This feature is a direct consequence of the analytical form of the spectrum derived in Eq.~\ref{eq:e_ii_approx}. Given the definitions (recall Eq.~\ref{eqapp:alphabeta_xc_param})
    \begin{align}
        \alpha=(1+x)c,\quad\beta=(1-x)c,
    \end{align}
     this parameter swap is equivalent to the transformation $x\to -x$. Indeed, our analytical solution Eq.~\ref{eq:e_ii_approx} predicts that this transformation leaves the magnitude of the energies $|E|$ invariant but change the signs of positive energies in the region $\beta>\alpha$ $(x<0)$ (note that $U<0$). This matches the numerical results in Fig.~\ref{fig:2d_spectrum}(b).
\end{enumerate}

Finally, we comment on the structure of the approximate eigenstates. In our leading-order analysis, the $i$-th eigenstate is just $\ket{e_{m,i}}$ (Eq.~\ref{eq:gs_ort_2}). The Gram-Schmidt procedure (Eqs.~\ref{eq:gs_ort_1} and \ref{eq:gs_ort_2}) constructs each orthonormal vector $\ket{e_{m,i}}$ as a specific linear combination of the original (non-orthogonal) basis vectors $\{\ket{\psi_{m,0}},\cdots,\ket{\psi_{m,i}}\}$. We can express this relationship as
\begin{equation}
\ket{e_{m,i}} = \sum_{j=0}^{i} c_{ij}^m \ket{\psi_{m,j}}=\sum_{j=0}^{i} c_{ij}^m\int_0^{k_b} d{k}f_{m,j}(k)\ket{k,m},
\label{eq:e_as_linear_combo}
\end{equation}
where the real coefficients $c_{ij}^m$ are uniquely determined by the orthogonalization and normalization procedures.
With $f_{m,i}$ given in Eq.~\ref{eq:f_mj(k)}, we can write down the general form of $\ket{e_{m,i}}$ as
\begin{align}
    \ket{e_{m,i}}=\sum_{j=0}^{i} c_{ij}^m\int_0^{k_b} d{k}\frac{e^{-\alpha k^2}k^{2j+|m|+\frac{1}{2}}}{\sqrt{2\pi j!(j+|m|)!}}\ket{k,m}.
\end{align}
For the ground state (denoted as $\ket{\psi_{m}}$) in a given angular momentum $m$ channel, the (unnormalized) eigenstate is simply $\ket{e_{m,0}}$
\begin{align}
    \ket{\psi_{m}}\approx\ket{e_{m,0}} &=\int_0^{k_b} dke^{-\alpha k^2}\frac{k^{|m|+\frac{1}{2}}}{\sqrt{2\pi j!(j+|m|)!}}\ket{k,m}\\
    &=\int_{|\bm{k}|\leq k_b} d^2\bm{k}e^{-\alpha k^2}\frac{k^{|m|}e^{im\varphi_{\bm{k}}}}{2\pi\sqrt{2\pi j!(j+|m|)!}}\ket{\bm{k}},\label{eq:gs_general_alpha_beta}
\end{align}
 which has the same functional form as the ground state in the limit $\alpha=\beta$ (see Eq.~\ref{eq:GMP_LM_WF}), even for the general case with $x\ne0$ (i.e.~$\alpha\ne\beta$).


\subsubsection{$v=\infty$, $\bm{p}\neq0$}\label{appsubsec:2d_vFinf_finitep}

\begin{figure}[t]
    \centering
    \includegraphics[width=0.4\linewidth]{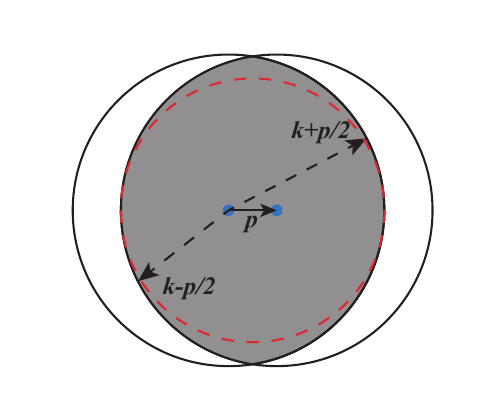}
    \caption{Grey shading denotes the allowed set of single-particle momenta for the two-electron problem for the 2D Berry Trashcan model with total momentum $\bm{p}$ and $v=\infty$ (see App.~\ref{appsubsec:2d_vFinf_finitep}). The black circles have radius $k_b$.}
    \label{fig:finite_mom_region}
\end{figure}

In this section, we study the finite-momentum two-electron spectrum of the 2D Berry Trashcan model. The primary goal is to determine the dispersion of the ground state two-electron branch as a function of total momentum $\bm{p}$. Similar to the 1D case in App.~\ref{secapp:1d_Hamiltonian_2e_spectrum}, we define a two-electron basis state with total momentum $\bm{p}$ as
\begin{align}
\ket{\bm{k},\bm{p}}=\frac{1}{\Omega_{tot}}\sum_{\bm{k}}^{\{\bm{k+\frac{\bm{p}}{2}},\bm{k-\frac{\bm{p}}{2}}\}}\gamma^\dagger_{\bm{k+\frac{\bm{p}}{2}}}\gamma^\dagger_{\bm{-k+\frac{\bm{p}}{2}}}|\text{vac}\rangle.
\end{align}
The interaction Hamiltonian is expressed in this basis as
\begin{gather}
    \hat{H}^{\text{int}}\ket{\Psi}=\int' \frac{d^2\bm{k} }{(2\pi)^2}\int' \frac{d^2\bm{k}' }{(2\pi)^2}f_{\bm{k}}U_{\bm{k}'-\bm{k}}e^{-2i\beta\bm{k}'\times \bm{k}}\ket{\bm{k}',\bm{p}},\label{eq:finite_ham}
\end{gather}
where
\begin{align}
        \ket{\Psi}=\int' \frac{d^2\bm{k} }{(2\pi)^2}f_{\bm{k}}\ket{\bm{k},\bm{p}}.
\end{align}
Eq.~\ref{eq:finite_ham} is identical to Eq.~\ref{app:eq:h_int_uk'k}, except that the primes on the integrals denote that the momenta are now restricted within the gray region in Fig.~\ref{fig:finite_mom_region}. Although the rotational $SO(2)$ symmetry is broken, we can recover an approximate symmetry in the small momentum limit $p=|\bm{p}|\to0$. In this limit, we approximate the true integration domain (the gray region) with a circular one (the red dashed circle in Fig.~\ref{fig:finite_mom_region}). This restores an approximate $SO(2)$ symmetry for the relative momentum $\bm{k}$, allowing us to decompose the interaction into distinct angular momentum channels. In this case, our previous analysis for the zero-momentum ($\bm{p}=0$) two-electron state from App.~\ref{app:sec:vf_inf_p_0} can be directly applied here, with the simple replacement of the momentum cutoff $k_b\to k_b-p/2$ and a shift of the total momentum from $\bm{0}\to\bm{p}$. To verify this approach, we numerically calculated the overlap between our ansatz and the true ground state obtained from ED. For a total momentum $\bm{p}=k_b/3 $ on a system with $N_{k_b}=37$, the fidelity is exceptionally high, with the overlap $|\bra{\Psi^{ED}}\Psi^A\rangle|$ deviating from unity by less than $10^{-4}$. This confirms that our approximate finite-momentum ansatz provides a remarkably accurate description of the ground state at $p\neq0$.

\begin{figure}[h]
    \centering
    \includegraphics[width=1.0\linewidth]{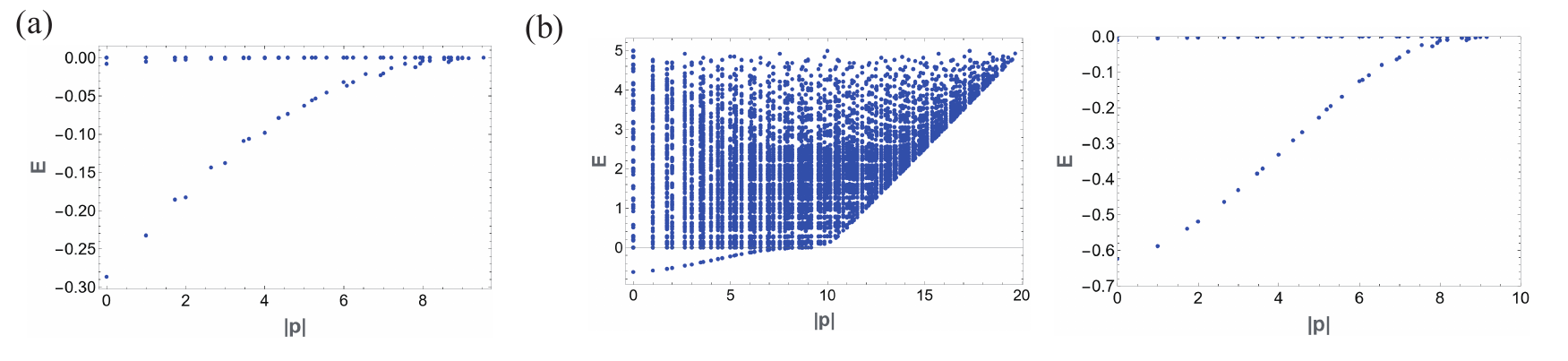}
    \caption{The two-electron energy spectrum of the attractive 2D Berry Trashcan model with $U=-2/A_b$, $\alpha=\beta, \varphi_{\text{BZ}}=\pi/2$ (We absorb a factor of $\Omega_{tot}$ in $U$).  (a) Spectrum for $v=\infty$ with $N_{k_b}=91$ points. (b) Spectrum for $v=5$ with $\Lambda=k_b$, calculated using a momentum mesh with $N_{k_b}=355$ total points. The right panel of (b) is a zoom in view of the left panel of (b).}
    \label{fig:2d_2e_spectrum}
\end{figure}

With this approximation, the energy of the two-electron ground state with total momentum $\bm{p}$ is given by
\begin{align}
    E_{2,\bm{p}}&=U\frac{1 - \Gamma(2, 2\alpha (k_b-p/2))}{8\pi\alpha}\\
    &\approx\frac{\alpha U(k_b-p/2)^{4}}{4\pi}\approx\frac{\alpha U(k_b^4-2pk_b^3)}{4\pi} \quad \quad \text{for } \alpha\to0.\label{app:eq:2d_dispersion_2e}
\end{align}
The expression reveals that the dispersion is linear for small $p$, analogous to the 1D case. This theoretical prediction is confirmed by our numerical results shown in  Fig.~\ref{fig:2d_2e_spectrum}(a). For the parameters $U=-2/A_b,\varphi_{\text{BZ}}=\pi/2,v=\infty$, the calculated ground state energy clearly exhibits linear dispersion near $p=0$. A more general discussion regarding the finite-momentum ground states is provided in App.~\ref{app:sec:rsga_2d_generalization}.

\subsubsection{Finite $v$, $\bm{p}=0$}\label{appsubsec:2d_vFfinite_p0}

In this subsection, we consider the two-electron problem for finite $v$ and $\bm{p}=0$.
In the angular momentum basis $|k,m\rangle$, the action of the total Hamiltonian is
\begin{align}
    \hat{H}\ket{k,m}=\theta({k-k_b})v(k-k_b)\ket{k,m}+U\int_{k'\leq k_b+\Lambda}\frac{dk'}{2\pi}{\sqrt{kk'}}e^{-\alpha(k^2+k'^2)}\sum_{j=0}^{\infty}\frac{(2\alpha k k')^j}{j!}J_{m+j}(2(\alpha+\beta) k k')\ket{k',m}.\label{app:eq:DPR1_Hamiltonian_v}
\end{align}
When $\alpha=|\beta|$, the interaction Hamiltonian $\hat{H}^\text{int}$ is rank-1, so that the total Hamiltonian for angular momentum $m$ (Eq.~\ref{app:eq:DPR1_Hamiltonian_v}) is a symmetric DPR1 matrix (like in the 1D case in App.~\ref{appsubsec:1d_2e_finitevF}) with the form 
\begin{align}
    H_m(k,k')=d(k)\delta_{k,k'}+Uu(k)u(k'),
\end{align}
where

\begin{align}
    u(k)=\frac{(\sqrt{2\alpha})^{|m|}}{\sqrt{2\pi(|m|!)}}k^{\frac{1}{2}+|m|}e^{-\alpha k^2}.
\end{align}
Its eigenvalues $\lambda$ can be solved by the secular equation (see App.~\ref{appsec:1d_2h})
\begin{align}
    1=\int_0^{k_b+\Lambda}dk\frac{Uu(k)^2}{\lambda-d(k)}\label{eq:secular_equation_0},
\end{align}
and for $U<0$ satisfy the following Weyl's inequality
\begin{align}
    & d(k_1)+U\int_0^{k_b+\Lambda}dku^2(k)\le\lambda_{{1}}\le d(k_1)\label{eq:weyl_largest_trashcan}\\
    &d(k_{i})\leq\lambda_{i+1}\leq d(k_{i+1}),\quad i \in[1, {i_\text{max}}-1].
\end{align}
Here for convenience, we discretize the momentum $k$ so it takes discrete values $k_i$, where $i=1,\ldots,{i_\text{max}}$. Assuming $k_1=0,k_{i_b}=k_b, k_{i_{\text{max}}}=k_b+\Lambda$, then $d(k)$ is
\begin{align}
    d(k_i)=\begin{cases}
   0, & 1\leq i \leq i_b,\\
   2v(k_i-k_b),  & i_b \leq i \leq i_{\text{max}}.
\end{cases}
\end{align}
$i_b$ parametrizes the momentum index above which the kinetic energy is finite. 
In fact, a tighter upper bound on the ground state energy $\lambda_1$  can be established for our specific case by considering a variational state restricted to the momentum interval $k\leq k_b$, yielding $\lambda_1\leq U\int_0^{k_b}dku^2(k)$.
This result guarantees that the ground state energy for two electrons for a finite trashcan velocity $v$ is always gapped. The spectrum also features $i_b-1$ zero-energy modes. Introducing a finite velocity increases the energy gap between the ground state and the zero-energy states.

In the continuum limit, where momentum $k$ is a continuous variable, the two-electron spectrum coincides with the spectrum consisting of the sum of the kinetic energies
\begin{align}
\lambda=\{2d(k), \text{ for }k \in [0,k_b+\Lambda]\},\label{eq:trashcan_spectrum}
\end{align} 
with an additional gapped ground state energy satisfying Eqs.~\ref{eq:secular_equation_0} and \ref{eq:weyl_largest_trashcan}. The full energy spectrum with $U=-2/A_b,v=5,\alpha=\beta,\varphi_{\text{BZ}}=\pi/2$ and $\Lambda=k_b$ is shown in Fig.~\ref{fig:2d_2e_spectrum}(b). The GS energy of -0.62 is lower than the -0.29 found for $v=\infty$ as expected. Despite the different velocities, the ground state dispersion remains linear for small momentum $\bm{p}\to 0$, similar to the $v=\infty$ case shown in panel (a). As for the GS wavefunction $\ket{\psi_m}$ for angular momentum $m$, it can be parameterized as
\begin{align}
    \ket{\psi_{m}}=&\frac{1}{\Omega_{tot}}\sum_{k}^{\{\bm{k}\}}\frac{u(k)}{d(k)-E_{g,m}}\ket{k,m}\\
=&\begin{cases}
       \frac{1}{\Omega_{tot}}\sum_{k}^{\{\bm{k}\}}\frac{(\sqrt{2\alpha})^{|m|}}{-E_{g,m}\sqrt{2\pi(|m|!)}}k^{\frac{1}{2}+|m|}e^{-\alpha k^2}\ket{k,m}, & k<k_b,\\
       \frac{1}{\Omega_{tot}}\sum_{k}^{\{\bm{k}\}}\frac{(\sqrt{2\alpha})^{|m|}}{(2v(k-k_b)-E_{g,m})\sqrt{2\pi(|m|!)}}k^{\frac{1}{2}+|m|}e^{-\alpha k^2}\ket{k,m},  & k_b\leq k \leq k_b+\Lambda,
\end{cases}
\end{align}
where $E_{g,m}$ is the ground state energy. For $k<k_b$ the wavefunction has a form identical to that of the $v=\infty$ case. For $k_b\leq k \leq k_b+\Lambda$, the wavefunction extends into the trashcan wall and decays rapidly, with the velocity $v$ controlling the decay rate.

\subsubsection{Two-hole spectrum}\label{app:sec:hole_doping}


In this section, we discuss the problem of adding two holes $(2h)$ to the fully filled trashcan bottom for $v=\infty$ with the Hamiltonian given in Eq.~\ref{eq:2dham_trashcan_0}. 
To study the hole doped region, we follow a similar analysis to that in App.~\ref{appsec:1d_2h} for the 1D case, and first rewrite the interacting Hamiltonian $\hat{H}^{\text{int}}$ so that it is normal-ordered with respect to the fully filled trashcan bottom $\ket{\text{full}}$. In other words, we reorder the four-fermion operator to bring the annihilation operators to the left of creation operators
\begin{align}
         \gamma_{\mbf{k}+ \mbf{q}}^\dagger \gamma_{\mbf{k}'- \mbf{q}}^\dagger \gamma_{\mbf{k}'} \gamma_{\mbf{k}}&= \gamma_{\mbf{k}'} \gamma_{\mbf{k}}
         \gamma_{\mbf{k}+ \mbf{q}}^\dagger \gamma_{\mbf{k}'- \mbf{q}}^\dagger-\delta_{\bm{q},0}(\gamma_{\bm{k}}\gamma_{\bm{k}}^\dagger+\gamma_{\bm{k}'}\gamma_{\bm{k}'}^\dagger) +\delta_{\bm{k}',\bm{{k}+\bm{q}}}(\gamma_{\bm{k}+\bm{q}}\gamma_{\bm{k}+\bm{q}}^\dagger+\gamma_{\bm{k}}\gamma_{\bm{k}}^\dagger)+\delta_{\bm{q},0}-\delta_{\bm{k}',\bm{k}+\bm{q}}.
\end{align}
The interaction Hamiltonian can then be rewritten as
\begin{align}
    \hat{H}^{\text{int}}&= \frac{1}{2\Omega_{tot}}  \sum_{\mbf{k}, \mbf{k'}, \mbf{q}}^{\{ \mbf{k}, \mbf{k'}, \mbf{k} + \mbf{q}, \mbf{k'} - \mbf{q}\} }  V_{\bm q} M_{\bm k,\bm q}M^*_{\bm k',\bm q}     \gamma_{\mbf{k}'} \gamma_{\mbf{k}}
         \gamma_{\mbf{k}+ \mbf{q}}^\dagger \gamma_{\mbf{k}'- \mbf{q}}^\dagger+\frac{1}{\Omega_{tot}}\sum_{\mbf{k},\bm{q}}^{\{ \mbf{k},\bm{k}+\bm{q}\} }V_{\bm{q}}|M_{\bm{k},\bm{q}}|^2\gamma_{\mbf{k}} \gamma_{\mbf{k}}^\dagger \nonumber\\
         &-\frac{N_{k_b}V_0}{\Omega_{tot}}\sum_{\mbf{k}}^{\{ \mbf{k}\} }\gamma_{\mbf{k}} \gamma_{\mbf{k}}^\dagger
         +\frac{N_{k_b}^2V_{0}}{2\Omega_{tot}}-\frac{1}{2\Omega_{tot}}\sum_{\mbf{k},\bm{q}}^{\{ \mbf{k},\bm{k}+\bm{q}\} }V_{\bm{q}}|M_{\bm{k},\bm{q}}|^2,\label{eq:ph_hint}
\end{align}
where $N_{k_b}=\sum_{\mbf{k}}^{\{ \mbf{k}\}}$ is the number of momenta in the cutoff. Eq.~\ref{eq:ph_hint} is an effective Hamiltonian for the holes on top of the fully filled trashcan bottom. The first term is the interaction between holes, which we note takes the \textit{same} sign as that between electrons. The second term and third terms, containing $\gamma_{\bm{k}} \gamma_{\bm{k}}^\dagger$ operators, combine to form an effective interaction-induced dispersion for holes. The last two terms are constant energy shifts reflecting the total energy of $\ket{\text{full}}$.

In the following, we still restrict to the interaction potential that corresponds to
\begin{equation}
    U_{\bm{q}}=Ue^{-\alpha q^2},
\end{equation}
with $U$ negative, which leads to a purely attractive interaction if $\alpha\geq|\beta|$. The effective hole dispersion can be explicitly evaluated as (we neglect the constant term of dispersion for simplicity)
\begin{align}
    E_{\bm{k}}^{h}&=\frac{1}{\Omega_{tot}}\sum_{\bm{q}}^{\{\bm{k}+\bm{q}\}  }V_{\bm{q}}|M_{\bm{k},\bm{q}}|^2=U\int_{|\bm{q}|\leq k_b}\frac{d^2\bm{q}}{(2\pi)^2}e^{-\alpha |\bm{q-k}|^2}\nonumber\\
    &=\frac{Ue^{-\alpha k^2}}{(2\pi)^2}\int_0^{k_b} dq \, q e^{-\alpha q^2}\int_0^{2\pi}d\theta e^{2\alpha q k \cos \theta}\nonumber\\
    &=\frac{Ue^{-\alpha k^2}}{2\pi}\int_0^{k_b} dq\,q e^{-\alpha q^2}I_0(2\alpha kq)\label{eq:2d_hole_disperion}\\
    &\approx\begin{cases}
    \displaystyle \frac{U}{4\pi\alpha}=\frac{Uk_b^2}{2|\varphi_{\text{BZ}}|} & \quad \text{when }\alpha k_b^2 \to \infty \\
    \\
    \displaystyle \frac{Ue^{-\alpha k^2}}{2\pi}(\frac{k_b^2}{2}-\frac{\alpha k_b^4}{4}+\frac{\alpha^2k^2k_b^4}{4})= \frac{Ue^{-\frac{|\varphi_{\text{BZ}|}k^2}{2\pi k_b^2} }}{2\pi}(\frac{k_b^2}{2}-\frac{|\varphi_{\text{BZ}}|k_b^2}{8\pi}+\frac{\varphi_{\text{BZ}}^2k^2}{16\pi ^2})\label{eq:kinetic_hole} & \quad \text{when } \alpha k_b^2 \to 0
    \end{cases}.
\end{align}
In the large $\alpha k_b^2$ limit, the hole dispersion is exactly flat and the system restores particle-hole symmetry (with a chemical potential shift). In the $\alpha k_b^2 \to 0$ limit, the holes experience a quadratic dispersion which has a minimum at $\bm{p}=0$. Therefore the ground state is in the zero momentum sector. 

The four-fermion interaction of the holes $\hat{H}^{\text{int,hole}}$ in Eq.~\ref{eq:ph_hint} is identical to that of the electrons, except that the particle-hole transformation converts $\beta\to-\beta$ for the holes. We now consider the limit where $\alpha=\beta$ (effectively $-\beta$ for interaction between holes) and perform an angular momentum decomposition. According to the analysis in App.~\ref{app:sec:vf_inf_p_0}, we find that the Hamiltonian is only nonzero if angular momentum $m<0$, and in the angular momentum sector $m$, the interacting Hamiltonian for two holes is a rank-1 matrix as discussed in App.~\ref{app:sec:vf_inf_p_0} with the form
\begin{align}
    \hat{H}^{\text{int,hole}}\ket{k,m}&=U\int_0^{k_b}\frac{dk'}{2\pi}\sqrt{kk'}e^{-\alpha(k^2+k'^2)}\frac{(2\alpha kk')^{-m}}{(-m)!}\ket{k',m}\nonumber\\
    &=U\int_0^{k_b} dk' u(k)u(k')\ket{k',m}
\end{align}
with $u(k)=\frac{(\sqrt{2\alpha})^{-m}}{\sqrt{2\pi (-m)!}}k^{\frac{1}{2}-m}e^{-\alpha k^2}$.

The corresponding `kinetic' Hamiltonian for two holes with total momentum $\bm{p}=0$ is a diagonal Hamiltonian with diagonal terms $d(k)=E_{\bm{k}}^{h}+E_{-\bm{k}}^{h}\approx\frac{Uk_b^2}{2\pi}e^{-\alpha k^2}$, where we expand $E_{\bm{k}}^h$ to zeroth order in $\alpha k_b^2$ in Eq.~\ref{eq:kinetic_hole}. Therefore, the total Hamiltonian for two holes is a symmetric DPR1 matrix (just like the 1D case in App.~\ref{appsec:1d_2h}), such that its eigenvalues ($\lambda$) can be solved by the secular equation
\begin{align}
    1=\int_0^{k_b}dk\frac{Uu(k)^2}{\lambda - d(k)}\label{eq:secular_equation_1}.
\end{align}
The eigenvalues also satisfy Weyl's inequality
\begin{align}
    &d(k_1) + U\int_0^{k_b} dku^2(k)\le\lambda_1\le d({k_{1}})\label{eq:weyl_largest}\\
    &d({k_i})\le\lambda_{i+1}\le d({k_{i+1}}),\quad i \in[1, i_{\text{max}}-1],
\end{align}
where we discretize the momenta $k$ for convenience, so it takes discrete values $k_i$, with $i=1,\ldots,{i_\text{max}}$. In particular, $k_1=0$ and $k_{i_{\text{max}}}=k_b$. Since $U<0$, we have $d(k_1)\le d(k_2)\le\cdots\le d(k_{i_{\text{max}}})$. In the continuum limit, $k$ is treated as a continuous variable, and the many-body spectrum coincides with the interaction-induced kinetic term (which forms the `kinetic continuum')
\begin{align}
E=\{d(k), \text{ for }k \in [0,k_b]\},
\end{align}
 with an additional ground state whose energy satisfies both Eqs.~\ref{eq:secular_equation_1} and \ref{eq:weyl_largest}. To constrain the ground state energy within the angular momentum sector $m=-1$, we now demonstrate that it is not gapped from the kinetic continuum. To see this, we evaluate
 \begin{align}
     \int^{k_b}_0dk\frac{Uu(k)^2}{d(0)-d(k)}-1=&\int_0^{k_b}dk\frac{2\alpha k^{3}e^{-2\alpha k^2}}{k_b^2(1-e^{-\alpha k^2})}-1\nonumber\\
     =&\frac{1}{x_b}\int_0^{x_b}dk\frac{xe^{-2x}}{1-e^{-x}}-1\quad\text{where }x_b=\alpha k_b^2\nonumber\\
     =&\frac{1}{x_b}\int_0^{x_b}dk\left(\frac{xe^{-2x}}{1-e^{-x}}-1\right)\nonumber\\
     =&\frac{1}{x_b}\int_0^{x_b}dk\frac{x+e^{x}-e^{2x}}{e^{2x}-e^{x}}.\label{eq:bound_2h_2d}
 \end{align}
 Since $e^{2x}>e^x+x$ for all $x>0$, Eq.~\ref{eq:bound_2h_2d} is always smaller than zero. For $U<0$ and any $\lambda<d(0)$, we have
 \begin{align}
     \int^{k_b}_0dk\frac{Uu(k)^2}{\lambda-d(k)}<\int^{k_b}_0dk\frac{Uu(k)^2}{d(0)-d(k)}.
 \end{align}
 We thus we conclude that Eq.~\ref{eq:secular_equation_1} is never satisfied with any $\lambda_1<d(0)$. Combining this result with the interlacing, we therefore prove that
the spectrum is gapless, and no gapped state can form below the continuum of energies $d(k)$.
The corresponding ground state wavefunction is 
\begin{align}
    \ket{\psi_{m}}&=\frac{1}{Z\Omega_{tot}}\sum_{k}^{\{k\}}\frac{u(k)}{d(k)-E_g}\ket{k,m}\nonumber\\
    &=\frac{1}{(2\pi)^2Z}\int_{|\bm{k}|\leq k_b} d^2\bm{k}\frac{\sqrt{2\alpha}^{-m}}{\sqrt{2\pi(-m)!}}\frac{k^{-m}e^{-\alpha k ^2+im\varphi_k}}{\frac{Uk_b^2}{2\pi}(e^{-\alpha k^2}-1)}\ket{\bm{k}},
\end{align}
where $Z$ is a normalization factor, and $\ket{\bm{k}}=\gamma_{\bm{k}}^\dagger\gamma_{-\bm{k}}^\dagger$ as defined in Eq.~\ref{app:eq:definiton_of_psi}. The two holes primarily occupy momenta near zero for the ground state. Based on this, we can approximate the ground state wavefunction with $m=-1$ as 
\begin{align}
    \ket{\psi_{1}}
    &=\frac{\sqrt{2\alpha}}{(2\pi)^{5/2}Z}\int_{|\bm{k}|\leq k_b} d^2\bm{k}\frac{ke^{-\alpha k ^2-i\varphi_k}}{\frac{Uk_b^2}{2\pi}(e^{-\alpha k^2}-1)}\ket{\bm{k}}\propto  \int_{|\bm{k}|\leq k_b} d^2\bm{k}\frac{e^{-\alpha k ^2}}{k_+}\ket{\bm{k}}.
\end{align}

\begin{figure}[H]
    \centering
    \includegraphics[width=0.75\linewidth]{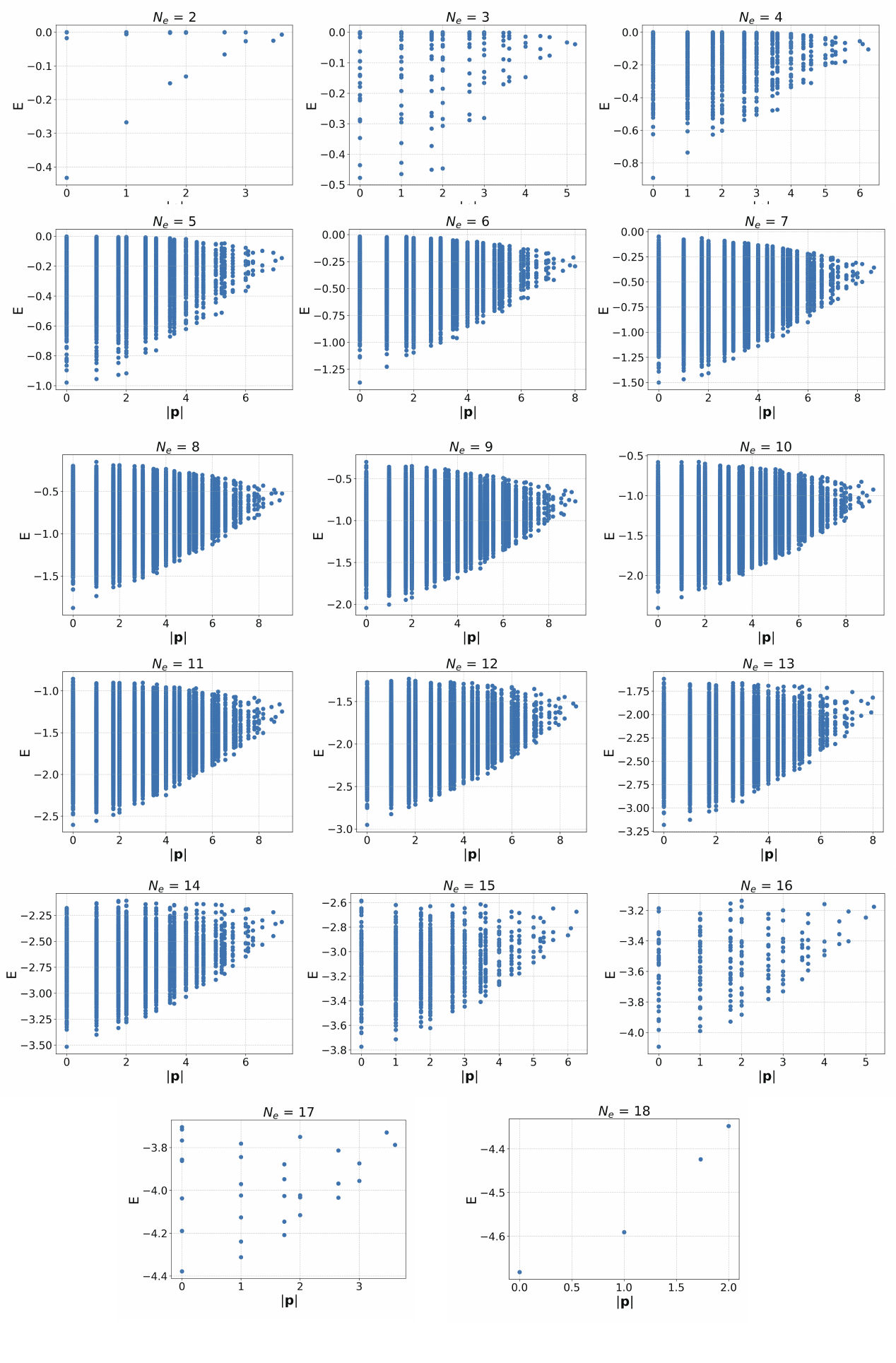}
    \caption{The full ED energy spectrum of the attractive 2D trashcan model with $U=-2/A_b,N_{k_b}=19,\alpha=\beta,\varphi_{\text{BZ}}=\pi/2$ and varying particle number $N_e$. The results are plotted as a function of the magnitude of the momentum $|\bm{p}|$. (We absorb a factor of $\Omega_{tot}$ in $U$).}
    \label{fig:2d_dipsersion_all_mom}
\end{figure}

We now turn to the finite momentum case, and study the dispersion of the two-hole ground state. We recall that we have proved that the two-hole spectrum is continuous and determined by the interaction-induced single-hole dispersion for $\bm{p}=0$
\begin{align}
    E_{\bm{k}}^h&=\frac{1}{\Omega_{tot}}\sum_{\bm{q}}^{\{ \mbf{k},\bm{k}+\bm{q}\}  }V_{\bm{q}}=\int_{\mathcal{H}_{\bm{q}}}\frac{Ud^2\bm{q}}{(2\pi)^2}e^{-\alpha |\bm{q-k}|^2}\approx \frac{Uk_b^2}{4\pi}e^{-\alpha k^2}\quad \text{when }\alpha k_b^2 \ll1.\label{app:eq:2d_hole_dispersion}
\end{align}
Since, the case of finite $\bm{p}$ is approximately identical to the zero-momentum case except a shift of the total momentum and a decrease of momentum cutoff $k_b$ (App.~\ref{appsubsec:2d_vFinf_finitep}), such proof still holds for the finite $\bm{p}$ case. The ground state energy (for attractive $U<0$) for two holes within the total momentum sector $\bm{p}$ is thus
\begin{align}
    E_g^h(\bm{p})=2E_{\bm{p}/2}^h=\frac{Uk_b^2}{2\pi}e^{-\alpha (p/2)^2}\approx\frac{Uk_b^2}{2\pi}(1-\frac{\alpha p^2}{4}),
\end{align}
which exhibits quadratic dispersion for small $p$, similar to the 1D case. Numerically, we also observe a linear to quadratic crossover in the small momentum dispersion with increasing $N_e$ (Fig.~\ref{fig:2d_dipsersion_all_mom}).

\newpage
\newpage

\subsection{Many-body Ground States For Attractive Interaction With $\alpha=\beta$}\label{app:sec:2d_many_body}

Inspired by the analytic many-body ground state wavefunction of the 1D trashcan model (see App.~\ref{sec:1d_manybody_gs}), we propose the following many-body ansatz in the total momentum $\bm{p}=0$ sector for $\alpha=\beta$
\begin{equation} \label{eq:GMPAnsatzGround}
\ket{\phi_{N_e}^{A}}
\propto 
\begin{cases} 
\bigl(\hat{O}_{2,m}^\dagger\bigr)^N\ket{0}, 
& N_{e} = 2N,\\[6pt]
\gamma_{0}^{\dagger}\,\bigl(\hat{O}_{2,m}^\dagger\bigr)^N\ket{0}, 
& N_{e} = 2N + 1,
\end{cases}
\end{equation}
where $\hat{O}_{2,m}^\dagger$ is the creation operator of the two-electron state with angular momentum $m$
\begin{equation}
    \hat{O}_{2,m}^\dagger =\frac{1}{(2\pi)^2 Z}\int_{|\bm{k}|\leq k_b}d^2\bm{k}k_{+}^{m}\,e^{-\alpha\mbf{k}^2}
    \hat{\gamma}^{\dagger}_{\mbf{k}}\,
    \hat{\gamma}^{\dagger}_{-\mbf{k}}= \frac{1}{Z\Omega_{tot}}\sum^{\{\bm{k}\}}_{\mbf{k}}k_{+}^{m}\,e^{-\alpha\mbf{k}^2}\,
    \hat{\gamma}^{\dagger}_{\mbf{k}}\,
    \hat{\gamma}^{\dagger}_{-\mbf{k}}.
\end{equation}
For the ground state, we consider $m=1$.

\subsubsection{Ground State Ansatz For Even $N_e$, $v=\infty$, $\bm{p}=0$}\label{app:sec:gs_even_2d}
To verify the ansatz for even-particle number, we compute the commutator between the interacting Hamiltonian $\hat{H}^\text{int}$
\begin{equation}
     \hat{H}^\text{int}= \frac{U}{2\Omega_{tot}}  \sum^{\{ \mbf{k}, \mbf{k'}, \mbf{k} + \mbf{q}, \mbf{k'} - \mbf{q}\}  }_{\bm{q},\bm{k},\bm{k'}}  e^{-\alpha \bm{q}^2-i \beta (  \mbf{q}\times (\mbf{k}- \mbf{k'})) }     \gamma_{\mbf{k}+ \mbf{q}}^\dagger \gamma_{\mbf{k}'- \mbf{q}}^\dagger \gamma_{\mbf{k}'} \gamma_{\mbf{k}},
\end{equation}
and $\hat{O}_{2,m}^\dagger$. We keep $m$ as a general positive odd integer for now. The commutator can be divided into two parts:
\begin{align}
    [\hat{H}^\text{int}, \hat{O}_{2,m}^\dagger]=[\hat{H}^\text{int}, \hat{O}_{2,m}^\dagger]_1+[\hat{H}^\text{int}, \hat{O}_{2,m}^\dagger]_2,
\end{align}
where $[\hat{H}^\text{int}, \hat{O}_{2,m}^\dagger]_1$ collects all terms consisting of four fermionic operators
\begin{align} \label{eq:myO2CommuPart1}
[\hat{H}^\text{int}, \hat{O}_{2,m}^\dagger]_1 &= \frac{U}{2\Omega_{tot}^2Z}\sum_{\mbf{k}, \mbf{k'}, \mbf{q}}^{\{ \mbf{k}, \mbf{k'}, \mbf{k} + \mbf{q}, \mbf{k'} - \mbf{q}\} }\sum_{\mbf{k_1}}^{\{\mbf{k_1} \}} e^{-\alpha \bm{q}^2-i\beta(\mbf{q} \times  (\mbf{k}-\bm{k}'))} k_{1,+}^{m}e^{-\alpha \bm{k}_1^2} (-\delta_{-\mbf{k_1}, \mbf{k'}} \gamma_{\mbf{k} + \mbf{q}}^{\dagger} \gamma_{\mbf{k'} - \mbf{q}}^{\dagger} \gamma_{\mbf{k_1}}^{\dagger}\gamma_{\mbf{k}} \nonumber\\&+ \delta_{\mbf{k}, -\mbf{k_1}}\gamma_{\mbf{k} + \mbf{q}}^{\dagger} \gamma_{\mbf{k'} - \mbf{q}}^{\dagger} \gamma_{\mbf{k_1}}^{\dagger} \gamma_{\mbf{k'}}+\delta_{\mbf{k'}, \mbf{k_1}} \gamma_{\mbf{k} + \mbf{q}}^{\dagger} \gamma_{\mbf{k'} - \mbf{q}}^{\dagger} \gamma_{-\mbf{k_1}}^{\dagger} \gamma_{\mbf{k}} - \delta_{\mbf{k}, \mbf{k_1}} \gamma_{\mbf{k} + \mbf{q}}^{\dagger} \gamma_{\mbf{k'} - \mbf{q}}^{\dagger} \gamma_{-\mbf{k_1}}^{\dagger} \gamma_{\mbf{k'}})\nonumber\\
&=\frac{U}{2\Omega_{tot}^2Z}\sum_{\mbf{k}, \mbf{k'}, \mbf{q}}^{\{\mbf{k}, \mbf{k'}, \mbf{k} + \mbf{q}, \mbf{k'} - \mbf{q}\} } e^{-\alpha \bm{q}^2-i\beta(\mbf{q} \times  (\mbf{k}-\bm{k}'))} (k_{+}'^{m}e^{-\alpha \bm{k}'^2} \gamma_{\mbf{k} + \mbf{q}}^{\dagger} \gamma_{\mbf{k'} - \mbf{q}}^{\dagger} \gamma_{-\mbf{k'}}^{\dagger} \gamma_{\mbf{k}} - k_{+}^{m}e^{-\alpha \bm{k}^2}\gamma_{\mbf{k} + \mbf{q}}^{\dagger} \gamma_{\mbf{k'} - \mbf{q}}^{\dagger} \gamma_{-\mbf{k}}^{\dagger} \gamma_{\mbf{k'}}\nonumber\\
&+k_{+}'^{m}e^{-\alpha \bm{k}'^2}\gamma_{\mbf{k} + \mbf{q}}^{\dagger} \gamma_{\mbf{k'} - \mbf{q}}^{\dagger} \gamma_{-\mbf{k'}}^{\dagger} \gamma_{\mbf{k}} - k_{+}^{m}e^{-\alpha \bm{k}^2} \gamma_{\mbf{k} + \mbf{q}}^{\dagger} \gamma_{\mbf{k'} - \mbf{q}}^{\dagger} \gamma_{-\mbf{k}}^{\dagger} \gamma_{\mbf{k'}})\nonumber\\
&=\frac{U}{2\Omega_{tot}^2Z}\sum_{\mbf{k}, \mbf{k'}, \mbf{q}}^{\{ \mbf{k}, \mbf{k'}, \mbf{k} + \mbf{q}, \mbf{k'} - \mbf{q}\} }2e^{-\alpha \bm{q}^2-i\beta(\mbf{q} \times  (\mbf{k}-\bm{k}'))}(k_{+}'^{m}e^{-\alpha \bm{k}'^2} \gamma_{\mbf{k} + \mbf{q}}^{\dagger} \gamma_{\mbf{k'} - \mbf{q}}^{\dagger} \gamma_{-\mbf{k'}}^{\dagger} \gamma_{\mbf{k}} - k_{+}^{m}e^{-\alpha \bm{k}^2} \gamma_{\mbf{k} + \mbf{q}}^{\dagger} \gamma_{\mbf{k'} - \mbf{q}}^{\dagger} \gamma_{-\mbf{k}}^{\dagger} \gamma_{\mbf{k'}}).
\end{align}
We relabel the summation variables as $\mbf{k}\rightarrow \mbf{k'}$, $\mbf{k'} \rightarrow \mbf{k}$, $\mbf{q}\rightarrow -\mbf{q}$ in the second term above, leading to
\begin{align} 
[\hat{H}^\text{int}, \hat{O}_{2,m}^\dagger]_1 &=\frac{U}{2\Omega_{tot}^2Z}\sum_{\mbf{k}, \mbf{k'}, \mbf{q}}^{\{ \mbf{k}, \mbf{k'}, \mbf{k} + \mbf{q}, \mbf{k'} - \mbf{q}\}}2e^{-\alpha \bm{q}^2-i\beta(\mbf{q} \times  (\mbf{k}-\bm{k}'))}(k_{+}'^{m}e^{-\alpha \bm{k}'^2} \gamma_{\mbf{k} + \mbf{q}}^{\dagger} \gamma_{\mbf{k'} - \mbf{q}}^{\dagger} \gamma_{-\mbf{k'}}^{\dagger} \gamma_{\mbf{k}} + k_{+}'^{m}e^{-\alpha \bm{k}'^2}  \gamma_{\mbf{k} + \mbf{q}}^{\dagger}\gamma_{\mbf{k}' - \mbf{q}}^{\dagger} \gamma_{-\mbf{k}'}^{\dagger} \gamma_{\mbf{k}})\nonumber\\
&=\frac{2U}{\Omega_{tot}^2Z}\sum_{\mbf{k}, \mbf{k'}, \mbf{q}}^{\{ \mbf{k}, \mbf{k'}, \mbf{k} + \mbf{q}, \mbf{k'} - \mbf{q}\}}e^{-\alpha \bm{q}^2-i\beta(\mbf{q} \times  (\mbf{k}-\bm{k}'))}k_{+}'^{m}e^{-\alpha \bm{k}'^2} \gamma_{\mbf{k} + \mbf{q}}^{\dagger} \gamma_{\mbf{k'} - \mbf{q}}^{\dagger} \gamma_{-\mbf{k'}}^{\dagger} \gamma_{\mbf{k}}. 
\end{align}
In general, $[\hat{H}^\text{int}, \hat{O}_{2,m}^\dagger]_1$ is non-zero. Only when it acts on the vacuum state $\ket{\text{vac}}$, do we obtain a vanishing result
\begin{align}
    [\hat{H}^\text{int}, \hat{O}_{2,m}^\dagger]_1\ket{\text{vac}}=0.
\end{align}

As for $[\hat{H}^\text{int}, \hat{O}_{2,m}^\dagger]_2$, which collects terms with two fermionic operators, we have
\begin{align} 
[\hat{H}^\text{int}, \hat{O}_{2,m}^\dagger]_2&= \frac{U}{2\Omega_{tot}^2Z}\sum_{\mbf{k}, \mbf{k'}, \mbf{q}}^{\{ \mbf{k}, \mbf{k'}, \mbf{k} + \mbf{q}, \mbf{k'} - \mbf{q}\} }\sum_{\mbf{k_1}}^{\{\mbf{k_1} \}} e^{-\alpha \bm{q}^2-i\beta(\mbf{q} \times  (\mbf{k}-\bm{k}'))} k_{1,+}^{m}e^{-\alpha \bm{k}_1^2} (-\delta_{\mbf{k}, -\mbf{k_1}} \delta_{\mbf{k'}, \mbf{k_1}} \gamma_{\mbf{k} + \mbf{q}}^{\dagger} \gamma_{\mbf{k'} - \mbf{q}}^{\dagger} + \delta_{\mbf{k}, \mbf{k_1}} \delta_{-\mbf{k_1}, \mbf{k'}} \gamma_{\mbf{k} + \mbf{q}}^{\dagger} \gamma_{\mbf{k'} - \mbf{q}}^{\dagger})\nonumber\\
&=\frac{U}{2\Omega_{tot}^2Z}\sum_{\mbf{k_1}, \mbf{q}}^{\{ \mbf{k_1}, \mbf{k_1} - \mbf{q}\} } e^{-\alpha \bm{q}^2-i\beta(\mbf{q} \times  (-2\bm{k}_1))}(-k_{1,+}^me^{-\alpha \bm{k}_1^2} \gamma_{-\mbf{k_1} + \mbf{q}}^{\dagger} \gamma_{\mbf{k}_1 - \mbf{q}}^{\dagger} )\nonumber\\
&+ \frac{U}{2\Omega_{tot}^2Z}\sum_{\mbf{k_1}, \mbf{q}}^{\{ \mbf{k_1}, \mbf{k_1} + \mbf{q}\} }e^{-\alpha \bm{q}^2-i\beta(\mbf{q} \times  2\bm{k}_1)}(k_{1,+}^me^{-\alpha \bm{k}_1^2}\gamma_{\mbf{k}_1 + \mbf{q}}^{\dagger} \gamma_{-\mbf{k}_1 - \mbf{q}}^{\dagger})\nonumber\\
&=\frac{U}{\Omega_{tot}^2Z}\sum_{\mbf{k}, \mbf{q}}^{\{ \mbf{k}, \mbf{k} + \mbf{q}\} }e^{-\alpha \bm{q}^2-i\beta(\mbf{q} \times  2\bm{k})}(k_{+}^me^{-\alpha \bm{k}^2}\gamma_{\mbf{k} + \mbf{q}}^{\dagger} \gamma_{-\mbf{k} - \mbf{q}}^{\dagger}).
\end{align}
Changing variables $\bm{k}+\bm{q}\to\bm{k}$ and then $\bm{q}\to\bm{k-k'}$, we obtain
\begin{align}
    [\hat{H}^\text{int}, \hat{O}_{2,m}^\dagger]_2&=\frac{U}{\Omega_{tot}^2Z}\sum_{\mbf{k,k'}}^{\{ \mbf{k,k'}\} }e^{-\alpha \bm{|k-k'|}^2+i2\beta(\mbf{k'} \times  \bm{k})}k_{+}'^me^{-\alpha \bm{k}'^2}\gamma_{\mbf{k}}^{\dagger} \gamma_{-\mbf{k}}^{\dagger}.
\end{align}
To perform the summation, we adopt the continuum limit and convert the summations into integrals
\begin{align}
    [\hat{H}^\text{int}, \hat{O}_{2,m}^\dagger]_2=&\frac{U}{\Omega_{tot}^2Z}\sum_{\mbf{k,k'}}^{\{ \mbf{k,k'}\}  }k_{+}'^me^{-2\alpha k_+'k_-'+(\alpha-\beta)k_-k_+'+(\alpha+\beta)k_+k_-'}e^{-\alpha \bm{k}^2}\gamma_{\mbf{k}}^{\dagger} \gamma_{-\mbf{k}}^{\dagger}\nonumber\\
    =&\frac{U}{(2\pi)^4Z}\int_0^{k_b} kdk\int_{0}^{2\pi}d\theta_k\int_0^{k_b} k'^{m+1}dk'e^{-2\alpha k'^2}\int_{0}^{2\pi} d\theta_{k'}e^{im\theta_{k'}+(\alpha-\beta)kk'e^{i(\theta_{k'}-\theta_{k})}+(\alpha+\beta)kk'e^{i(\theta_{k}-\theta_{k'})}}e^{-\alpha k^2}\gamma_{\bm{k}}^{\dagger} \gamma_{-\bm{k}}^{\dagger}\nonumber\\
    =&\frac{U}{(2\pi)^3Z}\int_0^{k_b} kdk\int_{0}^{2\pi}d\theta_k\left(\frac{\alpha+\beta}{\alpha-\beta}\right)^{m/2}e^{im\theta_k}\int_0^{k_b} dk'k'^{m+1}e^{-2\alpha k'^2}I_m\left(2kk'\sqrt{\alpha^2-\beta^2}\right)e^{-\alpha k^2}\gamma_{\bm{k}}^{\dagger} \gamma_{-\bm{k}}^{\dagger}.
\end{align}
To obtain the third line, we have used the relation: $\int_0^{2\pi}e^{im\phi}e^{Ae^{i\phi}+Be^{-i\phi}}d\phi=2\pi\left(\sqrt{B/A}\right)^mI_m(2\sqrt{AB})$. The above integral of $k'$ generally has no closed-form expression, so we consider the $\alpha=\beta>0$ limit (the $\alpha=-\beta>0$ limit is the same except $m\to |m|$ and we only consider negative $m$)
\begin{align}
        [\hat{H}^\text{int}, \hat{O}_{2,m}^\dagger]_2&=\frac{U}{(2\pi)^4Z}\int_0^{k_b} kdk\int_{0}^{2\pi}d\theta_k\int_0^{k_b} k'^{m+1}dk'e^{-2\alpha k'^2}\int_{0}^{2\pi} d\theta_{k'}e^{im\theta_{k'}+2\alpha kk'e^{i(\theta_{k}-\theta_{k'})}}e^{-\alpha k^2}\gamma_{\bm{k}}^{\dagger} \gamma_{-\bm{k}}^{\dagger}\nonumber\\
        &=\frac{U}{(2\pi)^4Z}\int_0^{k_b} kdk\int_{0}^{2\pi}d\theta_k e^{im\theta_k}\frac{2\pi(2\alpha k)^m}{m!}\int_0^{k_b} dk'k'^{2m+1}e^{-2\alpha k'^2}e^{-\alpha k^2}\gamma_{\bm{k}}^{\dagger} \gamma_{-\bm{k}}^{\dagger}\nonumber\\
        &=\frac{U}{(2\pi)^2 Z}\int_{|\bm{k}|\leq k_b} d^2\bm{k}k_+^me^{-\alpha k^2}\frac{\left(\Gamma(1+m)-\Gamma(1+m,2\alpha k_b^2)\right)}{8\pi\alpha m!}\gamma_{\bm{k}}^{\dagger} \gamma_{-\bm{k}}^{\dagger}\nonumber\\
        &=E_{2,m}\hat{O}_{2,m}^\dagger,
\end{align}
where $E_{2,m}$ is the ground state energy within the sector of angular momentum $m$. With the above results, we prove that for $\alpha=\beta$, 
\begin{align}
    [\hat{H}^\text{int}, \hat{O}_{2,m}^\dagger]\ket{\text{vac}}= E_{2,m}\hat{O}_{2,m}^\dagger\ket{\text{vac}}.
\end{align}

Similar to the 1d case, to obtain the many-body ground states, we examine whether the interaction Hamiltonian exhibits a RSGA-1. $\hat{H}^\text{int}$ is of the form $\gamma^\dagger \gamma^\dagger \gamma\gamma$ while $\hat{O}_{2,m}^\dagger$ is constructed with $ \gamma^{\dagger}_{\mbf{k}}\gamma^{\dagger}_{-\mbf{k}}$, so it is straightforward to see that
\begin{align}
    \left[\left[\left[\hat{H}^\text{int},\hat{O}_{2,m}^\dagger\right],\hat{O}_{2,m}^\dagger\right],\hat{O}_{2,m}^\dagger\right]=0.
\end{align}
Hence, to demonstrate a RSGA-1, we only need to calculate $\left[\left[\hat{H}^\text{int},\hat{O}_{2,m}^\dagger\right],\hat{O}_{2,m}^\dagger\right]$. Note that $\left[\left[\hat{H}^\text{int},\hat{O}_{2,m}^\dagger\right]_2,\hat{O}_{2,m}^\dagger\right]$ vanishes, leading to
\begin{align}
    \left[\left[\hat{H}^\text{int},\hat{O}_{2,m}^\dagger\right],\hat{O}_{2,m}^\dagger\right]&=\left[\left[\hat{H}^\text{int},\hat{O}_{2,m}^\dagger\right]_1,\hat{O}_{2,m}^\dagger\right]\nonumber\\
    &=\frac{2U}{\Omega_{tot}^3Z^2}\sum_{\mbf{k}, \mbf{k'}, \mbf{q},\mbf{k}_1}^{\{ \mbf{k}, \mbf{k'}, \mbf{k} + \mbf{q}, \mbf{k'} - \mbf{q},\mbf{k}_1\} }e^{-\alpha \bm{q}^2-i\beta(\mbf{q} \times  (\mbf{k}-\mbf{k}'))}k_{+}'^{m}k_{1,+}^{m}e^{-\alpha \bm{k}'^2-\alpha\mbf{k}_1^2} \left[\gamma_{\mbf{k} + \mbf{q}}^{\dagger} \gamma_{\mbf{k'} - \mbf{q}}^{\dagger} \gamma_{-\mbf{k'}}^{\dagger} \gamma_{\mbf{k}}, 
    \gamma^{\dagger}_{\mbf{k}_1}\,
    \gamma^{\dagger}_{-\mbf{k}_1}\right]\nonumber\\
    &=\frac{2U}{\Omega_{tot}^3Z^2}\sum_{\mbf{k}, \mbf{k'}, \mbf{q},\mbf{k}_1}^{\{ \mbf{k}, \mbf{k'}, \mbf{k} + \mbf{q}, \mbf{k'} - \mbf{q},\mbf{k}_1\} }e^{-\alpha \bm{q}^2-i\beta(\mbf{q} \times  (\mbf{k}-\mbf{k}'))}k_{+}'^{m}k_{1,+}^{m}e^{-\alpha \bm{k}'^2-\alpha\mbf{k}_1^2} \gamma_{\mbf{k} + \mbf{q}}^{\dagger} \gamma_{\mbf{k'} - \mbf{q}}^{\dagger} \gamma_{-\mbf{k'}}^{\dagger} \left[\delta_{\mbf{k},\mbf{k}_1}\gamma_{-\mbf{k}_1}^\dagger-\delta_{\mbf{k},-\mbf{k}_1}\gamma_{\mbf{k}_1}^\dagger\right]\nonumber\\
    &=\frac{4U}{\Omega_{tot}^3Z^2}\sum_{\mbf{k}, \mbf{k'}, \mbf{q}}^{\{ \mbf{k}, \mbf{k'}, \mbf{k} + \mbf{q}, \mbf{k'} - \mbf{q}\} }e^{-\alpha \bm{q}^2-i\beta(\mbf{q} \times  (\mbf{k}-\mbf{k}'))}k_{+}'^{m}k_{+}^{m}e^{-\alpha \bm{k}'^2-\alpha\mbf{k}^2} \gamma_{\mbf{k} + \mbf{q}}^{\dagger} \gamma_{\mbf{k'} - \mbf{q}}^{\dagger} \gamma_{-\mbf{k'}}^{\dagger} \gamma_{-\mbf{k}}^\dagger\nonumber\\
    &=\frac{4U}{\Omega_{tot}^3Z^2}\sum_{\bm{k}_1,\bm{k}_2,\bm{k}_3,\bm{k}_4}^{\{ \bm{k}_1,\bm{k}_2,\bm{k}_3,\bm{k}_4\} }e^{-\alpha |\bm{k}_1+\bm{k}_4|^2-i\beta(\bm{k}_1+\bm{k}_4)\times  (\mbf{k}_3-\mbf{k}_4)-\alpha \bm{k}_3^2-\alpha\mbf{k}_4^2}k_{+,3}^{m}k_{+,4}^{m} \gamma_{\mbf{k}_1}^{\dagger} \gamma_{\mbf{k}_2}^{\dagger} \gamma_{\mbf{k}_3}^{\dagger} \gamma_{\mbf{k}_4}^\dagger\delta_{\mbf{k}_1+\mbf{k}_2+\mbf{k}_3+\mbf{k}_4=0}\label{eq:2d_commmutator_HOO}.
\end{align}

We will see below that while the above double commutator does not vanish generally, there is an emergent algebraic structure for small $\alpha k_b^2,\beta k_b^2$. We begin by rewriting the coefficient of the summand in the last line of above equation in terms of $k_+$ and $k_-$ and expanding to first order in $\alpha$ and $\beta$
\begin{align}
\Biggl\{ 
    &1-\alpha \left( k_{1,+}k_{1,-} + k_{3,+}k_{3,-} + 2k_{4,+}k_{4,-} \right) - \left(\alpha + \frac{\beta}{2}\right)k_{1,+}k_{4,-} 
    - \left(\alpha - \frac{\beta}{2}\right)k_{1,-}k_{4,+} \nonumber\\
    &+ \frac{\beta}{2}\left(k_{1,+}k_{3,-} - k_{1,-}k_{3,+}\right) 
    + \frac{\beta}{2}\left(k_{3,-}k_{4,+} - k_{3,+}k_{4,-}\right) 
\Biggr\}k_{+,3}^{m}k_{+,4}^{m}.
\end{align}
The resulting expression contains 10 terms in total. Upon summation against $\gamma_{\mbf{k}_1}^{\dagger} \gamma_{\mbf{k}_2}^{\dagger} \gamma_{\mbf{k}_3}^{\dagger} \gamma_{\mbf{k}_4}^\dagger\delta_{\mbf{k}_1+\mbf{k}_2+\mbf{k}_3+\mbf{k}_4=0}$, several terms vanish due to symmetry. The first two terms vanish under the exchange of $\bm{k}_3 \leftrightarrow \bm{k}_4$, while the third, fourth, and the final two terms vanish under the exchange of $\bm{k}_1 \leftrightarrow \bm{k}_2$. For $m=1$ (which will be relevant when considering the ground state for an attractive interaction), the fifth and seventh terms also vanish under the separate exchanges of $\bm{k}_1 \leftrightarrow \bm{k}_3$ and $\bm{k}_1 \leftrightarrow \bm{k}_4$, respectively.

After restricting to $m=1$ and accounting for these symmetries, Eq.~\ref{eq:2d_commmutator_HOO} reduces to (to first order in $\alpha,\beta$)
\begin{align}
\left[\left[\hat{H}^\text{int},\hat{O}_{2,m=1}^\dagger\right],\hat{O}_{2,m=1}^\dagger\right]&\approx-\frac{4U}{\Omega_{tot}^3Z^2}\sum_{\bm{k}_1,\bm{k}_2,\bm{k}_3,\bm{k}_4}^{\{ \bm{k}_1,\bm{k}_2,\bm{k}_3,\bm{k}_4\}  }k_{1,-}\left( \left(\alpha - \frac{\beta}{2}\right)k_{4,+}+ \frac{\beta}{2}k_{3,+}\right)k_{+,3}k_{+,4}\gamma_{\mbf{k}_1}^{\dagger} \gamma_{\mbf{k}_2}^{\dagger} \gamma_{\mbf{k}_3}^{\dagger} \gamma_{\mbf{k}_4}^\dagger\delta_{\mbf{k}_1+\mbf{k}_2+\mbf{k}_3+\mbf{k}_4=0}\nonumber\\
&=-\frac{4U}{\Omega_{tot}^3Z^2}\sum_{\bm{k}_1,\bm{k}_2,\bm{k}_3,\bm{k}_4}^{\{ \bm{k}_1,\bm{k}_2,\bm{k}_3,\bm{k}_4\} } \left(\alpha - \beta\right)k_{1,-}k_{+,3}k_{+,4}^2\gamma_{\mbf{k}_1}^{\dagger} \gamma_{\mbf{k}_2}^{\dagger} \gamma_{\mbf{k}_3}^{\dagger} \gamma_{\mbf{k}_4}^\dagger\delta_{\mbf{k}_1+\mbf{k}_2+\mbf{k}_3+\mbf{k}_4=0}.\label{eq:2d_commmutator_HOO_1}
\end{align}
In the limit $\alpha=\beta$, we have therefore shown to linear order in $\alpha=\beta$
\begin{align}
\left[\left[\hat{H}^\text{int},\hat{O}_{2,m=1}^\dagger\right],\hat{O}_{2,m=1}^\dagger\right] \approx 0.
\end{align}
(We also performed the second order expansion in $\alpha$ and $\beta$, and found that the commutator $\left[\left[\hat{H}^\text{int},\hat{O}_{2,m=1}^\dagger\right],\hat{O}_{2,m=1}^\dagger\right]$ does not vanish even for $\alpha=\beta$. The lengthy derivation is omitted for brevity.) This result, taken together with the condition
\begin{align}
    \left[\hat{H}^\text{int},\hat{O}_{2,m}^\dagger\right]\ket{\text{vac}}=E_{2,m}\hat{O}_{2,m}^\dagger\ket{\text{vac}},
\end{align}
establishes that the 2D Berry Trashcan Hamiltonian also exhibits a RSGA-1 \cite{PhysRevB.102.085140}, as in the 1D case analyzed in App.~\ref{sec:1d_manybody_gs}, when working at linear order in $\alpha=\beta$. This algebraic structure implies that the states $(\hat{O}_{2,m=1}^\dagger)^N\ket{0}$ are approximate eigenstates of the $2N$-particle system, with an energy spectrum that is approximately given by the equally spaced $E_N = NE_{2,m=1}$.


We now prove that the ansatz for even particle number in Eq.~\ref{eq:GMPAnsatzGround} constitutes the ground state to linear order in $\alpha=\beta$. We start by rewriting the interaction Hamiltonian 
\begin{align}
    \hat{H}^\text{int}=& \frac{U}{2\Omega_{tot}}  \sum^{\{\bm{k}_1,\bm{k}_2,\bm{k}_3,\bm{k}_4\}}_{\bm{k}_1,\bm{k}_2,\bm{k}_3,\bm{k}_4}  e^{-\alpha (\bm{k}_1-\bm{k}_4)^2-i \beta (  \mbf{k_1-k_4})\times (\mbf{k_4}- \mbf{k_3}) }     \gamma_{\mbf{k}_1}^\dagger \gamma_{\mbf{k}_2}^\dagger \gamma_{\mbf{k}_3} \gamma_{\mbf{k}_4}\delta_{\bm{k}_1+\bm{k}_2,\bm{k}_3+\bm{k}_4}\\
    \approx&\frac{U}{2\Omega_{tot}}  \sum^{\{\bm{k}_1,\bm{k}_2,\bm{k}_3,\bm{k}_4\}}_{\bm{k}_1,\bm{k}_2,\bm{k}_3,\bm{k}_4}  \Biggl\{ 
    1-\alpha \left( k_{1,+}k_{1,-} + k_{4,+}k_{4,-} \right) + \left(\alpha + \frac{\beta}{2}\right)k_{1,+}k_{4,-} 
     +\left(\alpha - \frac{\beta}{2}\right)k_{1,-}k_{4,+} \nonumber\\
    &- \frac{\beta}{2}\left(k_{1,+}k_{3,-} - k_{1,-}k_{3,+}\right) 
    + \frac{\beta}{2}\left(k_{3,-}k_{4,+} - k_{3,+}k_{4,-}\right) 
\Biggr\}     \gamma_{\mbf{k}_1}^\dagger \gamma_{\mbf{k}_2}^\dagger \gamma_{\mbf{k}_3} \gamma_{\mbf{k}_4}\delta_{\bm{k}_1+\bm{k}_2,\bm{k}_3+\bm{k}_4},\\
=&\frac{U}{2\Omega_{tot}}  \sum^{\{\bm{k}_1,\bm{k}_2,\bm{k}_3,\bm{k}_4\}}_{\bm{k}_1,\bm{k}_2,\bm{k}_3,\bm{k}_4}  \Biggl\{ 
    \left(\alpha + \frac{\beta}{2}\right)k_{1,+}k_{4,-} 
     +\left(\alpha - \frac{\beta}{2}\right)k_{1,-}k_{4,+} - \frac{\beta}{2}\left(k_{1,+}k_{3,-} - k_{1,-}k_{3,+}\right) 
\Biggr\}     \gamma_{\mbf{k}_1}^\dagger \gamma_{\mbf{k}_2}^\dagger \gamma_{\mbf{k}_3} \gamma_{\mbf{k}_4}\delta_{\bm{k}_1+\bm{k}_2,\bm{k}_3+\bm{k}_4}.
\end{align}
In the last equation, we have dropped terms that vanish under (anti)symmetry. If $\alpha=\beta$, then we can further simplify the Hamiltonian as
\begin{align}
    \hat{H}^\text{int}\approx&\frac{U}{2\Omega_{tot}}  \sum^{\{\bm{k}_1,\bm{k}_2,\bm{k}_3,\bm{k}_4\}}_{\bm{k}_1,\bm{k}_2,\bm{k}_3,\bm{k}_4}  \Biggl\{ 
    \frac{3\alpha}{2}k_{1,+}k_{4,-} 
    - \frac{\alpha}{2}k_{1,+}k_{3,-} 
\Biggr\}     \gamma_{\mbf{k}_1}^\dagger \gamma_{\mbf{k}_2}^\dagger \gamma_{\mbf{k}_3} \gamma_{\mbf{k}_4}\delta_{\bm{k}_1+\bm{k}_2,\bm{k}_3+\bm{k}_4}\nonumber\\
=&\frac{U}{\Omega_{tot}}  \sum^{\{\bm{k}_1,\bm{k}_2,\bm{k}_3,\bm{k}_4\}}_{\bm{k}_1,\bm{k}_2,\bm{k}_3,\bm{k}_4} 
    \alpha k_{1,+}k_{4,-} 
   \gamma_{\mbf{k}_1}^\dagger \gamma_{\mbf{k}_2}^\dagger \gamma_{\mbf{k}_3} \gamma_{\mbf{k}_4}\delta_{\bm{k}_1+\bm{k}_2,\bm{k}_3+\bm{k}_4},
\end{align}
which is similar to the 1D case in Eq.~\ref{eq:1d_ham_k1234}. 

We can then follow a similar analysis as in  App.~\ref{sec:1d_manybody_gs} and rewrite the Hamiltonian in different forms to bound the many-body energies. 
We define
\begin{align}
    R_{\bm{q}}=\sum^{\{\bm{k},\bm{k}-\bm{q}\}}_{\bm{k}}k_{-}\gamma_{\bm{q-k}}\gamma_{\bm{k}},
\end{align}
in terms of which the interaction reduces to a separable form
\begin{align}
    \hat{H}^\text{int}&\approx\frac{U}{\Omega_{tot}}  \sum_{\bm{k}_1,\bm{k}_4,\bm{q}}^{\{\bm{k}_1,\bm{k}_4,\bm{q}-\bm{k}_1,\bm{q}-\bm{k}_4\}} 
    \alpha k_{1,+}k_{4,-} 
   \gamma_{\mbf{k}_1}^\dagger \gamma_{\bm{q}-\bm{k}_1}^\dagger \gamma_{\bm{q}-\bm{k}_4} \gamma_{\mbf{k}_4}=\frac{\alpha U}{\Omega_{tot}}\sum_{\bm{q}}R^\dagger_{\bm{q}}R_{\bm{q}}.
\end{align}
For repulsive $U>0$, the above form demonstrates that the Hamiltonian is positive semi-definite, so that its ground state energy is bounded from below by zero. 

If $U<0$, which corresponds to an attractive interaction, we can instead reorder the interaction Hamiltonian (expanded to linear order in $\alpha=\beta$) as 
\begin{align}
    \hat{H}^\text{int}\approx&\frac{U}{\Omega_{tot}}  \sum^{\{\bm{k}_1,\bm{k}_2,\bm{k}_3,\bm{k}_4\}}_{\bm{k}_1,\bm{k}_2,\bm{k}_3,\bm{k}_4} 
    \alpha k_{1,+}k_{4,-} \left[-
   \gamma_{\mbf{k}_2}^\dagger \gamma_{\mbf{k}_4}\gamma_{\mbf{k}_1}^\dagger  \gamma_{\mbf{k}_3}\delta_{\bm{k}_1+\bm{k}_2,\bm{k}_3+\bm{k}_4}+\gamma_{\mbf{k}_2}^\dagger  \gamma_{\mbf{k}_3}\delta_{\bm{k}_1+\bm{k}_2,\bm{k}_3+\bm{k}_4}\delta_{\bm{k}_1,\bm{k}_4}\right]\nonumber\\
   =&-\frac{U}{\Omega_{tot}}\left[  \sum^{\{\bm{k}_1,\bm{k}_4,\bm{k}_1+\bm{q},\bm{k}_4+\bm{q}\}}_{\bm{k}_1,\bm{k}_4,\bm{q}} 
    \alpha k_{1,+}k_{4,-} 
   \gamma_{\mbf{k}_4+\bm{q}}^\dagger \gamma_{\mbf{k}_4}\gamma_{\mbf{k}_1}^\dagger  \gamma_{\mbf{k}_1+\bm{q}}-\sum^{\{\bm{k}_1,\bm{k}_2\}} _{\bm{k}_1,\bm{k}_2} 
    \alpha \bm{k}_1^2 \gamma_{\mbf{k}_2}^\dagger  \gamma_{\mbf{k}_2}\right]\nonumber\\
    =&-\frac{\alpha U}{\Omega_{tot}}\sum_{\bm{q}} 
     M_{\bm{q}}^\dagger M_{\bm{q}}+E_{2,m=1} \frac{N_e}{2}\label{eq:2d_semi_pos_form},
\end{align}
where in the last line, we assume that we work in a symmetry sector of fixed particle number $N_e$, and we define
\begin{align}
    M_{\bm{q}}=\sum^{\{\bm{k},\bm{k}+\bm{q}\}}_{\bm{k}}k_{+}\gamma_{\bm{k}}^\dagger\gamma_{\bm{k+q}},\quad E_{2,m=1}&=\frac{2}{\Omega_{tot}}\sum^{\{\bm{k}\}}_{\bm{k}}\alpha U \bm{k}^2\rightarrow \frac{\alpha U}{\pi}\int_0^{k_b}k^3dk.
\end{align}
Above, $E_{2,m=1}$ is consistent with the ground state energy for two particles given in Eq.~\ref{app:eq:E_m_2limit}. 
Eq.~\ref{eq:2d_semi_pos_form} implies that the ground state energy is bounded from below by $\frac{N_e E_{2,m=1}}{2}$, since $-UM^\dagger_{\bm{q}}M_{\bm{q}}$ is positive semi-definite. 
Previously, we have proved that the ansatz $\ket{\phi_{2N}^A}=(\hat{O}_{2,m=1}^\dagger)^N\ket{\text{vac}}$ is an eigenstate of the Hamiltonian with $N_e=2N$ particles and energy $N E_{2,m=1}$. Therefore, $\ket{\phi_{2N}^A}$ is a ground state of Eq.~\ref{eq:2d_semi_pos_form}, and satisfies $M_{\bm{q}}\ket{\phi_{2N}^A}=0$.

We test the validity of our even-electron ansatz (Eq.~\ref{eq:GMPAnsatzGround}) by calculating its wavefunction overlap, $\langle \phi^{ED} | \phi^{A}\rangle$, with the ground state $\ket{\phi^{ED}}$ obtained from ED.
\begin{figure}
    \centering
    \includegraphics[width=0.5\linewidth]{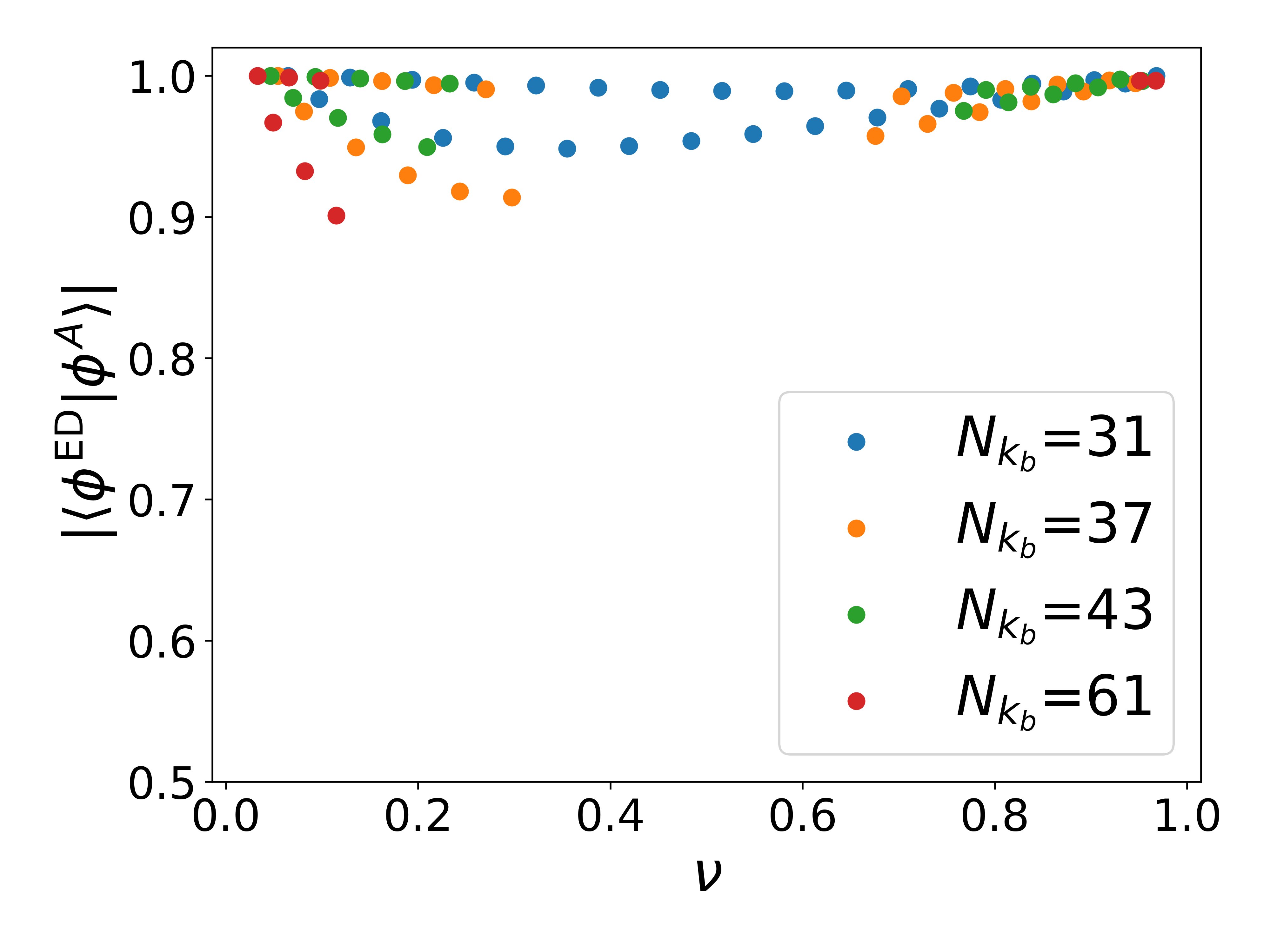}
    \caption{Wavefunction overlap between the analytical ansatz $\ket{\phi^A}$ (Eq.~\ref{eq:GMPAnsatzGround}) and the ED GS $\ket{\phi^{ED}}$ for the attractive 2D Berry Trashcan model with $v=\infty,U=-2/A_b$, $\alpha=\beta$ and $\varphi_{\text{BZ}}$ for different system sizes $N_{k_b}$.}
    \label{fig:overlap_2d_den}
\end{figure}
For the case where $\alpha=\beta$ and $\varphi_{\text{BZ}}=\pi/2$, the overlap between the ansatz and the exact ground state is nearly unity for small even particle numbers, as shown in Fig.~\ref{fig:overlap_2d_den}. For instance, with $N_e\leq 10$ and $N_{k_b}=37,43,61$, the overlap remains above $99\%$. This confirms that the ansatz accurately captures the true ground state in these cases. An overlap calculation across the full range of $\nu$ for $N_{k_b}=31$ reveals that the overlap for even $N_e$ is close to unity near empty and full filling, and has a minimum near half-filling, though the overlap remains large $>98\%$ throughout. This suggests that our ansatz most accurately describes the physics near the empty and full filling regimes. Such behavior is also observed in larger system sizes (Fig.~\ref{fig:overlap_2d_den}) and appears to be a robust feature that persists in the thermodynamic limit.

We discuss the above observations in light of the form of the ansatz. Near empty filling, the state $\hat{O}_{2,m=1}^\dagger\ket{\text{vac}}$ is the exact two-electron ground state, yielding an overlap of 1. Subsequent applications of $\hat{O}_{2,m=1}^\dagger$ generate states with more particles. However, deviations from the exact many-body ground state accumulate due to the non-vanishing second-order commutator $\left[\left[\hat{H}^\text{int},\hat{O}_{2,m=1}^\dagger\right],\hat{O}_{2,m=1}^\dagger\right]=\mathcal{O}((\alpha k_b)^2)$. This motivates why the overlap decreases as the particle number increases from empty filling.

On the hole-doped side, the single-hole state provides another exact reference point. We note that our finite-size momentum meshes, which obey $C_6$ rotation symmetry, all have an odd $N_{k_b}$ because we keep the $C_6$-invariant momentum $\bm{k}=0$. Hence, the single-hole sector has even $N_e$. The interaction-induced hole dispersion $E_{\bm{k}}^h$ (Eq.~\ref{app:eq:2d_hole_dispersion}), generated at full filling, has its minimum at $\bm{k}=\bm{0}$. The exact ground state therefore consists of a single hole at $\bm{k}=0$. In fact, the ansatz (Eq.~\ref{eq:GMPAnsatzGround}) is actually identical to the exact ground state, since the pairing operator $\hat{O}^\dagger_{2,m=1}$ only creates particles at non-zero $\bm{k}$, so $\bm{k}=0$ is always left unoccupied. 


We then extend our analysis to the case where $\alpha > \beta$ (we do not consider $\alpha<\beta$ since we are interested in a purely attractive interaction). As we showed in App.~\ref{app:sec:vf_inf_p_0}, the two-electron ground state in this regime is still well-approximated by the solution for $\alpha=\beta$. We therefore use the $\alpha=\beta$ ansatz and evaluate its overlap with the ED ground state for $\alpha>\beta$. The results are shown in Fig.~\ref{fig:overlap_main_2d}(c) of the main text. The high overlap (for example, the overlap remains $\gtrsim80\%$ for even $N_e\leq10$ and $N_{k_b}=43$, for $\alpha=2\beta$ with $\varphi_\text{BZ}=\pi/2$) demonstrates that the ansatz remains a robust approximation even when $\alpha$ becomes larger than $\beta$, which corresponds to a finite-range exponentially-decaying interaction.

Finally, we comment that when the band has completely trivial form factors ($\beta=0$), we obtain to first order in $\alpha$
\begin{align}
    \hat{H}^\text{int}
\approx&\frac{U\alpha}{2\Omega_{tot}}  \sum_{\bm{k}_1,\bm{k}_2,\bm{k}_3,\bm{k}_4}^{\{\bm{k}_1,\bm{k}_2,\bm{k}_3,\bm{k}_4\}}
    \left(
   k_{1,+}k_{4,-}+k_{1,-}k_{4,+} 
\right)
   \gamma_{\mbf{k}_1}^\dagger \gamma_{\mbf{k}_2}^\dagger \gamma_{\mbf{k}_3} \gamma_{\mbf{k}_4}\delta_{\bm{k}_1+\bm{k}_2,\bm{k}_3+\bm{k}_4},
\end{align}
which restores time-reversal symmetry in the Hamiltonian. 

\subsubsection{Ground State Ansatz For Odd $N_e$, $v=\infty$, $\bm{p}=0$}\label{app:sec:gs_ansatz_odd_2d}

Based on the even $N_e$ ansatz, we study the odd-particle ground states. To begin with, we compute the commutator between $\hat{H}^\text{int}$ and $\hat{\gamma}_0^\dagger$. Following the 1d discussion, we first consider the Hamiltonian with the form in Eq.~\ref{eq:2d_semi_pos_form}. Using the commutators $[M_{\bm{q}},\gamma_{\bm{p}}^\dagger]=\sum_{\bm{k}}^{\{\bm{k},\bm{k}+\bm{q}\}}k_+\gamma^\dagger_{\bm{k}}\delta_{\bm{p},\bm{k}+\bm{q}}$ and $[M^\dagger_{\bm{q}},\gamma_{\bm{p}}^\dagger]=\sum_{\bm{k}}^{\{\bm{k},\bm{k}+\bm{q}\}}k_-\gamma^\dagger_{\bm{k}+\bm{q}}\delta_{\bm{p},\bm{k}}$, we obtain
\begin{align}
    [\hat{H}^\text{int}, \gamma_{\bm{p}}^{\dagger}] &=-\frac{U}{\Omega_{tot}}\sum_{\bm{q}}\alpha\left([M_{\bm{q}}^\dagger,\gamma_{\bm{p}}^\dagger] M_{\bm{q}}+M_{\bm{q}}^\dagger[M_{\bm{q}},\gamma_{\bm{p}}^\dagger]\right)+\frac{E_{2,m=1}}{2}[\sum_{\bm{k}}\gamma_{\bm{k}}^\dagger\gamma_{\bm{k}},\gamma_{\bm{p}}^\dagger]\\
    &=-\frac{\alpha U}{\Omega_{tot}}\left(\sum^{\{\bm{p},\bm{p}+\bm{q}\}}_{\bm{q}}p_-\gamma_{\bm{p}+\bm{q}}^\dagger M_{\bm{q}}+\sum^{\{\bm{p},\bm{p}-\bm{q}\}}_{\bm{q}}(p-q)_{+}M_{\bm{q}}^\dagger\gamma_{\bm{p-q}}^\dagger)\right)+\frac{E_{2,m=1}}{2}\gamma_{\bm{p}}^\dagger.
\end{align}
Acting it on the even-electron ground state, which is annihilated by $M_{\bm{q}}$, we obtain
\begin{align}
    [\hat{H}^\text{int}, \gamma_{\bm{p}}^{\dagger}]\ket{\phi_{2N}^A}=-\frac{U}{\Omega_{tot}}\sum^{\{\bm{p},\bm{p}-\bm{q}\}}_{\bm{q}}\alpha(p-q)_{+}M_{\bm{q}}^\dagger\gamma_{\bm{p-q}}^\dagger\ket{\phi_{2N}^A}+\frac{E_{2,m=1}}{2}\gamma_{\bm{p}}^\dagger\ket{\phi_{2N}^A}.
\end{align}
Similar to the 1D case in App.~\ref{appsubsec:oddNe_vFinfty}, the first term is a complicated scattering term which makes odd-electron problem not exactly solvable.

To proceed, we propose the odd-particle ground state ansatz
\begin{align}
    \ket{\phi_{2N+1}^A}\propto\gamma_{\bm{0}}^\dagger(\hat{O}_2^\dagger)^N\ket{\text{vac}},
    \label{app:eq:odd_ansatz_2d}
\end{align}
which is built out of the even-particle ansatz created by $(\hat{O}_2^\dagger)^N$.
The choice of taking the momentum of the `unpaired electron' $\gamma^\dagger_{\bm{0}}$ as $\bm{p}=0$ can be motivated as follows. If the unpaired electron has momentum $\bm{p}$, this would pose an obstruction to forming electron pairs out of $\pm\bm{p}$ momenta. Since the even-electron ansatz has no pairing or occupation at zero momentum, creating an additional electron at $\bm{p}=0$ does not `disturb' the pairing of the other electrons.


To test the validity of our proposed odd-particle ansatz, we first calculate its overlap with the exact wavefunction from ED, as shown in Fig.~\ref{fig:overlap_2d_den}. While the overlap for odd $N_e$ is not as large as for even $N_e$, we find that it remains high. For example, for $N_e\leq 7$ and $N_{k_b}=37,43,61$, the overlap remains above $85\%$. An overlap calculation across the full range of $\nu$ for $N_{k_b}=31$ shows that the overlap for odd $N_e$ remains $>95\%$ throughout.

\begin{figure}
    \centering
    \includegraphics[width=1.0\linewidth]{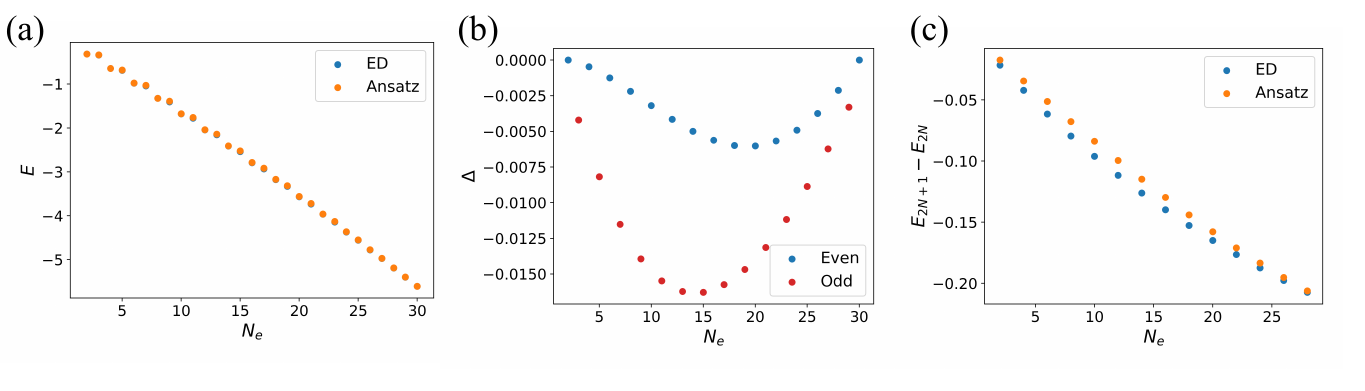}
    \caption{(a) Energy of the ED ground state $E^{ED}$ and energy expectation value of the ansatz $E^A=\bra{\phi^A }\hat{H}^{\text{int}}\ket{\phi^A}$ (which both have zero total momentum) for the attractive 2D Berry Trashcan with $v=\infty$, $\varphi_\text{BZ}=\pi/2$, and $N_{k_b}=31$. (b) The energy deviation, defined as $\Delta = E^{ED} - E^{A}$ , plotted as a function of the electron number $N_e$. Red and blue markers distinguish between systems with even and odd numbers of electrons, respectively. Note that our ansatz is is exact (and hence $\Delta=0$) for two electrons, and a single hole on top of full filling. (c) Comparison between the single-charge excitation energy $E_{2N+1}-E_{2N}$ extracted using ED, and using the energy expectation value of the ansatz.}
    \label{fig:expectation_e_2d}
\end{figure}

We also calculate the energy expectation value $E^A=\bra{\phi^A_{N_e}}\hat{H}^\text{int}\ket{\phi^A_{N_e}}$ of the ansatz, and compare it with the exact ground state energy $E^{ED}$ obtained in ED. Figs.~\ref{fig:expectation_e_2d}(a) and (b) show this comparison for various $N_e$. We find excellent agreement between the ansatz and the ED results. The energy deviation, defined as $\Delta = E^{\text{ED}} - E^{\text{A}}$, is orders of magnitude smaller than the total ground-state energy. This deviation $\Delta$ reaches a maximum near half-filling and becomes smallest near the empty- and full-filling limits. This observation suggests that our ansatz most accurately describes the ground state in these low- and high-density regimes. This behavior is consistent with the overlap calculations (Fig.~\ref{fig:overlap_2d_den}), where the odd-particle overlap is near-unity around empty- and full-filling, but minimal at half-filling. This contrasts with the 1D case, where the odd-particle ansatz performs best in the full-filling limit, but not so well near empty-filling.

In Fig.~\ref{fig:expectation_e_2d}(c), we also calculate the single-particle excitation energy, $E_{2N+1}-E_{2N}$, for adding a particle to the even-particle ground state with $N_e=2N$. We compare the result from ED, and the result from taking the energy expectation value of the ansatz. 
The results from our ansatz again exhibit good agreement with those of the exact ground state. The excitation energy increases with the electron number $N_e$ and eventually saturates to the value $E_{2,m=1}$ in the full-filling limit. This trend is qualitatively similar to the behavior observed in the 1D case.

In the following, we motivate why the odd-electron ansatz approximates well the exact ground state.
We begin by discussing the high accuracy of the ansatz in the full-filling limit. As shown in Fig.~\ref{fig:overlap_2d_den}, the overlap in this regime is notably better than that observed near empty filling.
To understand this, we first examine the state with $2n$ holes on top of full filling (note that this corresponds to odd $N_e$, since $N_{k_b}$ is always an odd integer). According to our ansatz (Eq.~\ref{app:eq:odd_ansatz_2d}), the wavefunction for $N_e=N_{k_b}-2n$ can be expressed as
\begin{align}
    \ket{\phi_{N_{k_b}-2n}^{A}}&=\frac{1}{Z}\sum_{\substack{\bm{k}_1, \cdots, \bm{k}_{(N_{k_b}-1-2n)/2}}}^{\{\bm{k}_1, \cdots, \bm{k}_{(N_{k_b}-1-2n)/2}\}}k_{1,+}e^{-\alpha \bm{k}_{1}^2}\cdots k_{{(N_{k_b}-1-2n)/2},+}e^{-\alpha \bm{k}_{(N_{k_b}-1-2n)/2}^2}\ket{\bm{0},\pm\bm{k}_1,\cdots,\pm\bm{k}_{(N_{k_b}-1-2n)/2}}
\end{align}
where $Z$ is a normalization factor, and $\ket{\bm{0},\pm\bm{k}_1,\cdots,\pm\bm{k}_{(N_{k_b}-1-2n)/2}}$ is a many-body Fock basis state where the occupied single-particle momenta are indicated. Following the strategy in App.~\ref{appsubsec:oddNe_vFinfty}, we can equivalently express this in terms of the unoccupied momenta (the `hole' momenta). To this end, we introduce the notation $\ket{\pm\bm{k}_1',\cdots,\pm\bm{k}'_{n}}_h$, which represents a many-body Fock basis state by the unoccupied momenta. In terms of the $\ket{\pm\bm{k}_1',\cdots,\pm\bm{k}'_{n}}_h$, we find
\begin{align}
    \ket{\phi_{N_{k_b}-2n}^{A}}&=\frac{1}{Z}\sum_{\bm{k}_1',\cdots\bm{k}'_n}^{\{\bm{k}_1',\cdots\bm{k}'_n\}}\frac{\prod_{\bm{k}_j \in \mathcal{H}_U}k_{j}e^{-\alpha \bm{k}_j^2}}{\prod_{i=1}^nk'_{i,+}e^{-\alpha \bm{k}_i'^2}}\ket{\pm\bm{k}_1',\cdots,\pm\bm{k}'_{n}}_h\nonumber\\
    &=\frac{1}{Z'}\sum_{\bm{k}_1',\cdots\bm{k}'_n}^{\{\bm{k}_1',\cdots\bm{k}'_n\}}\frac{1}{\prod_{i=1}^nk'_{i,+}e^{-\alpha \bm{k}_i'^2}}\ket{\pm\bm{k}_1',\cdots,\pm\bm{k}'_{n}}_h.
\end{align}
$\mathcal{H}_U$ represents the upper half of the momenta lying within the trashcan bottom (excluding $\bm{k}=0$), such that only one of $\pm\bm{k}$ is included.  

If $n=1$ (i.e.~two holes), then the amplitude in $\ket{\phi_{N_{k_b}-2}^{A}}$ for having a single pair of holes at $\pm\bm{k}'$ is
\begin{equation}
c_{\bm{k}'} \propto \frac{1}{k'_+e^{-\alpha \bm{k}'^2}} \approx \frac{1}{k'_+}, \quad \text{for small } \alpha k_b^2.
\end{equation}
Recall from App.~\ref{app:sec:hole_doping} that the actual ground state wavefunction for a single pair of holes, $\psi_{1,g}$, is approximately
\begin{equation}
\psi_{1,g}(\bm{k}') \propto \int_{|\bm{k}'|\leq k_b} d^2\bm{k}'\frac{e^{-\alpha k'^2-i\varphi_{k'}}}{ k'}\ket{\bm{k}} \propto \frac{1}{k_+'}.
\end{equation}
The above form matches our ansatz in for small $\alpha k_b^2$. This agreement for a single pair of holes explains the high overlap observed for states near full filling.

For the empty-filling side, we calculate the commutator of the interaction Hamiltonian $\hat{H}^\text{int}$ with the creation operator $\gamma_{\bm{0}}^\dagger$
\begin{align} \label{eq:GammaCom1_2d_1}
[\hat{H}^\text{int}, \gamma_{\bm{0}}^{\dagger}] &=\frac{U}{2\Omega_{tot}} \sum_{\mbf{k}, \mbf{k'}, \mbf{q}}^{\{\mbf{k}, \mbf{k'}, \mbf{k} + \mbf{q}, \mbf{k'} - \mbf{q} \} } e^{-\alpha \bm{q}^2-i \beta (  \mbf{q}\times (\mbf{k}- \mbf{k'})) }  [\gamma_{\mbf{k} + \mbf{q}}^{\dagger} \gamma_{\mbf{k'} - \mbf{q}}^{\dagger} \gamma_{\mbf{k'}} \gamma_{\mbf{k}}, \gamma_{\bm{0}}^{\dagger}]\nonumber\\
&=\frac{U}{2\Omega_{tot}} \sum_{\mbf{k}, \mbf{k'}, \mbf{q}}^{\{\mbf{k}, \mbf{k'}, \mbf{k} + \mbf{q}, \mbf{k'} - \mbf{q} \} } e^{-\alpha \bm{q}^2-i \beta (  \mbf{q}\times (\mbf{k}- \mbf{k'})) }  (-\delta_{\mbf{k'},\bm{0}} \gamma_{\mbf{k} + \mbf{q}}^{\dagger} \gamma_{\mbf{k'} - \mbf{q}}^{\dagger} \gamma_{\mbf{k}} + \delta_{\mbf{k},\bm{0}} \gamma_{\mbf{k} + \mbf{q}}^{\dagger} \gamma_{\mbf{k'} - \mbf{q}}^{\dagger} \gamma_{\mbf{k'}})\nonumber\\
&=-\frac{U}{\Omega_{tot}} \sum_{\mbf{k}, \mbf{q}}^{\{\mbf{k}, \mbf{k} + \mbf{q},  - \mbf{q} \} } e^{-\alpha \bm{q}^2-i \beta \mbf{q}\times \mbf{k} }   \gamma_{\mbf{k} + \mbf{q}}^{\dagger} \gamma_{ - \mbf{q}}^{\dagger} \gamma_{\mbf{k}}.
\end{align}
If the commutator above vanished, then acting on the even-particle ansatz with $\gamma^\dagger_{\bm{0}}$ would leave the energy unchanged. The fact that the commutator is non-zero leads to deviations in the energy of the ansatz for $2N$ and $2N+1$ particles. However, for small $N_e$, the summation over $\bm{k}$ above is restricted to only $\leq N_e$ momenta when acting on the even-particle ansatz. 
This suggests that the commutator above has a relatively small effect for small $N_e$. Near empty-filling, the energy of the odd-electron state would then be nearly degenerate with the even-electron ground state, $E_{2N+1}\approx E_{2N}$, which is consistent with our observed excitation energies which has minimum absolute value near empty filling.

\subsubsection{Generalization of the RSGA to 2D Trashcan Hamiltonians}\label{app:sec:rsga_2d_generalization}

Recall that in App.~\ref{app:sec:gs_even_2d}, we expanded the interaction Hamiltonian 
\begin{align}
    \hat{H}^\text{int}=& \frac{U}{2\Omega_{tot}}  \sum_{\bm{k}_1,\bm{k}_2,\bm{k}_3,\bm{k}_4}^{\{\bm{k}_1,\bm{k}_2,\bm{k}_3,\bm{k}_4\}}  e^{-\alpha (\bm{k}_1-\bm{k}_4)^2-i \beta (  \mbf{k_1-k_4})\times (\mbf{k_4}- \mbf{k_3}) }     \gamma_{\mbf{k}_1}^\dagger \gamma_{\mbf{k}_2}^\dagger \gamma_{\mbf{k}_3} \gamma_{\mbf{k}_4}\delta_{\bm{k}_1+\bm{k}_2,\bm{k}_3+\bm{k}_4}
\end{align}
 to the first order in $\alpha$ and $\beta$, and found that if $\alpha=\beta$, the Hamiltonian can be written in a negative semi-definite form (for attractive $U<0$)
\begin{align}
\hat{H}^\text{int}\approx&\frac{U}{\Omega_{tot}}  \sum_{\bm{k}_1,\bm{k}_2,\bm{k}_3,\bm{k}_4}^{\{\bm{k}_1,\bm{k}_2,\bm{k}_3,\bm{k}_4\}} 
    \alpha k_{1,+}k_{4,-} 
   \gamma_{\mbf{k}_1}^\dagger \gamma_{\mbf{k}_2}^\dagger \gamma_{\mbf{k}_3} \gamma_{\mbf{k}_4}\delta_{\bm{k}_1+\bm{k}_2,\bm{k}_3+\bm{k}_4}=\frac{\alpha U}{\Omega_{tot}}\sum_{\bm{q}}R^\dagger_{\bm{q}}R_{\bm{q}}\label{appeq:2d_Ham_RqRq_reminder}
\end{align}
where $R_{\bm{q}}=\sum^{\{\bm{k},\bm{q}-\bm{k}\}}_{\bm{k}}k_{-}\gamma_{\bm{q-k}}\gamma_{\bm{k}}$. We denote the antisymmetrized version of $R_{\bm{q}}^\dagger$ as $P^\dagger_{\bm{q}}$
\begin{align}
    P_{\bm{q}}^\dagger=\frac{1}{2}\sum_{\bm{k}}^{\{\bm{k},\bm{q}-\bm{k}\}}(2k_{+}-q_{+})\gamma_{\bm{k}}^\dagger\gamma_{\bm{q-k}}^\dagger=\frac{1}{2}\sum_{\bm{k}_1,\bm{k}_2}^{\{\bm{k}_1,\bm{k}_2\}}(k_{1,+}-k_{2,+})\gamma_{\bm{k}_1}^\dagger\gamma_{\bm{k}_2}^\dagger\delta_{\bm{k}_1+\bm{k}_2,\bm{q}},
\end{align}
which yields a similar form to that of the 1D case as discussed in App.~\ref{appsubsec:RSGA_generalization}. 

To study the generalized RSGA with finite momenta, we first prove that $P_{\bm{p}}^\dagger\ket{\text{vac}}$ is the ground state for two electrons in the sector with total momentum $\bm{p}$
\begin{align}
    P_{\bm{q}}P_{\bm{p}}^\dagger\ket{\text{vac}}&=\frac{1}{4}\sum_{\bm{k}_1,\bm{k}_2,\bm{k}_3,\bm{k}_4}^{\{\bm{k}_1,\bm{k}_2,\bm{k}_3,\bm{k}_4\}}(k_{1,-}-k_{2,-})(k_{3,+}-k_{4,+})\gamma_{\bm{k}_2}\gamma_{\bm{k}_1}\gamma_{\bm{k}_3}^\dagger\gamma_{\bm{k}_4}^\dagger\delta_{\bm{k}_3+\bm{k}_4,\bm{p}}\delta_{\bm{k}_1+\bm{k}_2,\bm{q}}\ket{\text{vac}}\nonumber\\
    &=\frac{1}{4}\sum_{\bm{k}_1,\bm{k}_2,\bm{k}_3,\bm{k}_4}^{\{\bm{k}_1,\bm{k}_2,\bm{k}_3,\bm{k}_4\}}(k_{1,-}-k_{2,-})(k_{3,+}-k_{4,+})(\delta_{\bm{k}_2,\bm{k}_4}\delta_{\bm{k}_1,\bm{k}_3}-\delta_{\bm{k}_2,\bm{k}_3}\delta_{\bm{k}_1,\bm{k}_4})\delta_{\bm{k}_3+\bm{k}_4,\bm{p}}\delta_{\bm{k}_1+\bm{k}_2,\bm{q}}\ket{\text{vac}}\nonumber\\
    &=\frac{1}{2}\sum_{\bm{k}_1,\bm{k}_2}^{\{\bm{k}_1,\bm{k}_2\}}(k_{1,-}-k_{2,-})(k_{1,+}-k_{2,+})\delta_{\bm{p},\bm{q}}\delta_{\bm{k}_1+\bm{k}_2,\bm{q}}\ket{\text{vac}}.
\end{align}
Acting the Hamiltonian on $P_{\bm{q}}^\dagger\ket{\text{vac}}$, we obtain
\begin{align}
    \hat{H}^\text{int}P_{\bm{p}}^\dagger\ket{\text{vac}}=\frac{\alpha U}{\Omega_{tot}}\sum_{\bm{q}}P_{\bm{q}}^\dagger P_{\bm{q}}P_{\bm{p}}^\dagger\ket{\text{vac}}=\frac{\alpha U}{2\Omega_{tot}}\sum_{\bm{k}_1,\bm{k}_2}^{\{\bm{k}_1,\bm{k}_2\}}(k_{1,-}-k_{2,-})(k_{1,+}-k_{2,+})\delta_{\bm{k}_1+\bm{k}_2,\bm{p}}P_{\bm{p}}^\dagger\ket{\text{vac}}\equiv E_{2,\bm{p}}P_{\bm{p}}^\dagger\ket{\text{vac}}.
\end{align}
Thus, we have proved that $P_{\bm{p}}^\dagger\ket{\text{vac}}$ is a 2-electron eigenstate with energy $E_{2,\bm{p}}=\frac{\alpha U}{2\Omega_{tot}}\sum_{\bm{k}_1,\bm{k}_2}(k_{1,-}-k_{2,-})(k_{1,+}-k_{2,+})\delta_{\bm{k}_1+\bm{k}_2,\bm{p}}$. 
Furthermore, the final form of the Hamiltonian in Eq.~\ref{appeq:2d_Ham_RqRq_reminder} implies that the Hamiltonian has rank 1 (i.e.~only one finite eigenvalue) within each momentum sector for two electrons.
Thus, if the interaction is attractive (i.e., $U<0$), then $P_{\bm{p}}^\dagger\ket{\text{vac}}$ is the ground state for the momentum sector $\bm{p}$. We can evaluate the energy of this state $E_{2,\bm{p}}$ in the continuum limit. For small total momentum $\bm{p}$, the integration domain for the relative momentum $\bm{k}_1-\bm{k}_2$ can be approximated as the area encircled by the red dashed line in Fig.~\ref{fig:finite_mom_region} (same approximation as in App.~\ref{appsubsec:2d_vFinf_finitep}),
yielding the energy
\begin{align}
    E_{2,\bm{p}}&\approx\frac{\alpha U}{2(2\pi)^2}\int_0^{k-\frac{p}{2}} dk4k^3\int_0^{2\pi}d\theta=\frac{\alpha U(k_b-p/2)^4}{4\pi}.
\end{align}
Notably, this expression is identical to the result in Eq.~\ref{app:eq:2d_dispersion_2e} when expanded to the first order in $\alpha$. This yields a linear dispersion at small momentum, which is consistent with the ED results in Fig.~\ref{fig:2d_dipsersion_all_mom}.

We now proceed to study the higher-order commutators. We first notice that similar to the 1D case, we trivially have
\begin{align}
\left[\left[\left[\hat{H}^\text{int},P_{\bm{q}_1}^\dagger\right],P_{\bm{q}_2}^\dagger\right],\gamma_{\bm{k}}^\dagger\right]=0\quad \Rightarrow  \quad \left[\left[\left[\hat{H}^\text{int},P_{\bm{q}_1}^\dagger\right],P_{\bm{q}_2}^\dagger\right],P_{\bm{q}_3}^\dagger\right]=0.
\end{align}
Therefore, we only need to study $\left[\left[\hat{H}^\text{int},P_{\bm{q}_1}^\dagger\right],P_{\bm{q}_2}^\dagger\right]$. To compute this, we study the higher-order commutators of the $P_{\bm{q}}$ operators. We trivially have $[P_{\bm{q}_1}^\dagger,P_{\bm{q}_2}^\dagger]=0$. We also find 
\begin{align}
    [P_{\bm{q}},P_{\bm{q}_1}^\dagger]=&\frac{1}{4}\sum^{\{\bm{k}_1,\bm{k}_2\}}_{\bm{k}_1,\bm{k}_2}\sum^{\{\bm{k}_3,\bm{k}_4\}}_{\bm{k}_3,\bm{k}_4}\left(k_{1,-}-k_{2,-}\right)\left(k_{3,+}-k_{4,+}\right)\delta_{\bm{k}_1+\bm{k}_2,\bm{q}}\delta_{\bm{k}_3+\bm{k}_4,\bm{q}_1}\left[\gamma_{\bm{k}_2}\gamma_{\bm{k}_1},\gamma_{\bm{k}_3}^\dagger\gamma_{\bm{k}_4}^\dagger\right]\nonumber\\
    =&\frac{1}{4}\sum^{\{\bm{k}_1,\bm{k}_2\}}_{\bm{k}_1,\bm{k}_2}\sum^{\{\bm{k}_3,\bm{k}_4\}}_{\bm{k}_3,\bm{k}_4}\left(k_{1,-}-k_{2,-}\right)\left(k_{3,+}-k_{4,+}\right)\delta_{\bm{k}_1+\bm{k}_2,\bm{q}}\delta_{\bm{k}_3+\bm{k}_4,\bm{q}_1}\Big(-\delta_{\bm{k}_1, \bm{k}_3}\gamma_{\bm{k}_4}^\dagger\gamma_{\bm{k}_2} + \delta_{\bm{k}_1, \bm{k}_4}\gamma_{\bm{k}_3}^\dagger\gamma_{\bm{k}_2} \nonumber\\
    &+ \delta_{\bm{k}_2 ,\bm{k}_3}\gamma_{\bm{k}_4}^\dagger\gamma_{\bm{k}_1} - \delta_{\bm{k}_2, \bm{k}_4}\gamma_{\bm{k}_3}^\dagger\gamma_{\bm{k}_1}+\delta_{\bm{k}_1, \bm{k}_3}\delta_{\bm{k}_2,\bm{k}_4}-\delta_{\bm{k}_1 \bm{k}_4}\delta_{\bm{k}_2, \bm{k}_3}\Big)
\end{align}
\begin{align}
    \left[\left[P_{\bm{q}},P_{\bm{q}_1}^\dagger\right],P_{\bm{q}_2}^\dagger\right]=&\frac{1}{8}\sum_{\bm{k}_{1},\cdots,\bm{k}_{6}}^{\{\bm{k}_1,\ldots,\bm{k}_6\}}\left(k_{1,-}-k_{2,-}\right)\left(k_{3,+}-k_{4,+}\right)\left(k_{5,+}-k_{6,+}\right)\delta_{\bm{k}_1+\bm{k}_2,\bm{q}}\delta_{\bm{k}_3+\bm{k}_4,\bm{q}_1}\delta_{\bm{k}_5+\bm{k}_6,\bm{q}_2}\nonumber\\
    &\Big[(\delta_{\bm{k}_2, \bm{k}_3}\delta_{\bm{k}_1,\bm{k}_5}-\delta_{\bm{k}_1, \bm{k}_3}\delta_{\bm{k}_2,\bm{k}_5})\gamma_{\bm{k}_4}^\dagger\gamma_{\bm{k}_6}^\dagger+(\delta_{\bm{k}_1, \bm{k}_3}\delta_{\bm{k}_2,\bm{k}_6}-\delta_{\bm{k}_2, \bm{k}_3}\delta_{\bm{k}_1,\bm{k}_6})\gamma_{\bm{k}_4}^\dagger\gamma_{\bm{k}_5}^\dagger\nonumber\\
    &+ (\delta_{\bm{k}_1, \bm{k}_4}\delta_{\bm{k}_2,\bm{k}_5}-\delta_{\bm{k}_2, \bm{k}_4}\delta_{\bm{k}_1,\bm{k}_5})\gamma_{\bm{k}_3}^\dagger\gamma_{\bm{k}_6}^\dagger+(\delta_{\bm{k}_2, \bm{k}_4}\delta_{\bm{k}_1,\bm{k}_6}-\delta_{\bm{k}_1, \bm{k}_4}\delta_{\bm{k}_2,\bm{k}_6})\gamma_{\bm{k}_3}^\dagger\gamma_{\bm{k}_5}^\dagger\Big]\nonumber\\
    &=\sum_{\bm{k}_{1},\cdots,\bm{k}_{6}}^{\{\bm{k}_1,\ldots,\bm{k}_6\}}\left(k_{1,-}-k_{2,-}\right)\left(k_{3,+}-k_{4,+}\right)\left(k_{5,+}-k_{6,+}\right)\delta_{\bm{k}_1+\bm{k}_2,\bm{q}}\delta_{\bm{k}_3+\bm{k}_4,\bm{q}_1}\delta_{\bm{k}_5+\bm{k}_6,\bm{q}_2}\delta_{\bm{k}_1,\bm{k}_5}\delta_{\bm{k}_2,\bm{k}_3}\gamma_{\bm{k}_4}^\dagger\gamma_{\bm{k}_6}^\dagger.
\end{align}
\begin{align}
    \left[\left[\hat{H}^\text{int},P_{\bm{q}_1}^\dagger\right],P_{\bm{q}_2}^\dagger\right]=&\frac{\alpha U}{\Omega_{tot}}\sum^{\{\bm{k}_1,\ldots,\bm{k}_6\}}_{\bm{k}_{1},\cdots,\bm{k}_{6},\bm{q}}\left(k_{1,-}-k_{2,-}\right)\left(k_{3,+}-k_{4,+}\right)\left(k_{5,+}-k_{6,+}\right)\delta_{\bm{k}_1+\bm{k}_2,\bm{q}}\delta_{\bm{k}_3+\bm{k}_4,\bm{q}_1}\delta_{\bm{k}_5+\bm{k}_6,\bm{q}_2}\delta_{\bm{k}_1,\bm{k}_5}\delta_{\bm{k}_2,\bm{k}_3}P_{\bm{q}}^\dagger\gamma_{\bm{k}_4}^\dagger\gamma_{\bm{k}_6}^\dagger\nonumber\\
    =&\frac{\alpha U}{2\Omega_{tot}}\sum^{\{\bm{k}_1,\ldots,\bm{k}_8\}}_{\bm{k}_{1},\cdots,\bm{k}_{8},\bm{q}}\left(k_{1,-}-k_{2,-}\right)\left(k_{3,+}-k_{4,+}\right)\left(k_{5,+}-k_{6,+}\right)\left(k_{7,+}-k_{8,+}\right)\delta_{\bm{k}_1+\bm{k}_2,\bm{q}}\delta_{\bm{k}_3+\bm{k}_4,\bm{q}_1}\delta_{\bm{k}_5+\bm{k}_6,\bm{q}_2}\delta_{\bm{k}_7+\bm{k}_8,\bm{q}}\nonumber\\
    &\delta_{\bm{k}_1,\bm{k}_5}\delta_{\bm{k}_2,\bm{k}_3}\gamma_{\bm{k}_7}^\dagger\gamma_{\bm{k}_8}^\dagger\gamma_{\bm{k}_4}^\dagger\gamma_{\bm{k}_6}^\dagger\nonumber\\
    =&\frac{\alpha U}{2\Omega_{tot}}\sum^{\{\bm{k}_7,\bm{k}_8,\bm{k}_4,\bm{k}_6\}}_{\bm{k}_7,\bm{k}_8,\bm{k}_4,\bm{k}_6}W_{\bm{k}_7,\bm{k}_8,\bm{k}_4,\bm{k}_6}^{\bm{q}_1,\bm{q}_2}\gamma_{\bm{k}_7}^\dagger\gamma_{\bm{k}_8}^\dagger\gamma_{\bm{k}_4}^\dagger\gamma_{\bm{k}_6}^\dagger,
\end{align}
with $W_{\bm{k}_7,\bm{k}_8,\bm{k}_4,\bm{k}_6}^{\bm{q}_1,\bm{q}_2}$ defined as
\begin{align}
    W_{\bm{k}_7,\bm{k}_8,\bm{k}_4,\bm{k}_6}^{\bm{q}_1,\bm{q}_2}&=\sum^{\{\bm{k}_1,\bm{k}_2,\bm{k}_3,\bm{k}_5\}}_{\bm{k}_1,\bm{k}_2,\bm{k}_3,\bm{k}_5,\bm{q}}\left(k_{1,-}-k_{2,-}\right)\left(k_{3,+}-k_{4,+}\right)\left(k_{5,+}-k_{6,+}\right)\left(k_{7,+}-k_{8,+}\right)\nonumber \\
    &\quad\quad\times\delta_{\bm{k}_1+\bm{k}_2,\bm{q}}\delta_{\bm{k}_3+\bm{k}_4,\bm{q}_1}\delta_{\bm{k}_5+\bm{k}_6,\bm{q}_2}\delta_{\bm{k}_7+\bm{k}_8,\bm{q}}\delta_{\bm{k}_1,\bm{k}_5}\delta_{\bm{k}_2,\bm{k}_3}
    \nonumber
    \\&=\left(k_{7,+}-k_{8,+}\right)\sum^{\{\bm{k}_{3},\bm{k}_{5}\}}_{\bm{k}_{3},\bm{k}_{5}}\left(k_{5,-}-k_{3,-}\right)\left(k_{3,+}-k_{4,+}\right)\left(k_{5,+}-k_{6,+}\right)\delta_{\bm{k}_7+\bm{k}_8,\bm{k}_5+\bm{k}_3}\delta_{\bm{k}_3+\bm{k}_4,\bm{q}_1}\delta_{\bm{k}_5+\bm{k}_6,\bm{q}_2}\nonumber\\
    &=\left(k_{7,+}-k_{8,+}\right)\left(q_{2,-}-q_{1,-}+k_{4,-}
    -k_{6,-}\right)\left(q_{1,+}-2k_{4,+}\right)\left(q_{2,+}-2k_{6,+}\right)\nonumber\\
    &\quad\quad\times \delta_{\bm{k}_7+\bm{k}_8+\bm{k}_4+\bm{k}_6,\bm{q}_1+\bm{q}_2}\delta_{\bm{q}_1-\bm{k}_4\in\mathcal{H}}\delta_{\bm{q}_2-\bm{k}_6\in\mathcal{H}}.
\end{align}
Above, the symbol $\delta_{\bm{k}\in\mathcal{H}}$ takes the value $1$ ($0$) if $\bm{k}$ lies inside (outside) the trashcan bottom. Note that since $W_{\bm{k}_7,\bm{k}_8,\bm{k}_4,\bm{k}_6}^{\bm{q}_1,\bm{q}_2}$ is contracted with a fully antisymmetric product $\gamma_{\bm{k}_7}^\dagger\gamma_{\bm{k}_8}^\dagger\gamma_{\bm{k}_4}^\dagger\gamma_{\bm{k}_6}^\dagger$, we are only interested in the fully antisymmetric part (i.e.~we allow ourselves to freely add/subtract partially symmetric components to $W$). 
A simple relabeling gives
\begin{align}
    W_{\bm{k}_4,\bm{k}_3,\bm{k}_2,\bm{k}_1}^{\bm{q}_1,\bm{q}_2}&=\left(k_{4,+}-k_{3,+}\right)\left(q_{2,-}-q_{1,-}+k_{2,-}
    -k_{1,-}\right)\left(q_{1,+}-2k_{2,+}\right)\left(q_{2,+}-2k_{1,+}\right)\nonumber\\
    &\quad\quad\times\delta_{\bm{k}_4+\bm{k}_3+\bm{k}_2+\bm{k}_1,\bm{q}_1+\bm{q}_2}\delta_{\bm{q}_1-\bm{k}_2\in\mathcal{H}}\delta_{\bm{q}_2-\bm{k}_1\in\mathcal{H}}\nonumber\\
    &=\left(k_{4,+}-k_{3,+}\right)\left(q_{1,-}-q_{2,-}+k_{2,-}
    -k_{1,-}\right)\left(q_{1,+}-2k_{1,+}\right)\left(q_{2,+}-2k_{2,+}\right)\nonumber\\
    &\quad\quad\times\delta_{\bm{k}_4+\bm{k}_3+\bm{k}_2+\bm{k}_1,\bm{q}_1+\bm{q}_2}\delta_{\bm{q}_1-\bm{k}_1\in\mathcal{H}}\delta_{\bm{q}_2-\bm{k}_2\in\mathcal{H}}.
\end{align}
In the second equation, exploited the antisymmetry to interchange $\bm{k}_1$ and $\bm{k}_2$. Analogous to the 1D case, if there is no cutoff on the single-particle momenta (i.e.~$\delta_{\bm{q}_1-\bm{k}_1\in\mathcal{H}}\delta_{\bm{q}_2-\bm{k}_2\in\mathcal{H}}$ is always 1), then $W_{\bm{k}_4,\bm{k}_3,\bm{k}_2,\bm{k}_1}^{\bm{q}_1,\bm{q}_2}$ vanishes for all pair momentum $\bm{q}_1$ and $\bm{q}_2$. 
As proved in App.~\ref{appsubsec:RSGA_generalization}, this condition implies the existence of exact towers of finite-momentum states. 

The imposition a sharp momentum cutoff at $k_b$ (arising from $v_F=\infty$) causes $W_{\bm{k}_4,\bm{k}_3,\bm{k}_2,\bm{k}_1}^{\bm{q}_1,\bm{q}_2}$ to no longer vanish for arbitrary $\bm{q}_1$ and $\bm{q}_2$. The only exception is the zero-momentum sector ($\bm{q}_1 = \bm{q}_2 = 0$), which aligns with our finding in App.~\ref{appsubsec:RSGA_generalization}. Despite this, our preliminary numerical simulations reveal approximate towers of states even with a cutoff as low as $N_{k_b}=31$ and $\varphi_{\text{BZ}}=\frac{\pi}{2}$. 
A detailed investigation into  these structures will be the subject of future work~\cite{repulsive_unpub}.

The approximate RSGA-1 structure for finite momenta can also be used to motivate the ground state dispersion for $N_e>2$. Fig.~\ref{fig:2d_dipsersion_all_mom} shows that the dispersion is linear at small momenta for $N_e$ even. Using analogous arguments as in App.~\ref{appsubsec:1d_2e_general_dispersion}, we find that this would be expected if the RSGA-1 for general momenta were exact, given that the two-body energy $E_{2,\bm{p}}$ is linear at small momenta. We note that the dispersion for odd $N_e$ appears qualitatively different from even $N_e$ in Fig.~\ref{fig:2d_dipsersion_all_mom}, but the small system sizes prevent us from determining the precise scaling of the dispersion for odd $N_e$.

\subsubsection{Binding Energies}\label{sec:app:2d_binding_energy}

\begin{figure}[t]
    \centering
    \includegraphics[width=\linewidth]{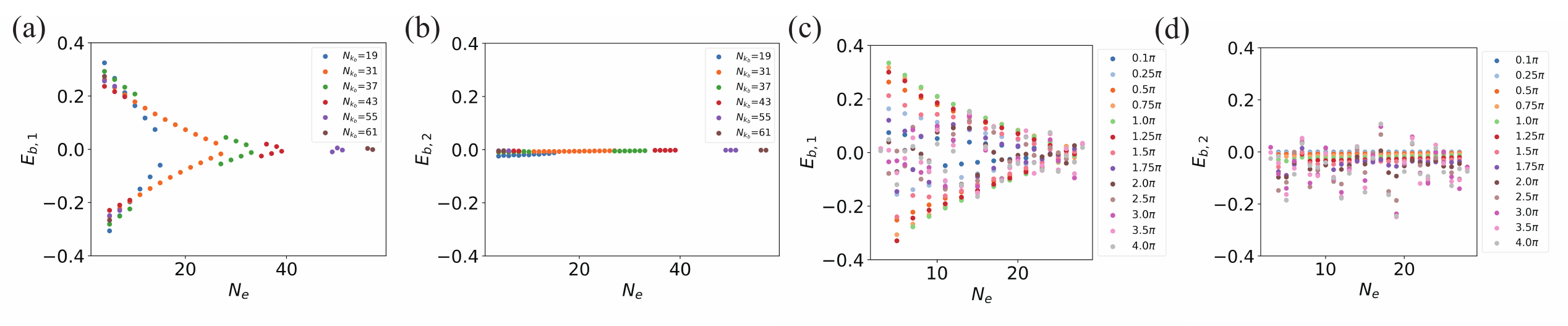}
    \caption{(a), (b) Binding energies $E_{b,1},E_{b,2}$ as a function of electron number $N_e$ with $U=-2/A_b, \bm{p}=0, \alpha=\beta$ and $\varphi_{\text{BZ}}=\pi/2$ for different system sizes $N_{k_b}$. (c), (d) Binding energies $E_{b,1},E_{b,2}$ as a function of electron number $N_e$ with $U=-2/A_b, \bm{p}=0, \alpha=\beta, N_{k_b}=31$ and different $\varphi_{\text{BZ}}$. The value of $\varphi_{\text{BZ}}$ is indicated in the legend. We absorb a factor of $\Omega_{tot}$ in $U$ in these calculations.}\label{fig:2d_binding_energy_attractive}
\end{figure}

Following a similar approach for the 1D toy model in App.~\ref{app:sec:1d_binding_energy}, we investigate presence of superconductivity in the attractive 2D Berry Trashcan model by numerically computing from ED the binding energy
\begin{align}
  E_{b,m}(N_e) \;=\; -2E(N_e)\;+\;E(N_e-m)\;+\;E(N_e+m)\,,
\end{align}
where $E(N_e)$ is the ground‐state energy of a system with $N_e$ particles.  Like for the 1D case in App.~\ref{app:sec:1d_binding_energy}, we concentrate on the cases $m=1$ and $m=2$, corresponding respectively to the pair binding energy $E_{b,1}$ and the quartet binding energy $E_{b,2}$.

As shown in Fig.~\ref{fig:2d_binding_energy_attractive} (a) and (b) for $\alpha=\beta$, $\varphi_\text{BZ}=\pi/2$ and different system sizes $N_{k_b}$, a clear even-odd staggering of $E_{b,1}$ is observed as a function of particle number $N_e$. In particular, $E_{b,1}$ is positive for an even $N_e$ and negative for an odd $N_e$, which demonstrates the energetic preference of the system towards forming electron pairs. 
Furthermore, $|E_{b,2}|$ remains close to zero for all electron number sectors.  
This enables spontaneous breaking of the global charge-$U(1)$ symmetry via a coherent superposition of different particle‐number sectors, a hallmark of a superconducting ground state.

We note that the RSGA-1, and hence our ansatz, is formally exact only to first order in $\alpha=\beta$, as detailed in App.~\ref{app:sec:gs_even_2d}. We therefore expect that a large enough $\varphi_{\text{BZ}}$ could invalidate our ansatz, and potentially destroy the superconducting phase. To test this hypothesis, we plot the binding energies for $\varphi_{\text{BZ}}$ ranging from $0.1\pi$ to $4\pi$ in Fig.~\ref{fig:2d_binding_energy_attractive} (c) and (d). The results clearly show that as $\varphi_{\text{BZ}}$ increases, the even-odd staggering in $E_{b,1}$ begins to break down and is effectively destroyed near $\varphi_{\text{BZ}}=2\pi$. At the same time, $|E_{b,2}|$ tends to increase with increasing $\varphi_{\text{BZ}}$. From these observations, we conclude that the superconducting phase is robust over a wide range of $\varphi_{\text{BZ}}$, but appears to be fragile against a sufficiently large Berry flux $\varphi_\text{BZ}$, which destabilizes the paired ground state.

\subsubsection{Pairing wavefunctions and Off-Diagonal-Long-Range-Order}\label{app:subsec:odlro_2d}
In App.~\ref{app:sec:2d_many_body}, we have shown that the ground state of the attractive 2D Berry Trashcan model with $\alpha=\beta$ can be expressed (approximately) as a condensate of paired electrons with the pairing operator defined as 
\begin{align}
    \hat{O}_2^\dagger=\int_{|\bm{k}|\leq k_b}\frac{d^2\bm{k}}{(2\pi)^2 Z}k_+^{m}e^{-\alpha \bm{k}^2}\gamma^{\dagger}_{\bm{k}}\gamma^{\dagger}_{-\bm{k}}=\frac{1}{\Omega_{tot}Z}\sum^{\{\bm{k}\}}_{\bm{k}}O_2(\bm{k})\gamma^{\dagger}_{\bm{k}}\gamma^{\dagger}_{-\bm{k}}.
\end{align}
To gain real-space insight into the electron pairing, we perform a Fourier transformation on the two-electron pairing operator
\begin{align}
    \hat{O}_2^\dagger&=\int d\bm{r}^2\int d\bm{r'}^2\gamma^\dagger_{\bm{r}} \gamma^\dagger_{\bm{r'}}\int_{|\bm{k}|\leq k_b}\frac{d^2\bm{k}}{4\pi^2Z}k_+^{m}e^{-\alpha \bm{k}^2}e^{-i\bm{k}\cdot(\bm{r-r'})}\nonumber\\
    &=\int d^2\bm{r}\int d^2\bm{R}\frac{e^{im\phi_{r}}(-i)^m}{2\pi Z}\int_{0}^{k_b}k^{m+1}e^{-\alpha k^2}J_{m}(kr)dk\gamma_{\bm{R}-\bm{r}/2}^\dagger\gamma_{\bm{R}+\bm{r}/2}^\dagger\nonumber\\
    &=\int d^2\bm{r}\int d^2\bm{R}O_2(\bm{r})\gamma_{\bm{R}-\bm{r}/2}^\dagger\gamma_{\bm{R}+\bm{r}/2}^\dagger,\label{eq:pairing_real_space_integral}
\end{align}
where we have parameterized $\bm{r-r'}\to\bm{r}$ and $\frac{\bm{r}+\bm{r'}}{2}\to\bm{R}$ in going from the first to the second line, and $O_2(\bm{r})$ captures the real-space pair wavefunction. Note that $O_2(\bm{r})$ does not depend on $\bm{R}$ because we have a total momentum $\bm{p}=0$ eigenstate. If we take $k_b\to\infty$, the integral above reduces to
\begin{align}
    O_2(\bm{r})=\frac{(-i\bm{r}_+)^m}{2\pi Z(2\alpha)^{m+1}}e^{-\frac{|\bm{r}|^2}{4\alpha}},
\end{align}
with $\bm{r}_+=r_x+ir_y$, which scales as $r^m$ at short distances and decays exponentially as a Gaussian at long distances. Since $m=1$ for the ground state, this pairing has $p+ip$ symmetry. 

On the other hand, in the small Berry flux limit with $\alpha k_b^2\ll1$ which is the relevant limit for the RnG system \cite{bernevig2025berrytrashcanmodelinteracting}, we expand the exponents in Eq.~\ref{eq:pairing_real_space_integral} to first order in $\alpha$. Then the integral reduces to
\begin{align}
    O_2(\bm{r}) \approx&\frac{e^{im\phi_{r}}(-i)^m}{2\pi Z}\left[\int_{0}^{k_b}k^{m+1}J_{m}(kr)dk-\alpha\int_{0}^{k_b}k^{m+3}J_{m}(kr)dk\right]\nonumber\\
    =&\frac{(-i)^m r_+^m}{2\pi r^m Z} \left[ \frac{k_b^{m+1}}{r}(1 - \alpha k_b^2)J_{m+1}(k_b r) + \frac{2\alpha k_b^{m+2}}{r^2}J_{m+2}(k_b r) \right] + \mathcal{O}(\alpha^2),
\end{align}
where we used the property $\int x^{\nu+1}J_{\nu}(x)dx=x^{\nu+1}J_{\nu+1}(x)$.
For short distances (small $k_b\bm{r}$), the pairing function
\begin{align}
   O_2(\bm{r}) \approx \frac{(-i)^m k_b^{2m+2}}{2\pi Z\cdot 2^{m+1}(m+1)!} \left[ 1 - \alpha k_b^2 \frac{m+1}{m+2} \right] r_+^m
\end{align}
scales as $r_+^m$ which is identical to the case with $k_b\to\infty$. For long distances (large $k_b r$), the pairing function
\begin{align}
   O_2(\bm{r}) \approx \frac{(-i)^m (1 - \alpha k_b^2) k_b^{m+1/2}}{2\pi Z} \sqrt{\frac{2}{\pi}} \frac{r_+^m}{r^{m+3/2}} \cos\left(k_b r - \frac{(m+1)\pi}{2} - \frac{\pi}{4}\right)
\end{align}
decays as $\sim r^{-3/2}$ which exhibits long-range behavior compared to the $\alpha k_b^2\to \infty$ limit.

The exponentially decaying (as a function of $r$) behavior of the pairing function in the limit of large $\alpha k_b^2\rightarrow\infty$ is analogous to the strong-pairing (topologically trivial) phase as discussed in Ref.~\cite{PhysRevB.61.10267}. The strong attractive interaction tightly binds the electron pairs and leads to a small spatial extension of the electron pair. Introducing a sharp momentum cutoff with small $\alpha k_b^2$ leads instead to pairing that decays algebraically $\sim r^{-3/2}$ with an oscillating envelope. The decay of the pairing function is intermediate between the strong-coupling phase and the weak coupling phase (where the decay would be $\sim r^{-1}$) in Ref.~\cite{PhysRevB.61.10267}. This suggests an unusual pairing behavior in the 2D attractive Berry Trashcan. A more detailed study is left for a future work~\cite{repulsive_unpub}.

To further characterize the superconducting physics of the ground state, we investigate the presence of ODLRO, which manifests as a non-vanishing value of the pairing correlation function
\begin{align}
    \rho^{(2)}_{(\bm{r}_1,\bm{r}_2),(\bm{r}_3,\bm{r}_4)}&=\bra{\text{GS}}\gamma_{\bm{r}_1}^{\dagger}\gamma_{\bm{r}_2}^{\dagger}\gamma_{\bm{r}_4}\gamma_{\bm{r}_3}\ket{\text{GS}}\nonumber\\
    &=\int_{|\bm{k}_i|\leq k_b}\frac{d^2\bm{k}_1d^2\bm{k}_2d^2\bm{k}_3 d^2\bm{k}_4}{(2\pi)^8} e^{-i(\bm{k}_1\cdot\bm{r}_1 + \bm{k}_2\cdot\bm{r}_2 - \bm{k}_3\cdot\bm{r}_3 - \bm{k}_4\cdot\bm{r}_4)} \bra{\text{GS}}\gamma^\dagger_{\bm{k}_1}\gamma^\dagger_{\bm{k}_2}\gamma_{\bm{k}_4}\gamma_{\bm{k}_3}\ket{\text{GS}} \nonumber\\
    &=\int_{|\bm{k}_i|\leq k_b}\frac{d^2\bm{k}_1d^2\bm{k}_2d^2\bm{k}_3 d^2\bm{k}_4}{(2\pi)^8} e^{-i(\bm{k}_1\cdot\bm{r}_1 + \bm{k}_2\cdot\bm{r}_2 - \bm{k}_3\cdot\bm{r}_3 - \bm{k}_4\cdot\bm{r}_4)}\rho^{(2)}_{(\bm{k}_1,\bm{k}_2),(\bm{k}_{3},\bm{k}_4)}
    \label{eq:ODLRO_def}
\end{align}
when the coordinates $\bm{r}_1,\bm{r}_2$ are infinitely far away from $\bm{r}_3,\bm{r}_4$.
To this end, we start with the momentum-space four-point correlator
\begin{align}
\rho^{(2)}_{(\bm{k}_1,\bm{k}_2),(\bm{k}_{3},\bm{k}_4)}=\bra{\text{GS}}\gamma_{\bm{k}_1}^{\dagger}\gamma_{\bm{k}_2}^{\dagger}\gamma_{\bm{k}_4}\gamma_{\bm{k}_3}\ket{\text{GS}}.
\end{align}
To evaluate this, we introduce the generating wavefunction in momentum space
\begin{align}
\ket{\xi,z}=\exp{\left(\sum_{\bm{k}}\xi_{\bm{k}}\gamma^{\dagger}_{\bm{k}}\right)}e^{z\hat{O}_2^{\dagger}}\ket{0},\label{app:eq:generating_func_2d}
\end{align}
which has norm
\begin{align}
    N(\xi,z)=\bra{\xi,z}\xi,z\rangle.
\end{align}
Here, $\xi_{\bm{k}}$ are anticommuting Grassmann variables with its Hermitian conjugate defined as $\overline{\xi}=\xi^{\dagger}$. Expanding Eq.~\ref{app:eq:generating_func_2d} in series of $z$, we find that the $z^n$ term corresponds to the component of $\ket{\xi_{\bm{k}}=0,z}$ with $2n$ particles. To study the expectation value of an observable $\hat{O}$ in the $2n$-particle ground state, we therefore need to expand $\bra{\xi,z}\hat{O}\ket{\xi,z}$ in $z,\bar{z}$, and isolate the coefficient of $|z|^{2n}$ term in the limit $\xi_{\bm{k}}=0$ (we also need the coefficient of $|z|^{2n}$ in $N(\xi,z)$ to determine the correct normalization).

We first calculate the norm $N(\xi,z)$. Since the $\xi_{\bm{k}}$'s are Grassmann numbers and $\left[\xi_{\bm{k}}\gamma_{\bm{k}}^{\dagger},\,\hat{O}_2^{\dagger}\right]=0$, we express $\ket{\xi,z}$ as 
\begin{align}
    \ket{\xi,z}=\prod^{\{\bm{k}\}}_{\bm{k}\in\text{UHP}}\left(1+\xi_{\bm{k}}\gamma^{\dagger}_{\bm{k}}+\xi_{-\bm{k}}\gamma^{\dagger}_{-\bm{k}}+(zO_2(\bm{k})+\xi_{-\bm{k}}\xi_{\bm{k}})\gamma_{\bm{k}}^{\dagger}\gamma_{-\bm{k}}^\dagger\right)\ket{0},
\end{align}
where UHP represents the upper half plane $\{\bm{k}:k_y>0\}$, and we have parameterized the two-particle operator as $\hat{O}_2^\dagger=\sum_{\bm{k}}^{\{\bm{k}\}}O_2(\bm{k})\gamma^\dagger_{\bm{k}}\gamma^\dagger_{-\bm{k}}$ so that $O_2(\bm{k})$ contains factors of e.g.~the normalization $Z$. This leads to
\begin{align}
N(\xi,z)=\bra{\xi,z}\xi,z\rangle=\prod^{\{\bm{k}\}}_{\bm{k}\in\text{UHP}}\left(1+\overline{\xi}_{\bm{k}}\xi_{\bm{k}}+\overline{\xi}_{-\bm{k}}\xi_{-\bm{k}}+(\overline{z}O_2(\bm{k})^*+\overline{\xi}_{\bm{k}}\overline{\xi}_{-\bm{k}})(zO_2(\bm{k})+\xi_{\bm{-k}}\xi_{\bm{k}})\right).
\end{align}
For $\xi=0$, we have
\begin{align}
    N(z)\equiv N(\xi=0,z)&=\prod^{\{\bm{k}\}}_{\bm{k}\in\text{UHP}}\left(1+|z|^2|O_2(\bm{k})|^2\right)=e^{\sum^{\{\bm{k}\}}_{\bm{k}\in\text{UHP}}\ln\left(1+|z|^2|O_2(\bm{k})|^2\right)}=e^{\frac{1}{(2\pi)^2}\int_{\bm{k}\in\text{UHP},|\bm{k}|\leq k_b}d^2\bm{k}\ln\left(1+|z|^2|O_2(\bm{k})|^2\right)}.
\end{align}
The integral in the exponential can be evaluated as
\begin{align}
    \frac{1}{(2\pi)^2}\int_{\bm{k}\in\text{UHP},|\bm{k}|\leq k_b}d^2\bm{k}\ln\left(1+|z|^2|O_2(\bm{k})|^2\right)&= \frac{1}{4\pi}\int_{0}^{k_b}kdk\ln\left(1+\frac{k^{2m}e^{-2\alpha k^2}|z|^2}{(2\pi)^4Z^2}\right)\nonumber\\
    &=\frac{1}{8\pi}\int_{0}^{k_b^2}dx\ln\left(1+\frac{x^{m}e^{-2\alpha x}|z|^2}{(2\pi)^4Z^2}\right)\nonumber\\
    &=\sum_{n=1}^{\infty} \frac{1}{8\pi} \frac{(-1)^{n-1}}{n} \left(\frac{|z|^2}{(2\pi)^4Z^2}\right)^n \int_{0}^{k_b^2}x^{nm}e^{-2n\alpha x}dx\nonumber\\
    &=\sum_{n=1}^{\infty} a_n|z|^{2n}
\end{align}
where $a_n=\frac{1}{8\pi} \frac{(-1)^{n-1}}{n} \left(\frac{1}{(2\pi)^4Z^2}\right)^n\left(\frac{1}{2n\alpha}\right)^{nm+1}\gamma(nm+1, 2n\alpha k_b^2)$, and $\gamma(s,x)$ is the lower incomplete Gamma function. This leads to
\begin{align}
    N(z)=\sum^{\infty}_{n=0}N_n|z|^{2n},
\end{align}
with the $n$th coefficient can be obtained by a recursion relation
\begin{align}
    N_n=\frac{1}{n}\sum_{\nu=1}^{n}\nu a_{\nu}N_{n-\nu}.
\end{align}

Having obtained the norm $N(z)$, we proceed to evaluate the correlators $\bra{\xi,z}\hat{O}\ket{\xi,z}$.Given the derivatives
\begin{align}
    \gamma_{\bm{k}}^\dagger\ket{\xi,z}=\partial_{\xi_{\bm{k}}}\ket{\xi,z},\quad \bra{\xi,z}\gamma_{\bm{k}}=-\partial_{\overline{\xi}_{\bm{k}}}\bra{\xi,z},
\end{align}
the correlation function $\bra{z}\gamma_{\bm{k}_1}\gamma_{\bm{k}_2}\cdots\gamma_{\bm{k}_i}^{\dagger}\gamma_{\bm{k}_{i+1}}^{\dagger}\ket{z}$ can be evaluated as
\begin{align}
\bra{z}\gamma_{\bm{k}_1}\gamma_{\bm{k}_2}\cdots\gamma_{\bm{k}_i}^{\dagger}\gamma_{\bm{k}_{i+1}}^{\dagger}\ket{z}=(-\partial_{\overline{\xi}_{\bm{k}_1}})(-\partial_{\overline{\xi}_{\bm{k}_2}})\cdots\bra{\xi,z}\cdots\partial_{\xi_{\bm{k}_i}}\partial_{\xi_{\bm{k}_{i+1}}}\ket{\xi,z}\bigg|_{\xi=0}=\cdots\partial_{\xi_{\bm{k}_{i}}}\partial_{\xi_{\bm{k}_{i+1}}}\partial_{\overline{\xi}_{\bm{k}_1}}\partial_{\overline{\xi}_{\bm{k}_2}}\cdots N(\xi,z)\bigg|_{\xi=0},
\end{align}
where $\ket{z}\equiv \ket{\xi=0,z}$. In particular, correlators of two fermion operators are
\begin{align}
    \bra{z}\gamma_{\bm{k}}\gamma_{\bm{k}'}^\dagger\ket{z}&=-\partial_{\overline{\xi}_{\bm{k}}}\bra{\xi,z}\partial_{\xi_{\bm{k}'}}\ket{\xi,z}\bigg|_{\xi=0}=\partial_{\xi_{\bm{k}'}}\partial_{\overline{\xi}_{\bm{k}}}N(\xi,z)\bigg|_{\xi=0}=\frac{\delta_{\bm{k},\bm{k}'}}{1+|z|^2|O_{2}(\bm{k})|^2}N(z)\\
    \\
    \bra{z}\gamma_{\bm{k}'}^\dagger\gamma_{\bm{k}}\ket{z}&=-\bra{z}\gamma_{\bm{k}}\gamma_{\bm{k}'}^\dagger\ket{z}+\delta_{\bm{k},\bm{k}'}=\frac{|z|^2|O_2(\bm{k})|^2\delta_{\bm{k},\bm{k}'}}{1+|z|^2|O_{2}(\bm{k})|^2}N(z)\\
    \bra{z}\gamma_{\bm{k}}\gamma_{\bm{k}'}\ket{z}&=\partial_{\overline{\xi}_{\bm{k}}}\partial_{\overline{\xi}_{\bm{k}'}}N(\xi,z)\bigg|_{\xi=0}=-\frac{\delta_{\bm{k},-\bm{k}'}zO_2(\bm{k})}{1+|z|^2|O_{2}(\bm{k})|^2}N(z)\\
    \bra{z}\gamma_{\bm{k}}^\dagger\gamma_{\bm{k}'}^\dagger\ket{z}&=\partial_{\xi_{\bm{k}}}\partial_{\xi_{\bm{k}'}}N(\xi,z)\bigg|_{\xi=0}=\frac{\delta_{\bm{k},-\bm{k}'}\overline{z}O_2(\bm{k})^*}{1+|z|^2|O_{2}(\bm{k})|^2}N(z).
\end{align}
According to Wick's theorem, the four fermion correlators can be evaluated as
\begin{align}
\frac{\bra{z}\gamma_{\bm{k}_1}\gamma_{\bm{k}_2}\gamma_{\bm{k}_3}^\dagger\gamma_{\bm{k}_4}^\dagger\ket{z}}{N(z)}=&\frac{1}{N(z)}\partial_{\xi_{\bm{k}_3}}\partial_{\xi_{\bm{k}_4}}\partial_{\overline{\xi}_{\bm{k}_1}}\partial_{\overline{\xi}_{\bm{k}_2}}N(\xi,z)\bigg|_{\xi=0}\nonumber\\
=&\frac{\bra{z}\gamma_{\bm{k}_1}\gamma_{\bm{k}_2}\ket{z}}{N(z)}\frac{\bra{z}\gamma_{\bm{k}_3}^\dagger\gamma_{\bm{k}_4}^\dagger\ket{z}}{N(z)}-\frac{\bra{z}\gamma_{\bm{k}_1}\gamma_{\bm{k}_3}^\dagger\ket{z}}{N(z)}\frac{\bra{z}\gamma_{\bm{k}_2}\gamma_{\bm{k}_4}^\dagger\ket{z}}{N(z)}+\frac{\bra{z}\gamma_{\bm{k}_1}\gamma_{\bm{k}_4}^\dagger\ket{z}}{N(z)}\frac{\bra{z}\gamma_{\bm{k}_2}\gamma_{\bm{k}_3}^\dagger\ket{z}}{N(z)}\nonumber\\
    =&-\frac{\delta_{\bm{k}_1,-\bm{k}_2}\delta_{\bm{k}_3,-\bm{k}_4}|z|^2O_2(\bm{k}_1)O_2^*(\bm{k}_3)}{\left(1+|z|^2|O_2(\bm{k}_1)|^2\right)\left(1+|z|^2|O_2(\bm{k}_3)|^2\right)}+\frac{\delta_{\bm{k}_1,\bm{k}_4}\delta_{\bm{k}_2,\bm{k}_3}}{\left(1+|z|^2|O_2(\bm{k}_1)^2|\right)\left(1+|z|^2|O_2(\bm{k}_2)|^2\right)}\nonumber\\
    &-\frac{\delta_{\bm{k}_1,\bm{k}_3}\delta_{\bm{k}_2,\bm{k}_4}}{\left(1+|z|^2|O_2(\bm{k}_1)^2|\right)\left(1+|z|^2|O_2(\bm{k}_2)|^2\right)}.\label{app:eq:4_point_kcorrelation_2d}
\end{align}
\begin{align}
\frac{\bra{z}\gamma_{\bm{k}_1}^\dagger\gamma_{\bm{k}_2}^\dagger\gamma_{\bm{k}_3}\gamma_{\bm{k}_4}\ket{z}}{N(z)}
=&\frac{\bra{z}\gamma_{\bm{k}_1}^\dagger\gamma_{\bm{k}_2}^\dagger\ket{z}}{N(z)}\frac{\bra{z}\gamma_{\bm{k}_3}\gamma_{\bm{k}_4}\ket{z}}{N(z)}-\frac{\bra{z}\gamma_{\bm{k}_1}^\dagger\gamma_{\bm{k}_3}\ket{z}}{N(z)}\frac{\bra{z}\gamma_{\bm{k}_2}^\dagger\gamma_{\bm{k}_4}\ket{z}}{N(z)}+\frac{\bra{z}\gamma_{\bm{k}_1}^\dagger\gamma_{\bm{k}_4}\ket{z}}{N(z)}\frac{\bra{z}\gamma_{\bm{k}_2}^\dagger\gamma_{\bm{k}_3}\ket{z}}{N(z)}\nonumber\\
    =&-\frac{\delta_{\bm{k}_1,-\bm{k}_2}\delta_{\bm{k}_3,-\bm{k}_4}|z|^2O_2^*(\bm{k}_1)O_2(\bm{k}_3)}{\left(1+|z|^2|O_2(\bm{k}_1)|^2\right)\left(1+|z|^2|O_2(\bm{k}_3)|^2\right)}+\frac{\delta_{\bm{k}_1,\bm{k}_4}\delta_{\bm{k}_2,\bm{k}_3}|z|^2|O_2(\bm{k}_1)|^2|z|^2|O_2(\bm{k}_2)|^2}{\left(1+|z|^2|O_2(\bm{k}_1)^2|\right)\left(1+|z|^2|O_2(\bm{k}_2)|^2\right)}\nonumber\\
    &-\frac{\delta_{\bm{k}_1,\bm{k}_3}\delta_{\bm{k}_2,\bm{k}_4}|z|^2|O_2(\bm{k}_1)|^2|z|^2|O_2(\bm{k}_2)|^2}{\left(1+|z|^2|O_2(\bm{k}_1)^2|\right)\left(1+|z|^2|O_2(\bm{k}_2)|^2\right)}.\label{app:eq:4_point_kcorrelation_2d_normal_ordered}
\end{align}

Recall that our objective is to obtain the two-particle real-space correlation function, which is a Fourier transform of the four-point momentum space correlator. As an intermediate step, we consider the Fourier transform of the non-number conserving and non-normalized state $\ket{z}$
\begin{align}
   \tilde{\rho}^{(2)}_{(\bm{r}_1,\bm{r}_2),(\bm{r}_3,\bm{r}_4)}
    &=\int_{|\bm{k}_i|\leq k_b}\frac{d^2\bm{k}_1d^2\bm{k}_2d^2\bm{k}_3 d^2\bm{k}_4}{(2\pi)^8} e^{-i(\bm{k}_1\cdot\bm{r}_1 + \bm{k}_2\cdot\bm{r}_2 - \bm{k}_3\cdot\bm{r}_3 - \bm{k}_4\cdot\bm{r}_4)} \bra{z}\gamma^\dagger_{\bm{k}_1}\gamma^\dagger_{\bm{k}_2}\gamma_{\bm{k}_4}\gamma_{\bm{k}_3}\ket{z}
    \label{eq:fourier_integral}.
\end{align}
The momentum-space expectation value in Eq.~\ref{app:eq:4_point_kcorrelation_2d_normal_ordered} is given as a sum of three terms, and we proceed by Fourier transforming each term separately. The contribution to the correlation function from the first term, denoted $G_1$, is
\begin{align}
    G_1 &= -|z|^2 N(z) \int_{|\bm{k}_i|\leq k_b}\frac{d^2\bm{k}_1d^2\bm{k}_2d^2\bm{k}_3 d^2\bm{k}_4}{(2\pi)^8} \frac{e^{-i(\bm{k}_1\cdot\bm{r}_1 + \bm{k}_2\cdot\bm{r}_2 - \bm{k}_3\cdot\bm{r}_3 - \bm{k}_4\cdot\bm{r}_4)} \delta_{\bm{k}_1,-\bm{k}_2}\delta_{\bm{k}_3,-\bm{k}_4} O_2(\bm{k}_1)O_2^*(\bm{k}_3)}{\left(1+|z|^2|O_2(\bm{k}_1)|^2\right)\left(1+|z|^2|O_2(\bm{k}_3)|^2\right)}\nonumber\\
    &= -|z|^2 N(z) \int_{|\bm{k}_i|\leq k_b} \frac{d^2\bm{k}_1 d^2\bm{k}_3}{(2\pi)^4} \frac{e^{-i\bm{k}_1\cdot(\bm{r}_1-\bm{r}_2)} e^{i\bm{k}_3\cdot(\bm{r}_3-\bm{r}_4)} O_2(\bm{k}_1)O_2^*(\bm{k}_3)}{\left(1+|z|^2|O_2(\bm{k}_1)|^2\right)\left(1+|z|^2|O_2(\bm{k}_3)|^2\right)}.
\end{align}
This integral is separable
\begin{align}
    G_1 = -|z|^2 N(z) \left[ \int_{|\bm{k}_1|\leq k_b} \frac{d^2\bm{k}_1}{(2\pi)^2} \frac{e^{-i\bm{k}_1\cdot(\bm{r}_1-\bm{r}_2)} O_2(\bm{k}_1)}{1+|z|^2|O_2(\bm{k}_1)|^2} \right] \left[ \int_{|\bm{k}_3|\leq k_b} \frac{d^2\bm{k}_3}{(2\pi)^2} \frac{e^{i\bm{k}_3\cdot(\bm{r}_3-\bm{r}_4)} O_2^*(\bm{k}_3)}{1+|z|^2|O_2(\bm{k}_3)|^2} \right].\label{appeq:2d_G1}
\end{align}
We define the anomalous propagator or pairing function $H(\bm{R})$
\begin{align}
    H(\bm{R}) \equiv \int_{|\bm{k}|\leq k_b} \frac{d^2\bm{k}}{(2\pi)^2} \frac{e^{-i\bm{k}\cdot\bm{R}} O_2(\bm{k})}{1+|z|^2|O_2(\bm{k})|^2}\label{app:eq:HR}.
\end{align}
The first bracketed integral in Eq.~\ref{appeq:2d_G1} is $H(\bm{r}_1 - \bm{r}_2)$. The second bracketed integral can be identified as the complex conjugate of $H(\bm{r}_3 - \bm{r}_4)$. Therefore, the final expression for $G_1$ is
\begin{align}
    G_1(\bm{r}_1, \bm{r}_2, \bm{r}_3, \bm{r}_4) = -|z|^2 N(z) H(\bm{r}_1 - \bm{r}_2) H(\bm{r}_3 - \bm{r}_4)^*.
\end{align}

The contributions from the second and third terms in Eq.~\ref{app:eq:4_point_kcorrelation_2d_normal_ordered}, which we collectively denote as $G_2$, are
\begin{align}
    G_2 =& N(z) \int_{|\bm{k}_i|\leq k_b}  \frac{d^2\bm{k}_1 d^2\bm{k}_2}{(2\pi)^4} \frac{|z|^4|O_2(\bm{k}_1)|^2|O_2(\bm{k}_2)|^2\left(e^{-i(\bm{k}_1\cdot\bm{r}_1 + \bm{k}_2\cdot\bm{r}_2 - \bm{k}_2\cdot\bm{r}_3 - \bm{k}_1\cdot\bm{r}_4)}-e^{-i(\bm{k}_1\cdot\bm{r}_1 + \bm{k}_2\cdot\bm{r}_2 - \bm{k}_1\cdot\bm{r}_3 - \bm{k}_2\cdot\bm{r}_4)}\right)}{\left(1+|z|^2|O_2(\bm{k}_1)|^2\right)\left(1+|z|^2|O_2(\bm{k}_2)|^2\right)}\nonumber\\
    =&N(z) \left[ \int_{|\bm{k}_1|\leq k_b}  \frac{d^2\bm{k}_1}{(2\pi)^2} \frac{|z|^2|O_2(\bm{k}_1)|^2e^{-i\bm{k}_1\cdot(\bm{r}_1-\bm{r}_4)}}{1+|z|^2|O_2(\bm{k}_1)|^2} \right] \left[ \int_{|\bm{k}_2|\leq k_b}  \frac{d^2\bm{k}_2}{(2\pi)^2} \frac{|z|^2|O_2(\bm{k}_2)|^2e^{-i\bm{k}_2\cdot(\bm{r}_2-\bm{r}_3)}}{1+|z|^2|O_2(\bm{k}_2)|^2} \right]\nonumber\\
    &-N(z) \left[ \int_{|\bm{k}_1|\leq k_b}  \frac{d^2\bm{k}_1}{(2\pi)^2} \frac{|z|^2|O_2(\bm{k}_1)|^2e^{-i\bm{k}_1\cdot(\bm{r}_1-\bm{r}_3)}}{1+|z|^2|O_2(\bm{k}_1)|^2} \right] \left[ \int_{|\bm{k}_2|\leq k_b}  \frac{d^2\bm{k}_2}{(2\pi)^2} \frac{|z|^2|O_2(\bm{k}_2)|^2e^{-i\bm{k}_2\cdot(\bm{r}_2-\bm{r}_4)}}{1+|z|^2|O_2(\bm{k}_2)|^2} \right].
\end{align}
We define $F(\bm{R})$ as the Fourier transform of the
\begin{align}
    F(\bm{R}) \equiv \int_{|\bm{k}|\leq k_b}  \frac{d^2\bm{k}}{(2\pi)^2} \frac{|z|^2|O_2(\bm{k})|^2e^{-i\bm{k}\cdot\bm{R}}}{1+|z|^2|O_2(\bm{k})|^2}\label{app:eq:FR}.
\end{align}
$G_2$ can then be written compactly as
\begin{align}
    G_2(\bm{r}_1, \bm{r}_2, \bm{r}_3, \bm{r}_4) = N(z)\left[ F(\bm{r}_1 - \bm{r}_4) F(\bm{r}_2 - \bm{r}_3)-F(\bm{r}_1 - \bm{r}_3) F(\bm{r}_2 - \bm{r}_4)\right].
\end{align}

Combining the contributions from $G_1$ and $G_2$, we find
\begin{align}
    \tilde{\rho}^{(2)}_{(\bm{r}_1,\bm{r}_2),(\bm{r}_3,\bm{r}_4)} = N(z) \Big[ F(\bm{r}_1 - \bm{r}_4) F(\bm{r}_2 - \bm{r}_3) -F(\bm{r}_1 - \bm{r}_3) F(\bm{r}_2 - \bm{r}_4)- |z|^2 H(\bm{r}_1 - \bm{r}_2) H(\bm{r}_3 - \bm{r}_4)^* \Big].
\end{align}
The functions $F(\bm{R})$ and $H(\bm{R})$ are defined in Eqs.~\ref{app:eq:HR} and \ref{app:eq:FR}, respectively. To analyze the presence of ODLRO, we evaluate the integrals in $F(\bm{R})$ and $H(\bm{R})$.
We start with the angular integral. For $F(\bm{R})$ the angular integral yields
\begin{align}
    \int_0^{2\pi} e^{-i\bm{k}\cdot\bm{R}} d\phi_k = \int_0^{2\pi} e^{-ikR\cos(\phi_k-\phi_R)} d\phi_k = 2\pi J_0(kR),
\end{align}
where $J_0$ is the Bessel function, and $\bm{R}=Re^{i\phi_R}$ with $R>0$. This reduces the expression for $F(\bm{R})$ to a single radial integral
\begin{align}
    F(\bm{R}) = \frac{1}{2\pi} \int_0^{k_b} \frac{C k^{2m+1}e^{-2\alpha k^2}J_0(kR)}{1 + C k^{2m}e^{-2\alpha k^2}} dk. \label{eq:F_exact}
\end{align}
where $C = |z|^2/|(2\pi)^2Z|^2$. Similarly, for $H(\bm{R})$, the angular integration over the term $k_+^m = k^m e^{im\phi_k}$ yields a factor of 
\begin{align}
    \int_0^{2\pi} e^{-i\bm{k}\cdot\bm{R}} k_+^m d\phi_k = k^m \int_0^{2\pi} e^{im\phi_k} e^{-ikR\cos(\phi_k-\phi_R)} d\phi_k = k^m \left( 2\pi (-i)^m e^{im\phi_R} J_m(kR) \right),
\end{align}
which reduces $H(\bm{R})$ to
\begin{align}
    H(\bm{R}) = \frac{(-i)^m e^{im\phi_R}}{2\pi Z} \int_0^{k_b} \frac{k^{m+1} e^{-\alpha k^2} J_m(kR)}{1 + C k^{2m}e^{-2\alpha k^2}} dk. \label{eq:H_exact}
\end{align}
While these expressions in Eqs. \eqref{eq:F_exact} and \eqref{eq:H_exact} lack a general closed-form solution, their asymptotic behavior for $R \to \infty$ can be extracted. For large $R$, the Bessel functions $J_m(kR)$ in the integrands oscillate rapidly, and such rapid oscillations render $H(\bm{R})$ and $F(\bm{R})$ negligible when $R\to \infty$. In particular, for such Fourier-type integrals with large $R$, the dominant contributions arise from the boundaries of the integration domain (in our case, at $k=0$ and $k=k_b$) where the rapid oscillations do not fully cancel out. Because the integrand vanishes at $k=0$, the asymptotic behavior is entirely governed by the upper boundary at $k=k_b$. Since $J_m(k_bR)$ decays as $(k_bR)^{-1/2}$ and the integral contributes an additional factor $R^{-1}$, the asymptotic behaviors for $F(R)$ and $H(R)$ for $R \to \infty$ are
\begin{align}
    |F(\bm{R})| \sim R^{-3/2} \quad \text{and} \quad |H(\bm{R})| \sim R^{-3/2} \quad (\text{for } R \to \infty).
\end{align}
We consider the case when the two pairs in the correlation function are infinitely far away from each other, i.e.~the intra-pair distances $k_b|\bm{r}_1-\bm{r}_2|$ and $k_b|\bm{r}_3-\bm{r}_4|$ are finite, while the inter-pair distances $k_b|\bm{r}_1-\bm{r}_4|$ and $k_b|\bm{r}_2-\bm{r}_3|$ are infinite. The four-point correlator reduces to 
\begin{align}
        \tilde{\rho}^{(2)}_{(\bm{r}_1,\bm{r}_2),(\bm{r}_3,\bm{r}_4)} =- N(z) |z|^2 H(\bm{r}_1 - \bm{r}_2) H(\bm{r}_3 - \bm{r}_4)^*,\label{app:eq:ODLRO_2d}
\end{align}
which takes a finite value and suggests the existence of ODLRO. In this limit
\begin{align}
    H(\bm{R}) \approx \frac{(-i)^m e^{im\phi_R}R^m}{2^{m+1}\pi Zm!} \int_0^{k_b} \frac{k^{2m+1} e^{-\alpha k^2} }{1 + C k^{2m}e^{-2\alpha k^2}} dk.
\end{align}
Since the ground state corresponds to the angular momentum sector $m=1$, $H(\bm{R})$ exhibits a $p+ip$ pairing symmetry as expected. While the (number-conserving) correlator $\rho^{(2)}_{(\bm{r}_1,\bm{r}_2),(\bm{r}_3,\bm{r}_4)}$ can, in principle, be determined analytically, the procedure is intricate. To compute $\rho^{(2)}_{(\bm{r}_1,\bm{r}_2),(\bm{r}_3,\bm{r}_4)}$ for a ground state of fixed particle number $N_e=2n$, one must expand Eq.~\ref{app:eq:ODLRO_2d} as a power series in $|z|$, and take the coefficient of the $|z|^{2n}$ term, multiplied by $(n!)^2$. However, since both the normalization factor $N(z)$ and $H(\bm{R})$ are complicated functions of $|z|$, obtaining the final analytical form is prohibitively difficult.

\begin{figure}[t]
    \centering
    \includegraphics[width=0.8\linewidth]{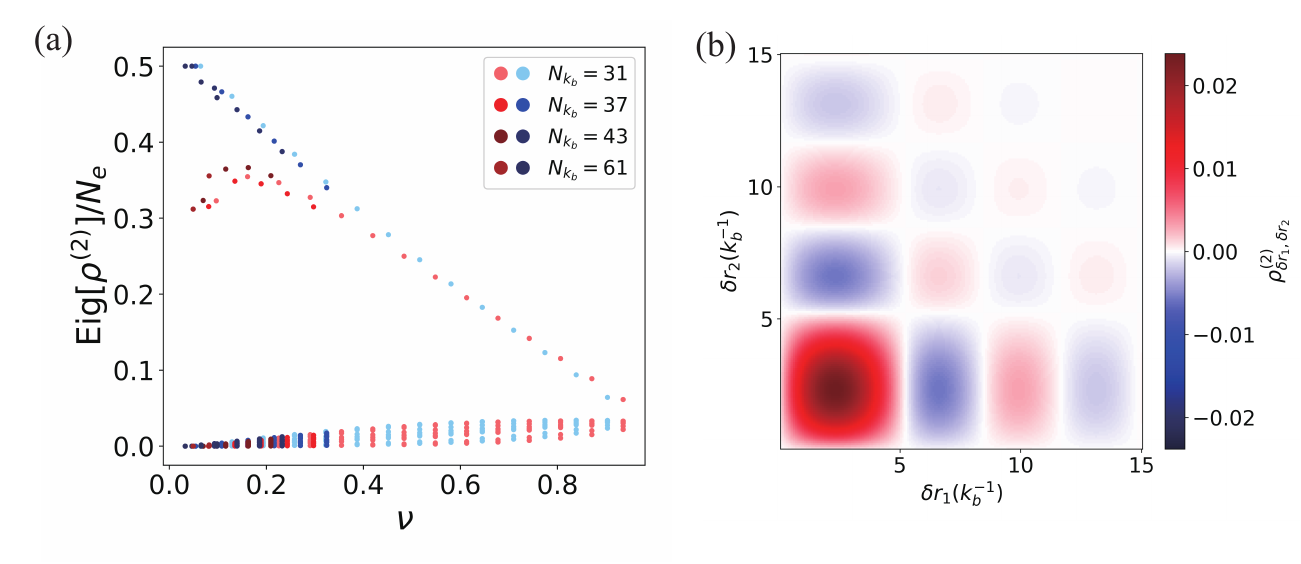}
    \caption{(a) Eigenvalues of the two-particle ground state density matrix (Eq.~\ref{app:eq:reduced_mom_odlro}) normalized by the electron number $N_e$ for  $\alpha=\beta,\varphi_\text{BZ}=\pi/2$, as a function of filling $\nu$, for different $N_{k_b}$. The results are obtained from ED calculations. Blue (red) dots indicate even (odd) $N_e$. The presence of a finite eigenvalue for finite filling factor $\nu$ indicates ODLRO. (b) Plot of the function $I(\delta r_1)  I(\delta r_2)$ in $\rho^{(2)}_{\delta r_1,\delta r_2}$ (Eq.~\ref{app:eq:odlro_2e_analytical}) for the two-electron ground state. }\label{fig:2d_odlro_sup}
\end{figure}

Given the complexity of performing this expansion for a many-body state, we turn to a more direct numerical method to verify ODLRO, similar to our approach in the 1D case. As discussed previously, the long-distance behavior of the two-particle correlator is dominated by the first term in Eq.~\ref{app:eq:4_point_kcorrelation_2d_normal_ordered}. Since this term restricts $\bm{k}_1=-\bm{k}_2$ and $\bm{k}_3=-\bm{k}_4$, we can rewrite $\rho^{(2)}_{(\bm{k}_1,-\bm{k}_1),(-\bm{k}_{2},\bm{k}_2)}$ as
\begin{align}
\rho^{(2)}_{(\bm{k}_1,-\bm{k}_1),(-\bm{k}_{2},\bm{k}_2)}=\rho^{(2)}_{\bm{k}_1,\bm{k}_{2}}=\bra{\text{GS}}\gamma_{\bm{k}_1}^{\dagger}\gamma_{-\bm{k}_1}^{\dagger}\gamma_{-\bm{k}_2}\gamma_{\bm{k}_2}\ket{\text{GS}}.\label{app:eq:reduced_mom_odlro}
\end{align}
We construct $\rho^{(2)}_{\bm{k}_1,\bm{k}_{2}}$ as a matrix and compute its eigenvalue spectrum. The results are shown in Fig.~\ref{fig:2d_odlro_sup}(a). The presence of a single, large eigenvalue separated from the rest of the spectrum, is a clear signature of ODLRO. In particular, this eigenvalue grows with $N_e$. Furthermore, the eigenvector corresponding to this dominant eigenvalue embodies the symmetry of the pairing. Fig.~\ref{fig:odlro_2d_main}(b) plots this dominant eigenvector, showing that its phase accumulates by $+2\pi$ upon encircling the origin in momentum space counterclockwise.  This behavior is the hallmark of a $p+ip$ state and directly confirms the predicted chiral nature of the superconductivity.

In the special case of two electrons ($N_e=2$), $\rho^{(2)}$ can be obtained analytically, providing further insight into the pairing structure.  In momentum space, it takes the form
\begin{align}
    \rho^{(2)}_{\bm{k}_1,\bm{k}_{2}}=\frac{1}{(2\pi)^4Z^2}k_{1,-}k_{2,+}e^{-\alpha (\bm{k}_1^2+\bm{k}_{2}^2)}.\label{app:eq:odlro_2e_2d}
\end{align}
This is manifestly a rank-1 matrix, which has only a single non-zero eigenvalue.  This finding is consistent with our numerical results in Fig.~\ref{fig:2d_odlro_sup}(a). To understand its spatial structure, we Fourier transform Eq.~\ref{app:eq:odlro_2e_2d} to real space
\begin{align}
    \rho^{(2)}_{(\bm{r}_1,\bm{r}_2),(\bm{r}_3,\bm{r}_4)}
    &=\frac{1}{(2\pi)^4Z^2}\int_{|\bm{k}_i|\leq k_b}\frac{d^2\bm{k}_1d^2\bm{k}_2 }{(2\pi)^4} e^{-i(\bm{k}_1\cdot(\bm{r}_1 -\bm{r}_2) - \bm{k}_2\cdot(\bm{r}_3-\bm{r}_4))} k_{1,-}k_{2,+}e^{-\alpha (\bm{k}_1^2+\bm{k}_{2}^2)}.
\end{align}
With $\delta r_1=r_1-r_2,\,\delta r_2=r_3-r_4$, the above equation reduces to
\begin{align}
    \rho^{(2)}_{\delta r_1,\delta r_2}&=\frac{1}{(2\pi)^8Z^2}\int_0^{k_b} k_1dk_1\int_0^{2\pi}d\phi_{k_1}\int_0^{k_b}k_2dk_2\int_0^{2\pi}d\phi_{k_2} e^{-ik_1\delta r_1\cos(\phi_{k_1}-\phi_{\delta r_1}) + ik_2\delta r_2\cos{(\phi_{k_2}-\phi_{\delta r_2})}} k_{1,-}k_{2,+}e^{-\alpha (k_1^2+k_{2}^2)}\nonumber\\
    &=\frac{1}{(2\pi)^8 Z^2} \left( -2\pi i e^{-i\phi_{\delta r_1}} \int_0^{k_b} k_1^2 e^{-\alpha k_1^2} J_1(k_1 \delta r_1) dk_1 \right) \left( 2\pi i e^{i\phi_{\delta r_2}} \int_0^{k_b} k_2^2 e^{-\alpha k_2^2} J_1(k_2 \delta r_2) dk_2 \right)\nonumber\\
    &=\frac{e^{i(\phi_{\delta r_2}-\phi_{\delta r_1})}}{(2\pi)^6 Z^2} \left(   \int_0^{k_b} k_1^2 e^{-\alpha k_1^2} J_1(k_1 \delta r_1) dk_1 \right) \left(  \int_0^{k_b} k_2^2 e^{-\alpha k_2^2} J_1(k_2 \delta r_2) dk_2 \right)\nonumber\\
    &=\frac{e^{i(\phi_{\delta r_2}-\phi_{\delta r_1})}}{(2\pi)^6 Z^2} I(\delta r_1)  I(\delta r_2).\label{app:eq:odlro_2e_analytical} 
\end{align}
The function $I(\delta r_1)  I(\delta r_2)$, which controls the spatial decay of the correlations, is plotted in Fig.~\ref{fig:2d_odlro_sup}(b). $\rho^{(2)}_{\delta r_1,\delta r_2}$ remains finite for $\delta r_1,\delta r_2\sim k_b^{-1}$.

